\pdfoutput=1
\documentclass[a4paper,11pt]{article}
\usepackage{amsmath,amssymb,amsfonts}
\usepackage[multiple,stable]{footmisc}
\usepackage[amsmath,amsthm,thmmarks]{ntheorem}
\allowdisplaybreaks[4]
\usepackage[noconfig]{refstyle}
\usepackage[dvipsnames]{xcolor}
\usepackage[hyperfootnotes=false, linktocpage=true, colorlinks, citecolor=blue, linkcolor=blue, urlcolor=Maroon]{hyperref}
\usepackage{array,tabularx,booktabs,geometry,subfigure,graphicx,longtable}
\usepackage[bf]{caption}
\usepackage{tikz}
\usepackage[all]{xy}
\usetikzlibrary{shapes,arrows}
\tikzstyle{block} = [rectangle, draw, text width=7em, text centered, rounded corners, minimum height=3em]
\usepackage{graphicx,color}
\usepackage{cite}
\usepackage{tikz}
\usepackage{tikz-cd}
\usetikzlibrary{cd}
\usepackage{makecell}

\let\eqref=\relax
\newref{eq}{name={eq.~},Name={Eq.~},names={eqs.~},Names={Eqs.~},rngtxt={-},refcmd=(\ref{#1})}
\newref{f}{name={footnote~},Name={Footnote~},names={footnotes~},Names={Footnotes~}}
\newref{app}{name={appendix~},Name={Appendix~},names={appendixes~},Names={Appendixes~}}
\newref{tab}{name={table~},Name={Table~},names={tables~},Names={Tables~}}
\newref{ch}{name={chapter~},Name={Chapter~},names={chapters~},Names={Chapters~}}
\newref{sec}{name={section~},Name={Section~},names={sections~},Names={Sections~}}
\newref{fig}{name={figure~},Name={Figure~},names={figures~},Names={Figures~}}
\numberwithin{equation}{section}
\newcommand{\eref}[1]{(\ref{#1})}

\newcommand{\ba}{\begin{array}}
\newcommand{\ea}{\end{array}}

\newcommand{\be}{\begin{equation}}
\newcommand{\ee}{\end{equation}}
\newcommand{\bea}{\begin{equation}\begin{aligned}}	
\newcommand{\eea}{\end{aligned}\end{equation}}		

\newcommand{\iddots}{\mathinner{\mkern2mu\raise1pt\hbox{.}\mkern2mu \raise4pt\hbox{.}\mkern2mu\raise7pt\hbox{.}\mkern1mu}}

\providecommand{\id}{\leavevmode\hbox{\small$\mathrm{1}$\kern-3.8pt\normalsize$\mathrm{1}$}}
\def\fnote#1#2{\begingroup\def\thefootnote{#1}\footnote{#2}
     \addtocounter{footnote}{-1}\endgroup}

\begin{document}

\vspace{1cm}

\title{
       \vskip 40pt
       {\huge \bf Yukawa Textures From Singular Spectral Data}}

\vspace{2cm}

\author{ Mohsen Karkheiran${}^{1}$}
\date{}
\maketitle
\begin{center} {\small${}^1$Center for Theoretical Physics of the Universe,
Institute for Basic Science, Daejeon 34051, South Korea}\\
\fnote{}{mohsenkar@ibs.re.kr}
\end{center}

\begin{abstract}
\noindent
The Yukawa textures of effective heterotic models are studied by using singular spectral data. One advantage of this approach is that it is possible to dissect the cohomologies of the bundles into smaller parts and identify the pieces that contain the zero modes, which can potentially have non-zero Yukawa couplings. Another advantage is the manifest relationship between the Yukawa textures in heterotic models and local F-theory models in terms of fields living in bulk or localized inside the 7-branes. We only work with Weierstrass elliptically fibered Calabi-Yau manifolds here.  The idea for generalizing this approach to every elliptically fibered Calabi-Yau with rational sections is given at the end of this paper. 
\end{abstract}

\thispagestyle{empty}
\setcounter{page}{0}
\newpage

\tableofcontents

\section{Introduction And Summary} \label{sec1}
One of the main goals of string phenomenology is to derive the standard model of particle physics in the low energy limit. Compactification of the heterotic string over Calabi-Yau manifolds from ten to four dimensions seems promising. In such models, the result will be an $\mathcal{N}=1$ super-GUT model in four dimensions with gauge groups $E_6$, $SO(10)$ and $SU(5)$. These models can be further broken to standard model gauge groups by using Wilson lines. For example see \cite{Braun:2005bw}. However, one of the significant challenges is the Yukawa couplings of the effective theory. The exact value of these couplings is tough to compute in heterotic string models. The reason is that one needs both to normalize the kinetic term of the four-dimensional action and, at the same time, the value of the holomorphic coupling in the superpotential. The main goal of this paper is to facilitate the computations. We will introduce new ways to compute the Yukawa textures, and we believe it can be further generalized to compute the Yukawa couplings. The secondary goal of this paper is to show the relationship between the Yukawa couplings in heterotic effective theory and the Yukawa couplings that can be derived from the dual F-theory. The reason that we are interested in this issue is that in other string theory models, there are always intersection branes, and there are fields that live over branes and the brane intersections. In such models, the appearance of Yukawa couplings in the effective theory is at least more intuitive. However, in perturbative heterotic models, there are no branes, and naively it seems there is no connection between the Yukawa couplings in intersection brane models and the Yukawa couplings in heterotic models. The approach that we choose to compute the Yukawa couplings makes this connection explicit. We hope we can use this to study the vanishing theorems \cite{Anderson:2021unr} and unification of Yukawa couplings in future works\cite{Buchbinder:2016jqr}.

Generally, the Yukawa couplings appear from a Chern-Simons term in the ten-dimensional action \cite{Green:1987mn}
\begin{eqnarray}\label{Chern-simons}
\int_X \Omega \wedge Tr(A\wedge A\wedge A),
\end{eqnarray}
where $X$ is the Calabi-Yau manifold, $\Omega$ is the holomorphic top form of $X$, and $A$ is the background gauge field living inside $X$. The fluctuations of $A$ around this background value appear as four-dimensional chiral multiplets, which are charged under the effective gauge group. Therefore these chiral multiplets are identified with the zero modes of the Dirac operator in $X$, which can be identified with the elements of the cohomologies of the vector bundle associated with the gauge field $A$, i.e., $H^*(V)$, $H^*(\Lambda^kV)$, $H^*(V^*\otimes V)$, etc. 

As an example consider $E_6$ models, there are chiral and anti-chiral fields in representations \textbf{27} and $\overline{\textbf{27}}$ (for a detailed explanation see \cite{Anderson:2008ex} and references there). So there are \textbf{27-27-27}, \textbf{$\overline{\textbf{27}}$-$\overline{\textbf{27}}$-$\overline{\textbf{27}}$}, and \textbf{1-27-$\overline{\textbf{27}}$} couplings. How they can be computed in terms of cohomologies? In principle one can use the cup product between the cohomologies to produce cohomologies of other bundles \cite{Anderson:2010vdj,Anderson:2010tc}. For example for a $SU(3)$ bundle
\begin{eqnarray}
H^1(V) \otimes H^1(V) \otimes H^1(V) \longrightarrow H^3(\Lambda^3V)\simeq H^3(\mathcal{O}_X)\simeq\mathbb{X}.
\end{eqnarray}
Elements of $H^1(V)$ correspond to the \textbf{27} chiral multiplets. If certain elements of $H^1(V)$ are in the image of the above homomorphism between cohomology groups, one can say the corresponding fluctuations of the gauge field can be ``multiplied" to form a $(0,3)$ form which can be contracted with $\Omega$ in the Chern-Simons action \eref{Chern-simons} to give a non-zero Yukawa coupling in four dimensions. Similarly the non-zero homomorphism
\begin{eqnarray}
H^1(V^*)\otimes H^1(V^*)\otimes H^1(V^*)\longrightarrow H^3(\Lambda^3V^*)\simeq \mathbb{C},
\end{eqnarray}
corresponds to non-zero \textbf{$\overline{\textbf{27}}$-$\overline{\textbf{27}}$-$\overline{\textbf{27}}$}, and 
\begin{eqnarray}
Ext^1(\mathcal{O}_X,V) \otimes Ext^1(V,V)\otimes Ext^1(V,\mathcal{O}_X)\longrightarrow Ext^3(\mathcal{O}_X,\mathcal{O}_X)\simeq \mathbb{C},
\end{eqnarray}
corresponds to \textbf{1-27-$\overline{\textbf{27}}$}. For future reference, let us mention the couplings for the $SO(10)$, and $SU(5)$ super-GUT models,
\begin{eqnarray}
\arraycolsep=1.4pt\def\arraystretch{2.2}
\begin{array}{ccccccccc}
H^1(V^*) &\otimes& H^1(V^*) &\otimes& H^1(\Lambda^2V^*)&\longrightarrow& \mathbb{C},& \textbf{16-16-10}, & SU(4), \\
H^1(V^*)&\otimes& H^1(V^*)&\otimes& H^1(\Lambda^3V^*)&\longrightarrow& \mathbb{C},& \textbf{10-10-5}, & SU(5), \\
H^1(V^*)&\otimes& H^1(\Lambda^2V^*)&\otimes& H^1(\Lambda^2V^*)&\longrightarrow& \mathbb{C},& \textbf{10-$\overline{\textbf{5}}$-$\overline{\textbf{5}}$}, & SU(5).
\end{array}
\end{eqnarray}
Generally, to compute the Yukawa couplings, these cohomologies can be represented by polynomials. Then it can be checked directly whether these polynomials can ``contract" to form an element of $H^3(\mathcal{O}_X)= H^0(\mathcal{O}_X)$ \cite{Anderson:2010vdj,Anderson:2010tc}. This type of calculation has been used recently to carefully study the Yukawa couplings in the heterotic line bundle models\cite{Gray:2019tzn}, which is already very complicated, needless to mention how tough it can be for more general bundles. 

As mentioned before, the primary goal of this paper is to give another approach to compute the Yukawa couplings. We do not claim that this approach makes the calculations much easier. However, at least, we can dissect the procedure and give smaller and simpler pieces that the chiral multiplets with potential non-zero Yukawa couplings can live there. Namely, instead of computing the cohomology of the bundle, which can be very messy, we give cohomologies of simpler sheaves like line bundles living on a curve or surface, such that the elements of these groups can correspond to chiral fields that have non-zero Yukawa couplings. However, the misery is preserved. One needs to check many ``coboundary maps" to see whether the elements of these groups contribute to the cohomologies of $V$ and the associated bundles. Nevertheless, as mentioned before, at least the procedure is broken into smaller and simpler pieces, and the connection to the intersecting brane models becomes transparent.
 
The setup that we are mainly going to use is $(X,V)$. Where $X$ is a Weierstrass elliptically fibered Calabi-Yau threefold. In other words, it is Calabi-Yau embedded in a $\mathbb{P}^{231}$ fibration over a surface (denoted by $B$) $\pi: \mathcal{A}\longrightarrow B$ with an algebraic equation
\begin{eqnarray}
Y^2+X^3+F X Z^4+G Z^6=0,
\end{eqnarray}
where $(X,Y,Z)$ are coordinates of $\mathbb{P}^{231}$ with degrees $(2,3,1)$. The elliptic fibration $\pi: X\longrightarrow B$ has a section denoted by $\sigma$
\begin{eqnarray}
i_{\sigma}: \sigma \hookrightarrow X,
\end{eqnarray}
such that $\pi \circ i_{\sigma}=id_B$. The vector bundle $V$ is supposed to be a stable degree zero holomorphic bundle over $X$. Such kind of bundles can be constructed by spectral data which introduced by Friedman, Morgan and Witten in \cite{Friedman:1997yq,Friedman:1997ih}.\footnote{When the spectral cover is reducible it is not guaranteed that the corresponding bundle is stable, and it must be checked case by case.} Spectral data is given by a doublet $(\mathcal{S}_n, \mathcal{L}_n)$. $\mathcal{S}_n\subset X$ is a finite cover of the base manifold $B$ with degree $n$ which is embedded inside $X$, and $\mathcal{L}_n$ is line bundle on $\mathcal{S}_n$.\footnote{$n$ is the rank of $V$.} We will use these spectral data to compute the Yukawa couplings. 

\begin{figure}
    \centering
    \includegraphics[width=0.618\textwidth]{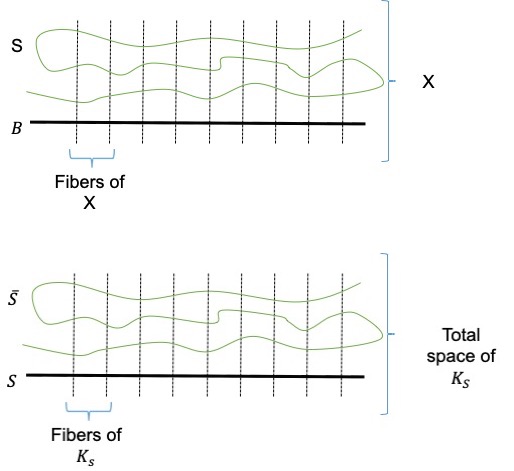}
    \caption{\textbf{Top:} Spectral cover embedded in elliptically fibered Calabi-Yau (compact). \textbf{Down:} Spectral cover embedded in the total space of canonical bundle (non-compact). The neighborhood of the zero section in $X$ is mapped to the total space of the canonical bundle \cite{Hayashi:2008ba}.}
    \label{fig:HitVSHet}
\end{figure}

On the other hand, local F-theory models can be defined in terms of intersection 7-branes, and Higgs bundles living over these branes \cite{Beasley:2008dc,Donagi:2009ra}. The Higgs bundles are defined by a doublet $(E,\Phi)$ where $E$ is ``flux" inside the 7-brane and $\Phi$ is 2-form with values in $ad(E)$. The Higgs bundle can also be defined by spectral data $(\overline{\mathcal{S}}_n,\overline{\mathcal{L}}_n)$. The spectral cover (embedded in the total space of the canonical bundle) is defined as the characteristic polynomial of $\Phi$, and is a finite cover of the 7-brane. $\overline{\mathcal{L}}_n$ is a line bundle over the spectral cover. These two spectral covers in heterotic string theory and local F-theory are closely related \cite{Donagi:2008kj,Donagi:2008ca}. Therefore with our approach, the Yukawa couplings in local F-theory models can also be calculated (see Figure \ref{fig:HitVSHet}). We are not going to study this side completely, but we will comment on it. 

We summarize the main results in the following tables.

\begin{table}[h!]
    \centering
    \begin{tabular}{||c | c |c ||}
        \hline
        \thead{Spectral Cover}     & \thead{Zero Modes}  & \thead{\textit{Possible} Yukawa Couplings} \\[6pt]
        \hline
        $S_3=a_3Y+a_2 XZ+a_0Z^3$ & \textbf{27} on $a_3=0$ & No Yukawa couplings\\[6pt]
        \hline
        $S_3=Z(a_2X+a_0Z^2)$ & 
           \makecell{\textbf{27} Bulk\\\textbf{27} $a_2=0$} & 
        \textbf{27-27-27} on $a_2=0$ \\[6pt]
        \hline
    \end{tabular}
    \caption{$E_6$ models Yukawa Couplings}
    \label{tab:SU(3)}
\end{table}
\begin{table}[h!]
    \centering
    \begin{tabular}{||c | c| c ||}
      \hline
        \thead{Spectral Cover}     & \thead{Zero Modes}  & \thead{\textit{Possible} Yukawa Couplings} \\[6pt]
        \hline
        $S_4=a_4X^2+a_3 YZ+a_2X Z^2+a_0 Z^4$ & \makecell{\textbf{16} $a_4=0$\\ \textbf{10} $a_3=0$} & \textbf{16-16-19} on $a_3=a_4=0$\\[6pt]
        \hline
        $S_4=Z (a_3Y+a_2XZ+a_0Z^3)$& \makecell{\textbf{16} Bulk \\ \textbf{16} $a_3=0$\\ \textbf{10} $a_3=0$} & \textbf{16-16-10} on $a_3=0$ \\[6pt]
        \hline
        $S_4=(a_2X+a_0Z^2)(b_2X+b_0Z^2)$ &\makecell{\textbf{16} $a_2=0$\\\textbf{16} $b_2=0$ \\\textbf{10} $a_2 b_0-a_0 b_2=0$} & \textbf{16-16-10} on $a_2=b_2=0$ \\[6pt]
        \hline
        $S_4=Z^2(a_2X+a_0Z^2)$& \makecell{\textbf{16} Bulk\\\textbf{16} Bulk\\\textbf{10} Bulk}& \textbf{16-16-10} Bulk \\[6pt] 
        \hline 
        $S_4=(a_2X+a_0Z^2)^2$& \makecell{\textbf{16} $a_2=0$\\\textbf{10} Bulk}& 3$\times$\textbf{16-16-10} on $a_2=0$\\[6pt] 
        \hline
    \end{tabular}
    \caption{$SO(10)$ models possible Yukawa couplings. The middle column corresponds to the zero modes that can have non-zero Yukawa couplings. Note that for the last case there are three different possibilities.}
    \label{tab:SU(4)}
\end{table}
\begin{table}[h!]
    \centering
    \begin{tabular}{||c|c|c||}
        \hline
        \thead{Spectral Cover}     & \thead{Zero Modes}  & \thead{\textit{Possible} Yukawa Couplings} \\[6pt]
        \hline
        \makecell{$S_5=$\\$a_5XY+a_4X^2Z+a_3YZ^2+a_2XZ^3+a_0Z^5$}& \makecell{\textbf{10} $a_5=0$\\\textbf{5}/$\overline{\textbf{5}}$\\ $a_4a_3^2-a_2a_3a_5+a_0a_5^2=0$} & \makecell{\textbf{10-10-5}/\textbf{10-$\overline{\textbf{5}}$-$\overline{\textbf{5}}$} on \\$a_5=a_4=0$,\\ $a_5=a_3=0$} \\[6pt] 
        \hline
        $S_5=Z(a_4X^2+a_3YZ+a_2 XZ^2+a_0Z^4)$& \makecell{\textbf{10} Bulk\\\textbf{10}  $a_4=0$\\\textbf{5}/$\overline{\textbf{5}}$ $a_4=0$\\\textbf{5}/$\overline{\textbf{5}}$  $a_3=0$} & \makecell{\textbf{10-10-5} on $a_4=0$\\\textbf{10-$\overline{\textbf{5}}$-$\overline{\textbf{5}}$} on $a_4=a_3=0$}\\[6pt]
        \hline
        $S_5=Z(a_2X+a_0Z^2)(b_2X+b_0Z^2)$ & \makecell{\textbf{10} Bulk\\\textbf{10} $a_2=0$\\ \textbf{10} $b_2=0$ \\ \textbf{5}/$\overline{\textbf{5}}$ $a_2=0$\\\textbf{5}/$\overline{\textbf{5}}$ $b_2=0$\\\textbf{5}/$\overline{\textbf{5}}$ $a_2b_0-a_0b_2=0$} & \makecell{\textbf{10-10-5} on\\ $a_2=0$ and $b_2=0$\\ \textbf{10-$\overline{\textbf{5}}$-$\overline{\textbf{5}}$} on\\ $a_2=b_2=0$} \\[6pt]
        \hline
        $S_5=(a_2X+a_0Z^2)(b_3Y+b_2XZ+b_0Z^3)$ & \makecell{\textbf{10} $a_2=0$\\\textbf{10} $b_3=0$ \\\textbf{5}/$\overline{\textbf{5}}$ $b_3=0$ \\\textbf{5}/$\overline{\textbf{5}}$ $c_{\Lambda^2V_5}$ \ref{sec332}} & \makecell{\textbf{10-10-5} on $a_2=b_3=0$ \\\textbf{10-5-$\overline{\textbf{5}}$} on\\ $c_{\Lambda^2V_5} \cap \lbrace b_3=0\rbrace$} \\[6pt]
        \hline
        $S_5=Z(a_2X+a_0Z^2)^2$& \makecell{\textbf{10} Bulk\\ \textbf{10} $a_2=0$\\\textbf{5}/$\overline{\textbf{5}}$ Bulk \\\textbf{5}/$\overline{\textbf{5}}$ $a_2=0$ } & \makecell{\textbf{10-10-5} on $a_2=0$ \\ 
        2$\times$\textbf{10-$\overline{\textbf{5}}$-$\overline{\textbf{5}}$} Bulk}\\[6pt]
        \hline
        $S_5=Z^2 (a_3Y+a_2XZ+a_0Z^3)$ &\makecell{\textbf{10} Bulk\\\textbf{10} $a_3=0$\\ \textbf{5}/$\overline{\textbf{5}}$ Bulk\\
        \textbf{5}/$\overline{\textbf{5}}$ $a_3=0$} & \makecell{\textbf{10-10-5} on $a_3=0$\\ \textbf{10-$\overline{\textbf{5}}$-$\overline{\textbf{5}}$} on $a_3=0$} \\[6pt]
        \hline
        $S_5=Z^3(a_2X+a_0Z^2)$&\makecell{\textbf{10} Bulk \\\textbf{5}/$\overline{\textbf{5}}$ Bulk} & \makecell{\textbf{10-10-5}  Bulk\\ \textbf{3$\times$10-$\overline{\textbf{5}}$-$\overline{\textbf{5}}$} Bulk} \\[6pt]
        \hline
    \end{tabular}
    \caption{$SU(5)$ models possible Yukawa couplings. The middle column corresponds to the zero modes that can have non-zero Yukawa couplings.}
    \label{tab:}
\end{table}
When there are vertical components, in addition to the Yukawa couplings that can emerge from the other components, there are possible extra contributions to the Yukawa couplings. However, for the $SU(5)$ models there are no new \textbf{10-$\overline{\textbf{5}}$-$\overline{\textbf{5}}$} couplings from the vertical components.

This paper is organized as follows. In section 2, we review and prove useful identities that will be used extensively in later sections. Then, we explain how to compute the Pontrjagin product for a simple degree two spectral covers. Section 3 and 4 contain the main body of calculations. In section 3, we focus only on the spectral data without vertical components. We also comment on the dual F-theory models. However, as mentioned before, our purpose is not to study the F-theory dual in detail. In section 4, we consider the effect of adding a smooth vertical component. Section 5 gives toy examples for $E_6$ and $SU(5)$ models with detailed calculations. In section 6, we will discuss how one can extend these techniques to Calabi-Yau manifolds without holomorphic sections. We also collect some useful terminologies and calculations in the appendix. 
\section{General Analysis} \label{sec2}
We start with a brief review of Fourier-Mukai transforms and spectral cover construction of stable degree zero holomorphic vector bundles. For more information, see \cite{BBRH}. Fourier-Mukai is a functor which is an auto equivalence of derived category of coherent sheaves on X,
\begin{eqnarray}
\Phi : D^b(X) \longrightarrow D^b(X).
\end{eqnarray}
It is shown that a Fourier-Mukai functor can be represented by an integral transform,
\begin{eqnarray}
\begin{tikzcd}[ampersand replacement=\&, column sep=small]
 \&  X\times_{B}X\arrow[dl,"\pi_1"']\arrow[dr,"\pi_2"]  \& \\
 X \& \& X
\end{tikzcd}\\
\phi(-) = R\pi_{2*} \left(\pi_1^*- \otimes \mathcal{P} \right),
\end{eqnarray}
where $\mathcal{P}$ is the kernel of the integral transform, and it must satisfy certain conditions in order to $\Phi$ be an auto-equivalence. For the purposes of this paper, we restrict $\mathcal{P}$ to the Poincare sheaf,
\begin{eqnarray}
\mathcal{P}=\mathcal{I}_{\Delta} \otimes \pi_1^* \mathcal{O}(\sigma)\otimes \pi_2^* \mathcal{O}(\sigma)\otimes  \rho^* K_B^{-1}.
\end{eqnarray}
When $V$ is a stable degree zero bundle the Fourier-Mukai transform of $V$ is of the following form,\footnote{Even when the restriction of $V$ on generic elliptic fiber is semistable and degree zero, the Fourier-Mukai transform of $V$ has this general form. But for physical models this is not enough.}
\begin{eqnarray}
\Phi(V) = i_{S*} \mathcal{L}[-1],
\end{eqnarray}
Where $S$ is a finite cover of the base i.e., spectral cover,
\begin{eqnarray}
\begin{tikzcd}[ampersand replacement=\&, column sep=small]
\pi: S \arrow[r,"n:1"] \& B,
\end{tikzcd}
\end{eqnarray}
and the sheaf $\mathcal{L}$ is called the spectral sheaf, which is a polarized rank one sheaf supported on $S$. Note that $S$ can be singular or even non-reduced. A more intuitive way to describe the Fourier-Mukai transform is as follows. When $V$ (which is stable, holomorphic and degree zero) restricted to a generic fiber (say $E$), it is equivalent to\footnote{Yo be precise S-equivalent \cite{Friedman}.}   
\begin{eqnarray}
V|_{E}\simeq \mathcal{O}_E(\sigma-p_1)\oplus \dots \oplus \mathcal{O}_E(\sigma-p_n),
\end{eqnarray}
where $\sigma$ is the zero section, and $p_n$ are points on $E$ \cite{atiyah}. Spectral cover is simply the surface which is made by the points $p_n$ as one mover over the base.

Our purpose in this paper is to determine the Yukawa couplings in the effective theory from the spectral data. Remember if the map,
\begin{eqnarray}
H^1(V) \otimes H^1(V) \longrightarrow H^2(\Lambda^2V)
\end{eqnarray}
is non-zero then the following Yukawa coupling is non-zero,
\begin{eqnarray}
H^1(V)\otimes H^1(V)\otimes H^1(\Lambda^2V^*) \longrightarrow \mathbb{C}.
\end{eqnarray}
So our main task is finding the cohomologies $H^*(V)$ and $H^*(\Lambda^2V)$ in terms of the spectral data. For smooth spectral cover $S$ this is already have been done \cite{Hayashi:2008ba,Hayashi:2009ge}. The first novelty in this paper is finding the cohomologies for the singular spectral covers. We will see in the future that for singular $S$ there are possibilities for non-vanishing Yukawa couplings. This depends on the behaviour of the spectral sheaves. 

To find $H^*(V)$ we can use the Leray spectral sequence, 
\begin{eqnarray}
E_2^{p,q} = H^p(R^q\pi_*V) \Rightarrow H^{p+q} (V).
\end{eqnarray}
So we should find $R^q\pi_* V$. This can be done using the following commutative diagram,
\begin{eqnarray}
\begin{tikzcd}[ampersand replacement=\&, column sep=small]
 \& X\times_B X \arrow[dl, "\pi_1"]\arrow[dr,"\pi_2"] \&X\times_B \sigma \arrow[l,"\Tilde{i}_{\sigma}"]\arrow[dll,"\Tilde{\pi}_1"]\arrow[dr,"\Tilde{\pi}"] \& \\
 X \& \& X \& \sigma \arrow[l,"i_{\sigma}"]
\end{tikzcd}
\end{eqnarray}
\begin{eqnarray}
R\pi_*V &=& R \Tilde{\pi}_* \left(\tilde{\pi}_1^*V \otimes \mathcal{O}_{\sigma} \right)\nonumber \\
 &=& R \Tilde{\pi}_* \left(\tilde{\pi}_1^*V \otimes L \tilde{i}_{\sigma}^* \mathcal{P} \right)\nonumber \\
 &=& R \Tilde{\pi}_* \circ L \tilde{i}_{\sigma}^* \left( \pi_1^* V \otimes \mathcal{P} \right)\nonumber \\
 &=& Li_{\sigma}^* R\pi_{2*} \left( \pi_1^* V \otimes \mathcal{P} \right)\nonumber \\
 &=& Li_{\sigma}^* \Phi(V) = Li_{\sigma}^* \left( i_{S*}\mathcal{L}\right)[-1],
\end{eqnarray}
where we have used the flattness of the map $\tilde{\pi}$ which is the consequence of the flattness of $\pi$. This is of course not a new result \cite{Donagi:2004ia}.  

To compute $H^*(\Lambda^2V)$ we need to know, as before, the derived pushforward $R\pi_* \Lambda^2V$. For this, we use a nice property of Fourier-Mukai transforms (defined via Poincare sheaf as the kernel) which exchanges the tensor product with Pontrjagin product. More concretely, consider a single smooth elliptic curve $E$. Then there is a morphism $m$ which corresponds to the abelian group action on the elliptic curve \cite{BBRH}, 
\begin{eqnarray}
\begin{tikzcd}[ampersand replacement=\&, column sep=small]
 \& E\times E\arrow[dl,"\pi_1"']\arrow[dr,"\pi_2"]\arrow[r,"m"] \& E \\
 E \& \& E
\end{tikzcd}\\
m(p,q) = p+q.
\end{eqnarray}
Then the Pontrjagin product is defined as the derived pushforward of $m$,
\begin{eqnarray}
V \ast W := R m_* (\pi_1^* V \otimes \pi_2^* W),
\end{eqnarray}
where $V$ and $W$ are sheaves over the elliptic curve $E$. Over $E$, Fourier-Mukai functor satisfy the following identity
\begin{eqnarray}
\Phi(V) \ast \Phi(W) = \Phi(V\otimes W) [-1].
\end{eqnarray}
One needs a similar identity over the elliptically fibered $X$. However, the morphism $m$ is not well defined for the total fibration. Because there are singular fibers, and the group action is not well defined when the points $p$ and $q$ correspond to the singular point of the elliptic fiber. Therefore if the spectral covers of $V$ and $W$ hit the singular point this identity will not work.

The good news is that we need the spectral cover of the antisymmetric product. So the addition of the singular point with itself doesn't contribute. In addition, for the cohomology of $\Lambda^2V$, only the restriction of the $\Phi(\Lambda^2V)$ on the zero section $\sigma$ contributes. So even if the spectral cover of $V$ hits the singular points of the fibers, they do not contribute in the $H^*(\Lambda^2V)$,\footnote{There is one caveat in this argument when the spectral cover has non-reduced vertical fiber. But again, we are only care about the restriction of the $\Phi(\Lambda^2V)$ to the zero section. So it still fine to use this formula.}  
\begin{eqnarray}
Li_{\sigma}^* \Phi(\Lambda^2V) &=& Li_{\sigma}^* \left(\Phi(V)\ast_A \Phi(V) \otimes K_{B}^{-1} [+1]\right)\nonumber\\
&=& Li_{\sigma}^* \left((i_{S*}\mathcal{L})\ast_A (i_{S*}\mathcal{L}) \otimes K_{B}^{-1} [-1]\right),
\end{eqnarray}
where by $\ast_A$ we mean ''antisymmetrized" version of the Pontrjagin product which hopefully will be more clear in the examples. 

To prove this we use the following commutative diagrams,\footnote{Here $\hat{X}$ is the Jacobian fibration, which is isomorphic to $X$.}
\begin{eqnarray}
\begin{tikzcd}[ampersand replacement=\&, column sep=small]
 \& \& X\times_B \hat{X}\times_B \hat{X} \arrow[dl,"1\times m"]\arrow[dr,"\pi_{23}"] \& \\
 \& X\times_B \hat{X} \arrow[dl,"\pi_X"]\arrow[dr,"\pi_{\hat{X}}"] \& \& \hat{X}\times_B\hat{X}\arrow[dl,"m"] \\
 X \& \& \hat{X}\& 
\end{tikzcd}
\end{eqnarray}
From here we can see,
\begin{eqnarray}
Rm_* \circ R\pi_{23*} \simeq R \pi_{\hat{X}*} \circ R (1\times m)_* .
\end{eqnarray}
The following diagram is also useful,
\begin{eqnarray}
\begin{tikzcd}[ampersand replacement=\&, column sep=small]
 \& \& X\times_B \hat{X}\times_B \hat{X} \arrow[dl,"1\times \tilde{\pi}_1"]\arrow[dr,"\pi_{23}"] \& \& \\
 \& X\times_B \hat{X} \arrow[dl,"\pi_X"]\arrow[dr,"\pi_{\hat{X}}"] \& \& \hat{X}\times_B\hat{X}\arrow[dl,"\tilde{\pi}_1"]\arrow[dr,"\tilde{\pi}_2"]\& \\
 X \& \& \hat{X}\& \& \hat{X}
\end{tikzcd}
\end{eqnarray}
Form here one can see,
\begin{eqnarray}
\tilde{\pi}_1^*\circ R\pi_{\hat{X}*} \simeq R\pi_{23*} \circ (1\times \tilde{\pi}_1)^*.
\end{eqnarray}
Therefore, 
\begin{eqnarray}
\tilde{\pi}_1^* \Phi(V) =  R\pi_{23*} \circ (1\times \tilde{\pi}_1)^* \left( \pi_X^* V \otimes \mathcal{P}\right).
\end{eqnarray}

So now let us compute $\Phi(V)\ast \Phi(W)$,
\begin{eqnarray}
\tilde{\pi}_1^* \Phi(V) \otimes \tilde{\pi}_2^* \Phi(W) &=& R\pi_{23*} \circ (1\times \tilde{\pi}_1)^* \left( \pi_X^* V \otimes \mathcal{P}\right) \otimes \tilde{\pi}_2^* \Phi(W)\nonumber \\
&=& R\pi_{23*} \left((1\times \tilde{\pi}_1)^* \left( \pi_X^* V \otimes \mathcal{P}\right) \otimes \pi_{23}^*\tilde{\pi}_2^* \Phi(W)  \right),
\end{eqnarray}
hence,
\begin{eqnarray}
Rm_* \left( \tilde{\pi}_1^* \Phi(V) \otimes \tilde{\pi}_2^* \Phi(W) \right) &=& Rm_* \circ R\pi_{23*} \left((1\times \tilde{\pi}_1)^* \left( \pi_X^* V \otimes \mathcal{P}\right) \otimes \pi_{23}^*\tilde{\pi}_2^* \Phi(W)  \right) \nonumber \\
&=& R \pi_{\hat{X}*} \circ R (1\times m)_* \left((1\times \tilde{\pi}_1)^* \left( \pi_X^* V \otimes \mathcal{P}\right) \otimes \pi_{23}^*\tilde{\pi}_2^* \Phi(W)  \right)\nonumber \\
&=& R \pi_{\hat{X}*} \left(\pi_X^* V\otimes R (1\times m)_* \left(\pi_{12}^*\mathcal{P}\otimes \pi_{23}^*\tilde{\pi}_2^* \Phi(W)  \right) \right).
\end{eqnarray}
Now there are two other relations that should be use,
\begin{eqnarray}
\pi_{23}^* \circ \tilde{\pi}_2^* \simeq \pi_{13}^* \circ \pi_{\hat{X}}^*, \\
\pi_{12}^* \mathcal{P} = (1\times m)^* \mathcal{P}\otimes \pi_{13}^* \mathcal{P},
\end{eqnarray}
where the first relation is coming from the simple relations between the projection morphisms, and the second one is true for the Poincare sheaf on any abelian variety. With the help of these two we simplify the last result above, 
\begin{eqnarray}
Rm_* \left( \tilde{\pi}_1^* \Phi(V) \otimes \tilde{\pi}_2^* \Phi(W) \right) &=& R \pi_{\hat{X}*} \left( \pi_X^* V \otimes \mathcal{P} \otimes R(1\times m)_* \pi_{13}^* (\mathcal{P}^* \otimes \pi_{\hat{X}}^* \Phi(W)) \right).
\end{eqnarray}
The following commutative diagram,
\begin{eqnarray}
\begin{tikzcd}[ampersand replacement=\&, column sep=small]
 \& X\times_B \hat{X}\times_B \hat{X} \arrow[dl,"\pi_{13}"]\arrow[dr,"1\times m"] \& \\
 X\times_B \hat{X}\arrow[dr,"\pi_X"] \& \& X \times_B \hat{X} \arrow[dl,"\pi_X"] \\
  \& X \&
\end{tikzcd}\nonumber \\
\Rightarrow R(1\times m)_* \circ \pi_{13}^* \simeq \pi_X^* \circ R\pi_{X*}.
\end{eqnarray}
Therefore,
\begin{eqnarray}
Rm_* \left( \tilde{\pi}_1^* \Phi(V) \otimes \tilde{\pi}_2^* \Phi(W) \right) &=& R \pi_{\hat{X}*} \left( \pi_X^* V \otimes \mathcal{P} \otimes pi_X^* \circ R\pi_{X*} (\mathcal{P}^* \otimes \pi_{\hat{X}}^* \Phi(W)) \right) \nonumber \\
&=& R \pi_{\hat{X}*} \left( \pi_X^* V \otimes \mathcal{P} \otimes \pi_X^* W \otimes  K_B[-1]\right), \nonumber
\end{eqnarray}
where in the last line we used the inverse Fourier-Mukai functor i.e., $R\pi_{X*} (\mathcal{P}^* \otimes \pi_{\hat{X}}^* \Phi(W)) = W \otimes K_B[-1]$. Then the final result is,
\begin{eqnarray}\label{tensorToPontrjiagin}
\Longrightarrow Rm_* \left( \tilde{\pi}_1^* \Phi(V) \otimes \tilde{\pi}_2^* \Phi(W) \right) = \Phi(V \otimes W) \otimes K_B[-1].
\end{eqnarray}
Of course the formula \eref{tensorToPontrjiagin} is wrong! But If we choose $W=V$ and and antisymmetrize the tensor product then it can be used for computing $R\pi_* \Lambda^2V$.

\subsection{Degree Two Spectral Cover}\label{sec21}
As a warm-up let us study the the spectral cover corresponding to $SU(2)$ bundles. Of course this theory doesn't have a Yukawa coupling, but it is useful for the later parts to work-out its Pontrjagin product.

Consider the smooth degree two spectral cover, 
\begin{eqnarray}
\Phi(V) &=& i_{S_2*}\mathcal{L}_2 [-1], \\
S_2 &=& a_2 X +a_0Z^2, 
\end{eqnarray}
where $a_2$ and $a_0$ are generic polynomials in the base $B_2$. Generally, we are interested in the cohomologies $H^*(V)$ and $H^*(\Lambda^2V)$. The former can be easily computed as, 
\begin{eqnarray}
H^i(V) = H^{i-1}(R^1\pi_*V) = H^{i-1} (\mathcal{L}_2|_{\sigma\cap S_2}).
\end{eqnarray}
For $H^*(\Lambda^2V)$, since $\Lambda^2V=\mathcal{O}_X$ (for any $SU(2)$ bundle) $H^i(\Lambda^2V)=H^i(\mathcal{O}_X)$. So in principle one doesn't need to use spectral data, however as mentioned before it is instructive to compute the Pontrjagin product. Since the antisymmetric Pontrjagin product is needed, one should ``multiply" the line bundles on the opposite sheets of the double cover to get a sheaf over the zero section. In other words, suppose the double cover intersects a generic fiber at two points $P_1$ and $P_2$, since the bundle is $SU(2)$ these points should ``add-up" to zero,
\begin{eqnarray}
P_1 \boxplus P_2 =0.
\end{eqnarray}
The antisymmetric Pontrjagin product basically take the sheaves supported over the points $P_1$ and $P_2$ and generates a sheaf over the ``zero section" $0$. Globally one can find this sheaf over the zero section as $Det(\pi_* i_{S_2*}\mathcal{L}_2)$. This is because the pushforward to the base is a direct sum of two line bundles corresponding to the two sheets of the double cover \cite{Friedman}, therefore the determinant gives the antisymmetrized product.\footnote{More precisely, the pushforward $\pi_*\mathcal{O}_{S_2} = \mathcal{O}_B\oplus \mathcal{O}_B (-\frac{R}{2})$ where $R$ is the branch locus. The first summand corresponds to the part of $\mathcal{O}_{S_2}$ which is even under the involution (exchange of the sheets) of the double cover $S_2$, and the second summand is the odd part.} So let us check this    
\begin{eqnarray}
i_{S_2*} \mathcal{L}_2\star_A i_{S_2*} \mathcal{L}_2 = i_{\sigma*}Det(\pi_* i_{S_2*}\mathcal{L}_2) \Longrightarrow \Phi(\Lambda^2V)= i_{\sigma*}Det(\pi_* i_{S_2*}\mathcal{L}_2) \otimes K_B^{-1}[-1].  
\end{eqnarray}
By the usual arguments one can show,
\begin{eqnarray}
R\pi_*\Lambda^2 V &=& Li_{\sigma}^* i_{\sigma*} Det(\pi_*i_{S_2*} \mathcal{L}_2)\otimes K_B^{-1}[-1]\\
 &=& Det(\pi_*i_{S_2*} \mathcal{L}_2) \otimes K_B^{-2} \oplus Det(\pi_*i_{S_2*} \mathcal{L}_2) \otimes K_B^{-1}[-1].
\end{eqnarray}
On the other hand it can be shown for a degree zero $SU(2)$ bundle one has the following relation \ref{Appendix},
\begin{eqnarray}
c_1(\pi_*i_{S_2*} \mathcal{L}_2 )= 2 c_1(K_B) \Longrightarrow  Det(\pi_*i_{S_2*} \mathcal{L}_2) = K_B^2.
\end{eqnarray}
Hence,
\begin{eqnarray}
R^0\pi_*\Lambda^2V &=& \mathcal{O}_B, \\
R^1\pi_*\Lambda^2V &=& K_B.
\end{eqnarray}
This is exactly the derived pushforward formula for the trivial bundle $\mathcal{O}_X$. Consistent with the expectation $\Lambda^2 V =\mathcal{O}_X$.

\section{Spectral Covers Without Vertical Components}\label{sec3}

In this section, the technology explained in the previous section is used to study more realistic models. For now let us restrict ourselves to spectral covers, either singular or not, that have not vertical components. Those cases are studied in the next section.
\subsection{$E_6$ GUT Models}\label{sec31}
\begin{figure}
    \centering
    \includegraphics[width=0.5\textwidth]{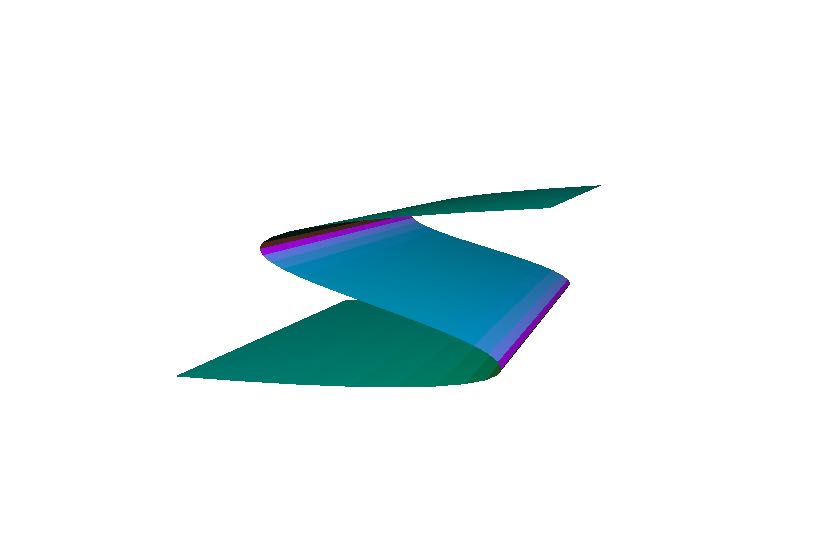}\includegraphics[width=0.5\textwidth]{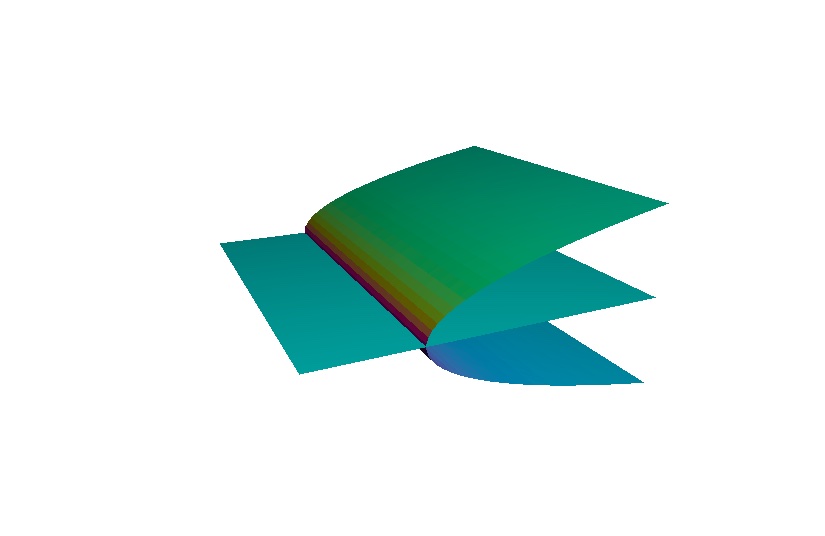}
    \caption{Left: Smooth degree three spectral cover. Right: Reducible degree three spectral cover, with one component isomorphic to the base manifold.}
    \label{tripleWOV}
\end{figure}

For $E_6$ GUT models one needs to put a $SU(3)$ bundle (denoted by $V$ as usual) in the internal space. Therefore the spectral cover in this case is a degree cover of the base surface. There only two possibility for the degree three spectral cover without vertical component, as shown in Figure \ref{tripleWOV}. 
\subsubsection{Smooth Spectral Cover}\label{sec311}
In this subsection we compute the cohomology of $V$ and $\Lambda^2V$ when spectral cover is given by,
\begin{eqnarray}
S_3= a_3 Y+a_2 XZ+a_0 Z^3.
\end{eqnarray}
As usual, for $H^*(V)$ we need $R\pi_*V$, which from our experience of smooth degree two spectral cover is,
\begin{eqnarray}
R\pi_*V =Li_{\sigma}^* \mathcal{L}_3 [-1]. 
\end{eqnarray}
Therefore $H^p(V)=H^{p-1}(R^1\pi_*V)=H^{p-1}(\mathcal{L}_3|_{S_3 \cap \sigma})$. Usually the curve $S_3 \cap \sigma$, which is given by the equation $a_3=0$ in the base of $X$, is called the matter curve for obvious reasons.

What can be said about $R\pi_* \Lambda^2V$? As for $V$, $\Lambda^2V$ also satisfy the following identity,
\begin{eqnarray}
R\pi_* \Lambda^2V = \Phi (\Lambda^2V)|_{\sigma}.
\end{eqnarray}
Now the formula of the previous section can be used,
\begin{eqnarray}
\Phi (\Lambda^2V)|_{\sigma} = Rm_* \left(\pi_1^* \Phi(V) \otimes_A \pi_2^* \Phi(V) \right)|_{\sigma} \otimes K_B^{-1} [+1],
\end{eqnarray}
where the subscript $A$ in the right hand side means antisymmetrized product. In addition, we are looking for restriction on the zero section. To see how this can happen, consider a generic elliptic fiber. Let $P_1$, $P_2$ and $P_3$ be the intersection of the fiber with $S_3$. Then by definition,
\begin{eqnarray}
P_1\boxplus P_2\boxplus P_3=0.
\end{eqnarray}
The points that correspond to the intersection of the same fiber with the spectral cover of $\Lambda^2V$ will be,
\begin{eqnarray}
P_1\boxplus P_2,\quad P_1\boxplus P_2,\quad P_2\boxplus P_3. 
\end{eqnarray}
So the restriction on the zero section $\sigma$ is non-zero if one of these points, say $P_1\boxplus P_2$, is the zero point of the elliptic curve so, 
\begin{eqnarray}
P_1\boxplus P_2=0, \quad P_3=0.
\end{eqnarray}
So $\Phi (\Lambda^2V)|_{\sigma}$ is only non-zero over the matter curve $a_3=0$.\footnote{This is of course not surprising, because for $SU(3)$ bundles $\Lambda^2 V = V^*$. However we prefer the approach mentioned above because it can be generalized to higher rank bundles. It is not very hard to show that  
\begin{eqnarray}
S_3(\Lambda^2V) = -a_3 Y+a_2 X Z+a_0 Z^3.
\end{eqnarray}
} So the bottom line is,
\begin{eqnarray}\label{interE6}
R\pi_* \Lambda^2V = \Phi (\Lambda^2V)|_{\sigma} =Rm_* (\pi_1^* \mathcal{L}_3|_{\tilde{c}}\otimes_A \pi_2^* \mathcal{L}_3|_{\tilde{c}})\otimes K_B^{-1}[-1],
\end{eqnarray}
where $\tilde{c}$ is the double cover of the matter curve $a_3=0$, and defined in the following way,
\begin{eqnarray}
S_3|_{a_3=0} = Z(a_2 X +a_0 Z^2)|_{a_3=0} = c \cup \tilde{c}.
\end{eqnarray}
So to evaluate the right hand side of \eref{interE6} one needs to multiply the line bundle $\mathcal{L}$ on different sheets of $\tilde{c}$. The final result is the following,
\begin{eqnarray}\label{Final}
R\pi_* \Lambda^2 V = Rm_* (\pi_1^* \mathcal{L}_3|_{\tilde{c}}\otimes_A \pi_2^* \mathcal{L}_3|_{\tilde{c}})\otimes K_B^{-1}[-1] = Det \left[\pi_* \mathcal{L}_3|_{\tilde{c}} \right]\otimes K_B^{-1}[-1].
\end{eqnarray}
To see this note that the pushforward $\pi_*\mathcal{O}_{\tilde{c}} = \mathcal{O}_c \oplus \mathcal{O}(-\frac{R}{2})$ locally. Where $R$ is the branch divisor of the $\tilde{c}$. Then each factor $\mathcal{O}_c$ and $\mathcal{O}(-\frac{R}{2})$ corresponds to line bundles over each sheet. Therefore taking the determinant of $\pi_*\mathcal{O}_{\tilde{c}}$ is the same as multiplying the line bundles over each sheet. 

So now it is clear that the Yukawa couplings \textbf{27-27-27} are zero. The reason is $R^1\pi_* V$ receives contribution from $\mathcal{L}|_c$, but $R^1\pi_* \Lambda^2V$ depends on $\mathcal{L}|_{\tilde{c}}$. So the following map must be vanishing,
\begin{eqnarray}\label{smoothSU3}
H^0(R^1\pi_*V) \otimes H^0(R^1\pi_* V) \longrightarrow_{0} H^0(R^1\pi_* \Lambda^2V).
\end{eqnarray}
\begin{figure}
    \centering
    \includegraphics[width=0.5\textwidth]{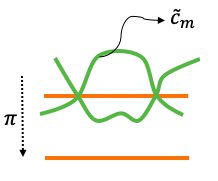}\includegraphics[width=0.5\textwidth]{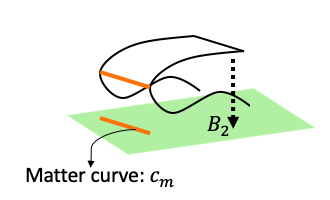}
    \caption{Left: the preimage of the matter curve inside the spectral cover. This is a reducible cover of $c_m$, where one component is isomorphic to the matter curve (the orange line) and one component is a double cover of the matter curve (green line). Right: the matter curve, i.e. the intersection locus of the zero section and the spectral cover.}
    \label{fig:my_label}
\end{figure}
In addition, the image of the product of the cohomologies on the left lives inside $H^0({\mathcal{L}_3|_{c_m}}^{\otimes 2})$, and the the cohomology on the right hand side is given by $H^0(\pi_*\mathcal{L}_3|_{\tilde{c}_m})$. However, it is shown in the \ref{Appendix} that $c_1(\pi_*V)=3K_B$ then
\begin{eqnarray}
c_1(i_{c_m}^*\pi_*V) = c_1({\mathcal{L}_3|_{c_m}})c_1(\pi_*\mathcal{L}_3|_{\tilde{c}_m})=3K_B|_{c_m},
\end{eqnarray}
where $i_{c_m}:c_m \hookrightarrow X$ is just the inclusion of the (preimage of) the matter curve inside the Calabi-Yau X. Using this relation it is clear that the map \eref{smoothSU3} is cannot be non-zero generically.

This has a clear F-theory interpretation. there are three ways that one get Yukawa couplings in local F-theory models \cite{Beasley:2008dc}: 1) triple intersection of 7-branes, 2) interaction of the bulk zero modes and the matter zero modes (which are localised over matter curves), 3) Interaction between bulk zero modes with each other. In the current model, all of the matter content of the effective theory are coming from the zero modes that are living over the curve $c_m$, and hence it is impossible to get non-zero Yukawa couplings from them. More precisely The F-theory geometry (after stable degeneration) is 
\begin{eqnarray}
Y^2+X^3+ \alpha_1 v X Y &+&  (\alpha_2 v^3+a_3 u v^2)Y +(\alpha_4v^4+a_2uv^3)X \nonumber \\
&+& (\alpha_6 v^6+a_0 u v^5),
\end{eqnarray}
where $(X,Y,u,v)$ are homogeneous coordinates of $\mathbb{WP}^{2,3,1,1}$, and $v=0$ is the locus of the GUT brane \cite{Donagi:2008ca}. The singularity over $v=0$ is generically $E_6$, but over the curve $a_3=0$ the singularity enhances to $E_7$, and directly from F-theory it can be seen there are \textbf{27} and $\bar{\textbf{27}}$ hypermultiplets living in this curve.  The singularity enhances to $E_8$ over the intersection locus $v=a_3=a_2=0$, however this doesn't correspond to Yukawa couplings \cite{Braun:2013cb}.  
\subsubsection{Reducible Reduced Spectral Cover}\label{sec312}
The spectral cover in this case is a triple cover of the base, and the only possible reducible spectral cover (with one component parallel to the base) is (Figure \ref{tripleWOV})
\begin{eqnarray}
S_3=Z(a_2X+a_0Z^2).
\end{eqnarray}
Without loss of generality the spectral sheaf can be written as \cite{Donagi:2011jy},
\begin{eqnarray}\label{redSU(3)sheaf}
\begin{tikzcd}[ampersand replacement=\&, column sep=small]
0\arrow[r]\& i_{\sigma*} \mathcal{L}_1 \arrow[r] \& i_{S*}\mathcal{L}_3\arrow[r] \& i_{S_2*}\mathcal{L}_2 \arrow[r] \& 0,
\end{tikzcd}
\end{eqnarray}
where $\mathcal{L}_1$ is a line bundle in base, $\mathcal{L}_2$ is a torsion free sheaf on $S_2$ with possible singularities over the intersection $\sigma\cap S_2 = \lbrace a_2=0\rbrace$ (but as mentioned before we assume there are no singularities in these sheaves). The extension morphism of this short exact sequence is localised over $\lbrace a_2=0\rbrace$,
\begin{eqnarray}
Ext^1(i_{S_2*}\mathcal{L}_2,i_{\sigma*} \mathcal{L}_1) = Hom (\mathcal{L}_2|_{\lbrace a_2=0\rbrace},\mathcal{L}_1|_{\lbrace a_2=0\rbrace}\otimes \mathcal{O}(-S_2)).
\end{eqnarray}
Next, one needs to find $\pi_1^* i_{S*}\mathcal{L}\otimes_A\pi_2^* i_{S*}\mathcal{L}$. To do this we can antisymmetrize the sequence above
\begin{eqnarray}
\begin{tikzcd}[ampersand replacement=\&, column sep=small]
0\arrow[r]\& \pi_1^* i_{\sigma*} \mathcal{L}_1 \otimes_S \pi_2^*i_{\sigma*} \mathcal{L}_1 \arrow[r] \& \pi_1^* i_{\sigma*} \mathcal{L}_1 \otimes \pi_2^*  i_{S*}\mathcal{L}\arrow[d] \& \& \\
\& \& \pi_1^* i_{S*} \mathcal{L} \otimes_A \pi_2^*  i_{S*}\mathcal{L}\arrow[r] \& \pi_1^* i_{S_2*} \mathcal{L}_2 \otimes_A \pi_2^*  i_{S_2*}\mathcal{L}_2\arrow[r] \& 0,  
\end{tikzcd}
\end{eqnarray}
or equivalently,
\begin{eqnarray}
\begin{tikzcd}[ampersand replacement=\&, column sep=small]
0\arrow[r] \& \pi_1^* i_{\sigma*} \mathcal{L}_1 \otimes \pi_2^*  i_{S_2*}\mathcal{L}_2\arrow[r]
\& \pi_1^* i_{S*} \mathcal{L} \otimes_A \pi_2^*  i_{S*}\mathcal{L}\arrow[r] \& \pi_1^* i_{S_2*} \mathcal{L}_2 \otimes_A \pi_2^*  i_{S_2*}\mathcal{L}_2\arrow[r] \& 0.
\end{tikzcd}
\end{eqnarray}
Applying the $Rm_*$ functor on this sequence gives a long exact sequence, but fortunately this is simple in the current situation,
\begin{eqnarray}\label{anti-sym seq}
\begin{tikzcd}[ampersand replacement=\&, column sep=small, row sep=small]
0\arrow[r] \& \mathcal{L}_1 \otimes i_{S_2*}\mathcal{L}_2 \arrow[r] \& R^0m_*( \pi_1^* i_{S*} \mathcal{L} \otimes_A \pi_2^*  i_{S*}\mathcal{L}) \arrow[r] \& i_{\sigma*} \left(Det\pi_* \mathcal{L}_2 \right) \arrow[r] \& 0,
\end{tikzcd}
\end{eqnarray}
Note that the spectral cover is unchanged,
\begin{eqnarray}
S_{\Lambda^2V} = S_3
\end{eqnarray}
Now let us apply the left derived functor $Li_{\sigma}^*$ to the sequence above
\begin{eqnarray}
\begin{tikzcd}[ampersand replacement=\&, column sep=small, row sep=small]
  0 \arrow[r] \& L^{-1}i_{\sigma}^* R^0m_*( \pi_1^* i_{S*} \mathcal{L} \otimes_A \pi_2^*  i_{S*}\mathcal{L}) \arrow[r] \arrow[d, phantom, ""{coordinate, name=Z}] \&  \left(Det\pi_* \mathcal{L}_2 \right)\otimes K_B^{-1}  \arrow[dll, to path={ -- ([xshift=2ex]\tikztostart.east)
  |- (Z)  [near end]\tikztonodes
-| ([xshift=-2ex]\tikztotarget.west)
-- (\tikztotarget)}]\& \\
i_{\lbrace a_2=0\rbrace*} (\mathcal{L}_1\otimes \mathcal{L}_2) \arrow[r] \& L^{0}i_{\sigma}^* R^0m_*( \pi_1^* i_{S*} \mathcal{L} \otimes_A \pi_2^*  i_{S*}\mathcal{L}) \arrow[r] \&  Det\pi_* \mathcal{L}_2  \arrow[r] \& 0.
\end{tikzcd}
\end{eqnarray}
Finding the exact result depends on the coboundary map which is induced by the extension of the \eref{anti-sym seq} restricted on $\lbrace a_2=0\rbrace$. As will be clear in the examples, the coboundary map is not necessarily zero, and the zero modes $i_{\lbrace a_2=0\rbrace*} (\mathcal{L}_1\otimes \mathcal{L}_2\otimes K_B^{-1})$ do not necessarily inject into $R^1\pi_*\Lambda^2V$ . To compute $H^1(V)$, $R^1\pi_*V$ is also needed
\begin{eqnarray}\label{V3pushforward}
&R\pi_*V= \mathcal{L}_1 \otimes K_B^{-1} \oplus \mathcal{F}[+1],& \\
&\begin{tikzcd}[ampersand replacement=\&, column sep=small]
0\arrow[r]\& \mathcal{L}_1 \arrow[r]\& \mathcal{F} \arrow[r] \& i_{\lbrace a_2=0 \rbrace*} \mathcal{L}_2 \arrow[r]\& 0.\nonumber
\end{tikzcd}&
\end{eqnarray}
By Leray spectral sequence, it is possible to compute the cohomologies in terms of these pushforwad results derived so far. For $V$ one gets
\begin{eqnarray}\label{LerayE6Red}
\begin{tikzcd}[ampersand replacement=\&, column sep=small, row sep=small]
0\arrow[r] \& H^1(\pi_*V) \arrow[r] \& H^1(V) \arrow[r] \& H^0(R^1\pi_*V)\arrow[r,"\alpha"] \& H^2(\pi_*V) \arrow[r] \& \dots 
\end{tikzcd}
\end{eqnarray}
and for $\Lambda^2V$,
\begin{eqnarray}
\begin{tikzcd}[ampersand replacement=\&,column sep= small]
H^0(R^1\pi_*\Lambda^2V) \arrow[r] \& H^2(\pi_*\Lambda^2V)\arrow[r] \& H^2(\Lambda^2V) \arrow[r]\& H^1(R^1\pi_*\Lambda^2V) \arrow[r] \& 0.
\end{tikzcd}
\end{eqnarray}
S0 $H^1(\pi_*V)$ gets injected into $H^1(V)$. In addition, the homomorphism $\alpha$ lives in
\begin{eqnarray}
\alpha\in Ext^2 (\mathcal{F},\mathcal{L}_1\otimes K_B^{-1}).
\end{eqnarray}
$\alpha$ should be zero, because at the end of the day, it should be induced by the extension morphism of the original defining sequence, and by checking the extension group above, $\alpha$ cannot be induced by the original sequence. Therefore,
\begin{eqnarray}
H^1(V)=H^1(\pi_*V) \oplus H^0(R^1\pi_* V).
\end{eqnarray}
On the other hand $H^1(R^1\pi_*\Lambda^2V)$ injects into $H^2(\Lambda^2V)$, so clearly $H^1(i_{\lbrace a_2=0\rbrace*} (\mathcal{L}_1\otimes \mathcal{L}_2\otimes K_B^{-1}))$ can contribute in $H^2(\Lambda^2V)$.\footnote{Remember that 
\begin{eqnarray}
R\pi_*\Lambda^2V = Li_{\sigma}^* \mathcal{L}_3\star_A\mathcal{L}_3 \otimes K_B [-1].\nonumber
\end{eqnarray}} After this point it is hard to say general statements about the Yukawa couplings, and the situation can defer case by case, depending on the morphisms between the cohomology groups. Here we only give the candidates for the Yukawa couplings, and after this point computing the actual couplings in a given model is usually simple.  

\begin{figure}
    \centering
    \includegraphics[width=0.5\textwidth]{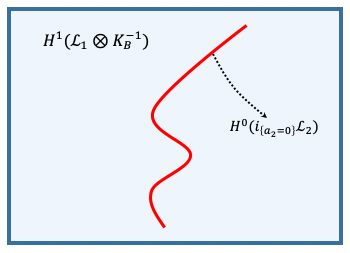}
    \caption{Some of \textbf{27} matter in the effective theory are from the bulk zero modes, and sime of them are zero modes on the intersection of the 7-branes (red curve).}
    \label{fig:Bulk-Matter SU(3)}
\end{figure}

Therefore we conclude that if $V$ is a stable, degree zero, holomorphic $SU(3)$ bundle, with reducible spectral cover (but reduced components), then the \textbf{27-27-27} Yukawa coupling can be non-zero, and is coming from the interaction of zero modes localised over $\lbrace a_2=0 \rbrace$ and bulk zero modes, and they correspond to the following morphism,
\begin{eqnarray}
H^1(\mathcal{L}_1\otimes K_B^{-1}) \otimes H^0(i_{\lbrace a_2=0 \rbrace*} \mathcal{L}_2) \longrightarrow H^1(i_{\lbrace a_2=0\rbrace*} (\mathcal{L}_1\otimes \mathcal{L}_2\otimes K_B^{-1})).
\end{eqnarray}
F-theory interpretation of this result is again easy. This type of Yukawa coupling are simply coming from the bulk and localised zero modes, see Figure \ref{fig:Bulk-Matter SU(3)}. Note that singularity of the F-theory Calabi-Yau fourfold corresponds to $E_7$. So one may naively expect to get $E_7$ gauge group in the effective theory. But in the point is that the rank of the Higgs field over the GUT 7-brane is dropped by one (T-brane setup) \cite{Cecotti:2010bp,Anderson:2013rka}
\begin{eqnarray}
\Phi=\left(\begin{array}{ccc}
   0  & a_2 & 0 \\
    -\frac{1}{a_0} & 0  &0 \\
     0 & -\frac{1}{a_0} & 0
\end{array}\right).
\end{eqnarray}
Therefore there will be a $U(1)$ gauge field living over the complex surface that 7-brane wraps (isomorphic to the base of $X$), and that breaks $E_7$ to $E_6$. Hence the cohomology $H^1(\mathcal{L}_1\otimes K_B^{-1})$ can be identified with the bulk zero modes inside the 7-brane (associated to the fluctuations of the $U(1)$ connection), and $H^0(i_{\lbrace a_2=0\rbrace}\mathcal{L}_2)$ is identified with the zero modes living over the intersection of the GUT 7-brane and a flavour 7-brane. 
Of course there are moduli dependent Yukawa couplings too:
\begin{eqnarray}
Ext_X^1(\mathcal{O}_X,V) \otimes Ext_X^1(V,V)\otimes Ext_X^1(V,\mathcal{O}_X)\longrightarrow H^3(O_X)\simeq \mathbb{C}.
\end{eqnarray}
Such morphisms\footnote{This is a result of Yoneda product.} correspond to \textbf{1-27-$\overline{\textbf{27}}$} couplings in the effective theory. The F-theory interpretation of this is well known \cite{Donagi:2003hh}. To connect this to F-theory we rewrite the above product in terms of spectral sheaves
\begin{eqnarray}
Ext^1 (i_{\sigma*}\mathcal{O},\mathcal{L}_3) \otimes Ext^1(\mathcal{L}_3,\mathcal{L}_3)\otimes Ext^1(\mathcal{L}_3,i_{\sigma*}\mathcal{O})\longrightarrow \mathbb{C}.
\end{eqnarray}
The first and the the third term are \textbf{27} and $\overline{\textbf{27}}$ as usual, and the middle term corresponds to the first order deformations of the spectral data (or the bundle $V$), and ``can" receive contributions from $Ext^1(\mathcal{L}_1,\mathcal{L}_3)$. This later group can be interpreted as the open strings that are stretched between the components of the spectral cover (which in the local F-theory correspond to components of intersecting 7-branes). We will see more details about this in examples.

\subsection{$SO(10)$ GUT Models}\label{sec32}
For $SO(10)$ models we need $SU(4)$ vector bundles, $V$ over the Calabi-Yau manifold $X$. Hence the spectral cover is a degree four cover of the base manifold. The number of the possibilities for spectral covers without vertical components in larger relative to the $E_6$ models, .

\subsubsection{Smooth Spectral Cover}\label{sec321}
\begin{figure}
    \centering
    \includegraphics[width=0.85\textwidth]{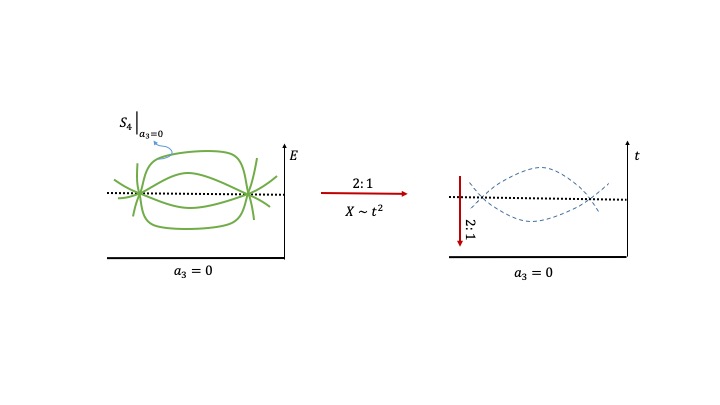}
    \caption{The green curve in the left picture is the restriction of $S_4$ over $a_3=0$. It is s degree 4 cover of the curve $a_3=0$ in the base manifold. Therefore if the two outer sheets correspond to $P_1$ and $P_2$ on a fiber such that $P_1\boxplus P_2=0$, then the inner sheets correspond to the points $P_3$ and $P_4$. By locally setting $X\sim \sqrt{t}$ it can be seen as the double cover of $a_3^2=0$.}
    \label{fig:mattercurve4}
\end{figure}

In this case $V$ is a $SU(4)$ vector bundle, and the spectral cover is a generic degree four smooth cover of the base,
\begin{figure}
    \centering
    \includegraphics[width=0.5\textwidth]{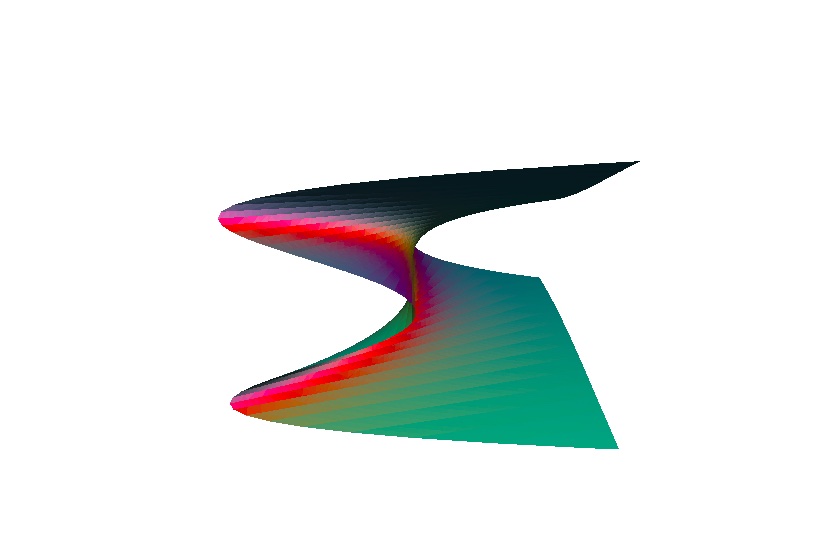}
    \caption{Smooth degree four cover of the base.}
    \label{fig:smooth4Pl}
\end{figure}
\begin{eqnarray}
S_4 = a_4 X^2 +a_3 Y Z +a_2 X Z^2 +a_0 Z^4.
\end{eqnarray}
As usual $R^1\pi_* V$ receives contribution from the matter curve $c :={a_4=0}$.\footnote{It is also easy to show that the spectral cover of $\Lambda^2 V$ (at least locally in the neighbourhood of $Z=0$) is given by the following equation,
\begin{eqnarray}
S_{\Lambda^2V} = -a_3^2 X^3+(a_2^2-4a_0a_4) X^2 Z^2 + a_0^2 Z^6.
\end{eqnarray}}
So the ``matter curve" for $\Lambda^2V$ is given by the non-reduced curve $a_3^2=0$. The reason (besides the the reason mentioned in the footnote), can be explained as in the smooth degree three spectral cover. Suppose $S_4$ intersect a generic fiber at 4 points $P_1$, $P_2$, $P_3$ and $P_4$ such that
\begin{eqnarray}\label{boxplus4}
P_1\boxplus P_2\boxplus P_3\boxplus P_4 =0.
\end{eqnarray}
Then $S_{\Lambda^2V}$ intersect the same fiber at six points
\begin{eqnarray}
P_1\boxplus P_2,\quad P_1\boxplus P_3,\quad P_1\boxplus P_4, \nonumber \\
P_2\boxplus P_3,\quad P_2\boxplus P_4,\quad P_3\boxplus P_4.
\end{eqnarray}
So when the zero section intersects with $S_{\Lambda^2V}$ (which corresponds to the locus of \textbf{10} of $SO(10)$), one of these six points (say $P_1\boxplus P_2$) becomes zero, and due to \eref{boxplus4}, $P_3\boxplus P_4$ will be zero too. So the matter curve of $\Lambda^2V$ is the locus where not only \eref{boxplus4} is true, but also sum of a pair of these four points is zero. The algebraic equation for such points on the fiber must involve only $X$ and $Z$. The reason is $X$ is a degree two variable, and any holomorphic equation that involves $X$ gives a pair of points such that their sum is zero. This happens when $a_3=0$ in $S_4$. In other words 
\begin{eqnarray}
S_4|_{a_3=0} = a_4 X^2 +a_2 X Z^2 +a_0 Z^4,
\end{eqnarray}
which is a degree four cover of $a_3=0$, or equivalently, a double (degree two) cover of the non-reduced curve $a_3^2=0$ (see Figure \ref{fig:mattercurve4}),
\begin{eqnarray}
\begin{tikzcd}[ampersand replacement=\&, column sep=small]
 P: S_4|_{a_3=0} \arrow[r,"2:1"] \& c_{\Lambda^2V} = {a_3^2=0}.
\end{tikzcd}
\end{eqnarray}
So as in the $SU(3)$ bundle one gets the following relation,
\begin{eqnarray}
R\pi_* \Lambda^2 V = Det \left[P_* \mathcal{L}|_{a_3=0} \right]\otimes K_B^{-1}[-1].
\end{eqnarray}
Note that $R^1\pi_*\Lambda^2V$ is non-zero over $a_3=0$, and $R^1\pi_*V$ is non-zero over the matter curve $a_4=0$, and their sections (i.e. the $H^0$ cohomology) inject into $H^1(\Lambda^2V)$ and $H^1(V)$ respectively. So the following map is non-zero
\begin{eqnarray}
H^0(R^1\pi_*V) \otimes H^0(R^1\pi_* V)\otimes H^0(R^1\pi_* \Lambda^2V) \longrightarrow H^0(p,\mathcal{O}_p) , 
\end{eqnarray}
where $p$ corresponds to the intersection locus of $c$ and $c_{\Lambda^2V}$ in the base. Then the Yukawa coupling \textbf{16-16-10} in non-vanishing over the intersection locus of the matter curves. 

The F-theory interpretation of this is again straightforward. The relevant part of the dual F-theory geometry (which can be derived by stable degeneration\cite{Donagi:2011jy}) is given by
\begin{eqnarray}
Y^2+X^3+ \alpha_1 v X Y &+& (\alpha_2 v^2+a_4 u v)X^2 + (\alpha_2 v^3+a_3 u v^2)Y \nonumber \\
&+& (\alpha_4v^4+a_2uv^3)X+(\alpha_6 v^6+a_0 u v^5).
\end{eqnarray}
Over $a_4=0$ the singularity enhances to $E_6$, so both \textbf{16} and $\overline{\textbf{16}}$ lives in there, and over $a_3=0$ the singularity enhances to $SO(11)$, which corresponds to the \textbf{10} hypermultiplets of $S(10)$ (``non-split" type singularity). At the intersection locus $a_3=a_4=0$ there are $E_7$ singularities enhancements which signal the original parent $E_7$ gauge theory that is broken to $SO(10)$ due to brane deformation. In the original $E_7$ theory there are triple interactions of the adjoint representation of $E_7$, and after deformation it reduces to \textbf{16-16-10} coupling. Also on the branching points of the double cover in Figure \ref{fig:mattercurve4}, which is given by the discriminant $a_2^2-4a_0 a_4=0$, corresponds to the points where the singularity enhances to $SO(14)$. From the heterotic string theory view it happens because the restriction of the bundle over these branch points becomes like $(\mathcal{O}_E(P_1-\sigma)\oplus\mathcal{O}_E(-P_1-\sigma))^{\oplus 2}$. This non-generic form of the bundle has a ``larger" commutant in $E_8$ ($SO(14)$) relative to direct sum of four generic degree zero line bundles. So the gauge group must enhance. However these points don't give any Yukawa couplings in F-theory, as expected from Heterotic view.  

\subsubsection{Reducible Reduced Spectral Covers}\label{sec322}
\begin{figure}
    \centering
    \includegraphics[width=0.5\textwidth]{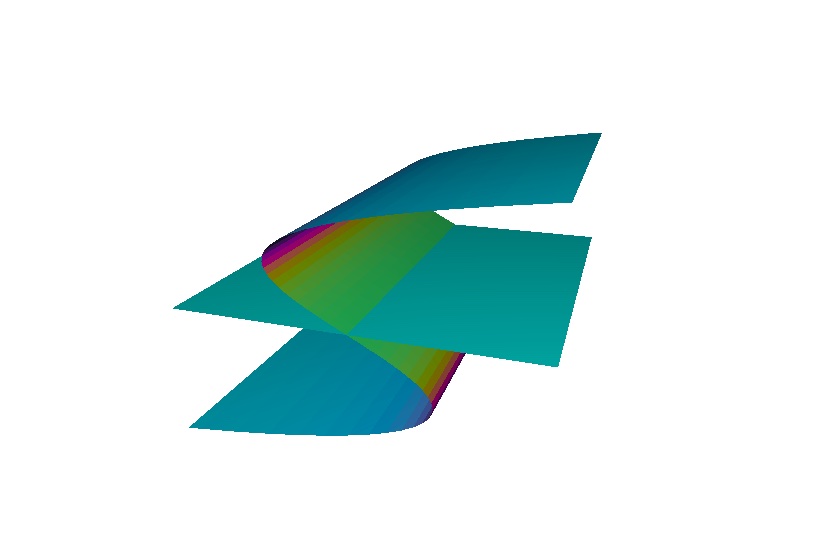}\includegraphics[width=0.5\textwidth]{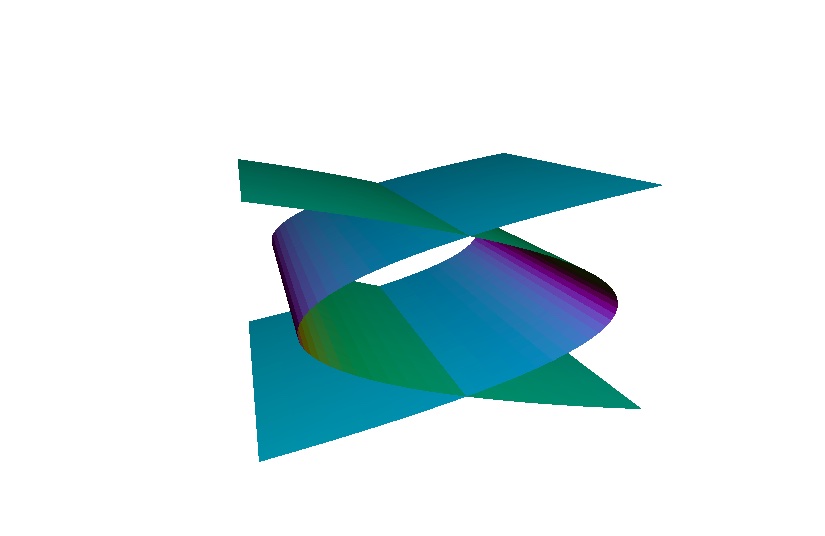}
    \caption{Reducible degree four spectral covers with reduced components.}
    \label{fig:redred4}
\end{figure}

Let us continue to $SU(4)$ bundles with reducible spectral covers such that each component is smooth, and there is no vertical components. In this case there are two possibilities (see Figure \ref{fig:redred4}),
\begin{eqnarray}
\begin{tikzcd}[ampersand replacement=\&, column sep=small]
S_4 = Z(a_3Y+ a_2X Z+a_0Z^3), \& 0\arrow[r]\& i_{\sigma*}\mathcal{L}_1 \arrow[r] \& \mathcal{L}_4 \arrow[r]\& i_{S_3*}\mathcal{L}_3\arrow[r]\& 0, \\
S_4 = (a_2X +a_0Z^2)(b_2 X+b_0Z^2),\& 0\arrow[r]\& i_{S_2*}\mathcal{L}_2 \arrow[r] \& \mathcal{L}_4 \arrow[r]\& i_{\tilde{S}_2*}\mathcal{L}'_2\arrow[r]\& 0,
\end{tikzcd}
\end{eqnarray}
where $a_i$ and $b_i$ are generic polynomials. 
\begin{itemize}
    \item $S_4 = Z(a_3Y+ a_2X Z+a_0Z^3)$
\end{itemize}
For the first case the calculation is similar to the $SU(3)$ bundle with reducible spectral cover, So generally the spectral sheaf $\mathcal{L}_4$, as an extension
\begin{eqnarray}\label{V4L1L3}
\begin{tikzcd}[ampersand replacement=\&,column sep=small]
0\arrow[r]\&\mathcal{L}_1\arrow[r]\& \mathcal{L}_4\arrow[r]\& \mathcal{L}_3\arrow[r]\& 0.
\end{tikzcd}
\end{eqnarray}
The associated spectral sheaf for $\Lambda^2V$ corresponds to the antisymmetrized Pontrjagin product of $L_4$,
\begin{eqnarray}
\begin{tikzcd}[ampersand replacement=\&, column sep=small]
0\arrow[r] \&\pi^*\pi_* \mathcal{L}_1\otimes i_{S_3*}\mathcal{L}_3 \arrow[r] \&  \mathcal{L}_4 \star_A \mathcal{L}_4 \arrow[r] \& \mathcal{L}_3 \star_A \mathcal{L}_3 \arrow[r] \& 0.
\end{tikzcd}
\end{eqnarray}
It is possible to read the algebraic equation for the spectral cover of $\Lambda^2V$ from this sequence
\begin{eqnarray}
S_{\Lambda^2V} = (a_3Y+ a_2X Z+a_0Z^3)(-a_3Y+ a_2X Z+a_0Z^3),
\end{eqnarray}
where the second factor is the support of $\mathcal{L}_3 \star_A \mathcal{L}_3$. As before the left derived functor $Li_{\sigma}^*$ should be applied on the sequence above,
\begin{eqnarray}
\begin{tikzcd}[ampersand replacement=\&, column sep=small,row sep=small ]
 0 \arrow[r] \& L^{-1}i_{\sigma}^* \mathcal{L}_4 \star_A \mathcal{L}_4 \arrow[r]\arrow[d, phantom, ""{coordinate, name=Z}] \&  \left(Det\pi_* \mathcal{L}_3|_{\tilde{c}_m} \right)\otimes K_B^{-1}  \arrow[dll,to path={ -- ([xshift=2ex]\tikztostart.east)
|- (Z) [near end]\tikztonodes
-| ([xshift=-2ex]\tikztotarget.west)
-- (\tikztotarget)}]\& \\
 \mathcal{L}_1\otimes i_{c_m}^* \mathcal{L}_3 \arrow[r] \& L^0i_{\sigma}^*  \mathcal{L}_4 \star_A \mathcal{L}_4 \arrow[r] \&  Det\pi_* \mathcal{L}_3|_{\tilde{c}_m} \arrow[r] \& 0,
\end{tikzcd}
\end{eqnarray}
where  $\tilde{c}_m$ is the double cover defined in section \ref{sec311},
\begin{eqnarray}
\begin{tikzcd}[ampersand replacement=\&, column sep=small]
\pi: \tilde{c}_m\cup c_m=S_3|_{\lbrace a_3=0\rbrace} \arrow[r,"3:1"]\& c_m = \lbrace Z = a_3=0\rbrace.
\end{tikzcd}\nonumber
\end{eqnarray}
As probably clear by now, for computing $H^1(V)$ through Leray spectral sequence, we need the derived pushforward of $V$
\begin{eqnarray}
R\pi_*V= \mathcal{L}_1\otimes K_B^{-1} \oplus \mathcal{F}[+1], \\
\begin{tikzcd}[ampersand replacement=\&, column sep=small]
0\arrow[r]\& \mathcal{L}_1 \arrow[r]\& \mathcal{F}\arrow[r]\& \mathcal{L}_3|_c \arrow[r]\& 0.
\end{tikzcd}
\end{eqnarray}
As before, one should be careful to check whether the cohomologies of the pushforwards actually contribute in the relevant cohomologies of $V$ and $\Lambda^2V$. In the present case, as in the $SU(3)$ bundle with reducible reduced spectral cover, the coboundary maps may not be zero because they can be induced by the extension group of the original sequence, which is the only defining parameter of this bundle (modulo the information needed to construct the constituent bundles). 

Therefore, let $V$ be a $SU(4)$ holomorphic stable bundle with degree zero with reducible spectral cover such that one component is zero section, and the other component a generic triple cover. Then there can be a non-vanishing \textbf{16-16-10} Yukawa coupling corresponding to the following map,
\begin{eqnarray}
H^1(\mathcal{L}_1\otimes K_B^{-1}) \otimes H^0(i_{\lbrace a_3=0 \rbrace*} \mathcal{L}_3) \longrightarrow H^1(i_{\lbrace a_3=0\rbrace*} (\mathcal{L}_1\otimes \mathcal{L}_3\otimes K_B^{-1})).
\end{eqnarray}

The F-theory interpretation is very similar to \ref{sec312}. So we ignore this case. We should just mention that as in the $SU(3)$ bundle, there are moduli dependent Yukawa couplings \textbf{1-16-$\overline{\textbf{16}}$}, which has open string interpretation as before. 
\begin{itemize}
    \item $S_4=(a_2 X +a_0Z^2)(b_2X+b_0Z^2)$
\end{itemize}
We proceed as usual. Consider the defining sequence,
\begin{eqnarray}
\begin{tikzcd}[ampersand replacement=\&, column sep=small]
0\arrow[r]\& \mathcal{L}_2\arrow[r]\& \mathcal{L}_4\arrow[r]\& \mathcal{L}'_2 \arrow[r]\& 0.
\end{tikzcd}
\end{eqnarray}
Where $\mathcal{L}_2$ is supported over $S_2=a_2 X +a_0Z^2$, and $\mathcal{L}'_2$ is another line bundle supported over $S_2'=b_2X+b_0Z^2$. The extension group of this sequence is given by
\begin{eqnarray}\label{extensionSO(12)split}
Hom_{S_2\cdot S_2'}\left(\mathcal{L}'_2,\mathcal{L}_2\otimes \mathcal{O}(S_2')\right)= H^0\left(\mathcal{L}_2\otimes \mathcal{L}_2^{*'}\otimes \mathcal{O}(S_2)\right),
\end{eqnarray}
where $S_2\cdot S_2'$ is the intersection of the two double covers, and itself is a double cover of the curve,
\begin{eqnarray}
a_2 b_0-a_0 b_2=0.
\end{eqnarray}
As usual the derived pushforward in this case is very simple to find
\begin{eqnarray}
& R\pi_*V=Li_{\sigma}^* \mathcal{L}_4[-1] = i_{\sigma}^* \mathcal{L}_4[-1],&
\begin{tikzcd}[ampersand replacement=\&, column sep=small]
0\arrow[r]\& \mathcal{L}_2|_c \arrow[r]\& i_{\sigma}^* \mathcal{L}_4 \arrow[r]\& \mathcal{L}'_2|_{c'} \arrow[r]\& 0,
\end{tikzcd}
\end{eqnarray}
where $c$ and $c'$ are curves given by $\lbrace a_2=0\rbrace$ and $\lbrace b_2=0\rbrace$ respectively. The matter curve is the union $c\cup c'$. The extension of this sequence is induced by the extension of the original sequence. However it can be zero, and it must be checked case by case. The next step is finding the antisymetrized Pontrjagin product,
\begin{eqnarray}\label{L4wL2L2}
\begin{tikzcd}[ampersand replacement=\&, column sep=small]
 \&0\arrow[d] \& \& \& \\
 \&\mathcal{L}_2\star_A  \mathcal{L}_2\arrow[d]  \& \& \& \\
0\arrow[r] \& Q\arrow[d]\arrow[r] \& \mathcal{L}_4\star_A\mathcal{L}_4\arrow[r] \&\mathcal{L}'_2 \star_A \mathcal{L}'_2 \arrow[r] \& 0 \\
 \&\mathcal{L}_2\star \mathcal{L}'_2\arrow[d] \& \& \& \\
 \&0 \& \& \& \\
\end{tikzcd}
\end{eqnarray}
Either from this diagram or directly, we can find the spectral cover of $\Lambda^2V$ which is a degree six cover of the base\footnote{We compute the spectral cover of the associated bundles by simply adding the roots of the spectral cover (as in \cite{Donagi:2009ra} for the Higgs bundle spectral cover). This is certainly true if we restrict $S_V$ to the neighbourhood of the zero section, and therefore can be used for the F-theory dual. But usually there is a caveat in finding the spectral cover of the associated bundles in heterotic side. The reason is that the fibers of the Calabi-Yau has torsion points relative to $\boxplus$ or the morphism $m$ (corresponding to the zeros of $Y$). These torsion points can slightly modify the spectral cover of the associated bundles in heterotic side. In the current example, over the points $a_0b_2-a_2b_0=a_0^3+F a_0 a_2^2-G a_2^3=0$ (where $F$, $G$ are Weierstrass defining polynomials) the double covers $S_2$ and $S_2'$ intersect the torsion points in the fibers, and these points are subset of their branching locus. So $S_{\Lambda^2V}$ too must be branched there, and this can checked to be true. The only feature that the author cannot explain are the vertical components over $a_0=b_0=0$. However they have to be there for correct divisor class.}
\begin{eqnarray}\label{S6V4S2S2'}
S_{\Lambda^2V} = Z^2 \left((a_2 b_0 -a_0b_2)^2X^2+2a_0b_0(a_2 b_0 +a_0b_2)X Z^2+a_0^2b_0^2 Z^4\right).
\end{eqnarray}
and the restriction over the zero section is
\begin{eqnarray}
Li_{\sigma}^* \mathcal{L}_4 \star_A \mathcal{L}_4 &=&\left(Det(\pi_*\mathcal{L}_2)\oplus Det(\pi_*\mathcal{L}'_2)\oplus \pi_*( \mathcal{L}_2\otimes \tau^*\mathcal{L}'_2) \right)\\
&\oplus & \left(Det(\pi_*\mathcal{L}_2)\oplus Det(\pi_*\mathcal{L}'_2) \right)\otimes K_B^{-1}[+1].\nonumber
\end{eqnarray}
Note that we expect the coboundary maps to be zero in this case. To find the relevant cohomologies we need to know the relations between the sheaves written above (see \ref{Appendix})
\begin{eqnarray}
Det(\pi_*\mathcal{L}_2)\otimes Det(\pi_*\mathcal{L}'_2) = K_B^4.
\end{eqnarray}
Unfortunately there are no \textbf{16-16-10} coupling from the interaction of bulk and localised zero modes. However there are Yukawa couplings from the triple intersection of 7-branes. The relevant cohomologies are
\begin{eqnarray}
H^1(V)\cap H^1(V) \cap H^1(\Lambda^2V) \longrightarrow H^0(\mathcal{L}_2|_{c})\cap H^0(\mathcal{L}'_2|_{c'})\cap H^0 \left(\pi_*( \mathcal{L}_2\otimes \tau^*\mathcal{L}'_2 )\right)\longrightarrow  \mathbb{C}.
\end{eqnarray}
So there are pointlike Yukawa couplings in this case. When the extension \eref{extensionSO(12)split} is non-zero, it induces a non zero morphisms on the intersection points of the matter curves of $V$. So, even though $H^0(\mathcal{L}_2|_c)$ always injects into $H^1(V)$, but some or all of the elements of $H^0(\mathcal{L}'_2|_{c'})$ (especially when $H^1(\mathcal{L}_2|_c)$) will be killed by the extension morphism. The physics of this is not very surprising, but interesting to know \cite{Anderson:2015cqy}. When we choose a zero extension, $V$ will be a $S(U(2)\otimes U(2))$ bundle (of course we should find a suitable point in Kahler moduli such that $V$ becomes poly stable). In this case the effective gauge group will be $SO(12)\times U(1)$. The extra $U(1)$ is generally anomalous and massive (for example see \cite{Anderson:2009sw,Anderson:2009nt}). The hypermultiplets corresponding to $H^0(\mathcal{L}_2|_c)$ and $H^0(\mathcal{L}'_2|_{c'})$ are in \textbf{32} and \textbf{32'} representations of $S(12)$, and the hypermultiplets corresponding to the extension morphism \eref{extensionSO(12)split} are \textbf{12} of $SO(12)$. On the other hand, $SO(12)$ can break to $SO(10)\otimes U(1)$. So to Higgs $SO(12)$ to $SO(10)$ we need to give a vev to a hypermultiplet charged under the $U(1)$ only. Note that under $SO(12)\rightarrow SO(10)\times U(1)$, the \textbf{12} breaks as 
\begin{eqnarray}
\textbf{12}\rightarrow \textbf{1}_2 \oplus \textbf{1}_{-2} \oplus \textbf{10}_0,
\end{eqnarray}
where the subscripts are the $U(1)$ charges. Therefore by giving vev to $\textbf{1}_2$ or $\textbf{1}_{-2}$ breaks the gauge group to $SO(10)$, and at the same time the Yukawa couplings \textbf{12-32-32'} gives mass to certain pairs of \textbf{32}-\textbf{32'} multiplets. Compare this with the action of non-zero extension morphisms in either $Ext^1(\mathcal{L}_2',\mathcal{L}_2)$ or $Ext^1(\mathcal{L}_2,\mathcal{L}_2')$. They also change the bundle from $S(U(2)\times U(2))$ to $SU(4)$, and hence they break the $SO(12)$ to $SO(10)$. Also the extension maps can kill (give mass) to some elements in $H^0(\mathcal{L}_2')$ (or $H^0(\mathcal{L}_2)$, depending on which extension morphism we choose). So we can identify the $\textbf{1}_2$ and $1_{-2}$ hypermultiplets with elements in $Ext^1(\mathcal{L}_2',\mathcal{L}_2)$ and $Ext^1(\mathcal{L}_2,\mathcal{L}_2')$ (or vise versa, depending on the initial choice that we've made). Note that, when the extension is zero, in addition to \eref{extensionSO(12)split} (which are part of the bundle moduli), there other hypermultiplets too, which are coming from $H^1(\Lambda^2V)$. These zero modes are also localised on $a_2b_0-a_0b_2$, and are given by $H^0(\pi_*(\mathcal{L}_2\otimes \tau^*\mathcal{L}_2'))$. These hypermultiplets also contribute to the \textbf{12-32-32'} couplings. After turning on the extensions \eref{extensionSO(12)split}, or equivalently, Higgsing $SO(12)$ to $SO(10)$, then the \textbf{12}s from $H^1(\Lambda^2V)$ will remain massless, and decompose into \textbf{10} of $SO(10)$. Similarly the corresponding Yukawa couplings decompose into \textbf{10-16-16} of the effective $SO(10)$ Yukawa couplings, if the \textbf{32} and \textbf{32'} hypermultiplets in those couplings remain massless after Higgsing (note that $\textbf{32}\rightarrow \textbf{16}\oplus \overline{\textbf{16}}$ likewise for \textbf{32'}).

The dual F-theory interpretation is also quite similar. In F-theory the geometry has split $I_2^*$ singularity (because the spectral cover is reducible). When the effective gauge group is really $SO(12)$, one can check the singularity of the F-theory dual, and over the curves $c$ and $c'$ inside the GUT 7-brane, enhances to $E_7$. The adgjoint of the $E_7$ reduces to $SU(2)\times SO(12)$ as
\begin{eqnarray}
\textbf{133}\rightarrow (\textbf{3},\textbf{1}) \oplus (\textbf{2},\textbf{32'}) \oplus (\textbf{1},\textbf{66}).
\end{eqnarray}
So over these $E_7$ enhancement one gets only $\textbf{32'}$. Where are the \textbf{12}s and \textbf{32}s? Note that $E_7$ enhancement means the parent gauge theory was $E_7$ and the brane deformation breaks this into $SO(10)$. So lets ``deform back" one of the components of the spectral cover that contains, say, $c$. Then the $E_7$ GUT singularity inside the 7-brane enhances to $E_8$ on the curve $c'$. So over this curve the \textbf{56} hypermultiplet of $E_7$ lives. Now we deform the component that contained the curve $c$ to the initial gegenric one. Therefore the \textbf{56} also breaks according ro the following rule
\begin{eqnarray}
\textbf{56}\rightarrow (\textbf{2},\textbf{12})\oplus (\textbf{1},\textbf{32}),
\end{eqnarray}
where $(\textbf{2},\textbf{12})$ corresponds to the strings that are charged relative to the gauge fields inside the GUT 7-brane. So after the deformation of the 7-brane these hypermultiplets will live on the intersection of the components of the spectral cover. Clearly \textbf{32} lives in the curve $c$. At this point, rest of the stories is basically the same as the heterotic view, and we ignore the details.

\begin{figure}
    \centering
    \includegraphics[width=0.5\textwidth]{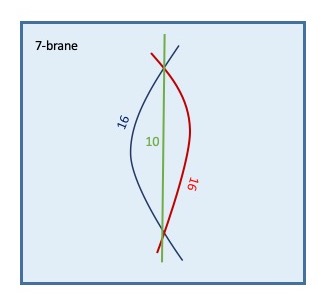}
    \caption{Yukawa couplings are coming from the intersection of the matter curves of $V$ and $\Lambda^2V$.}
    \label{fig:SO(12)split}
\end{figure}

\subsubsection{Spectral Covers With Non-Reduced Components}\label{sec323}
\begin{figure}
    \centering
    \includegraphics[width=0.5\textwidth]{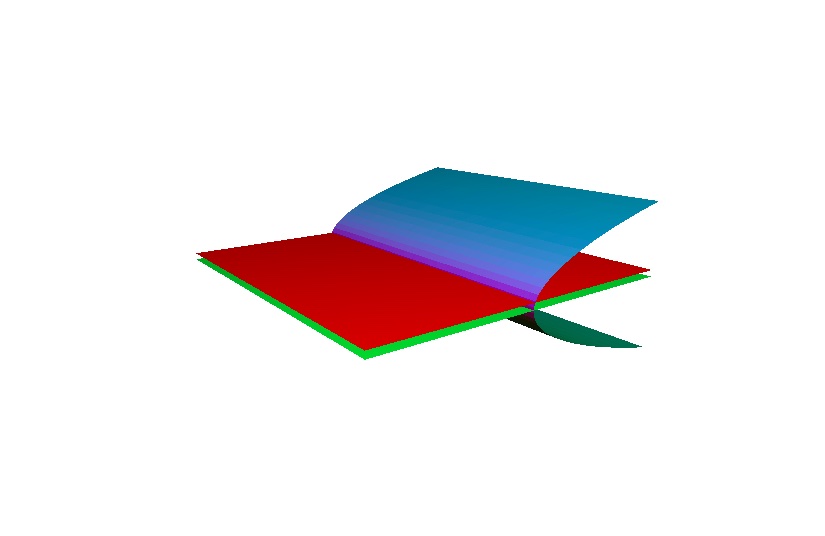}\includegraphics[width=0.5\textwidth]{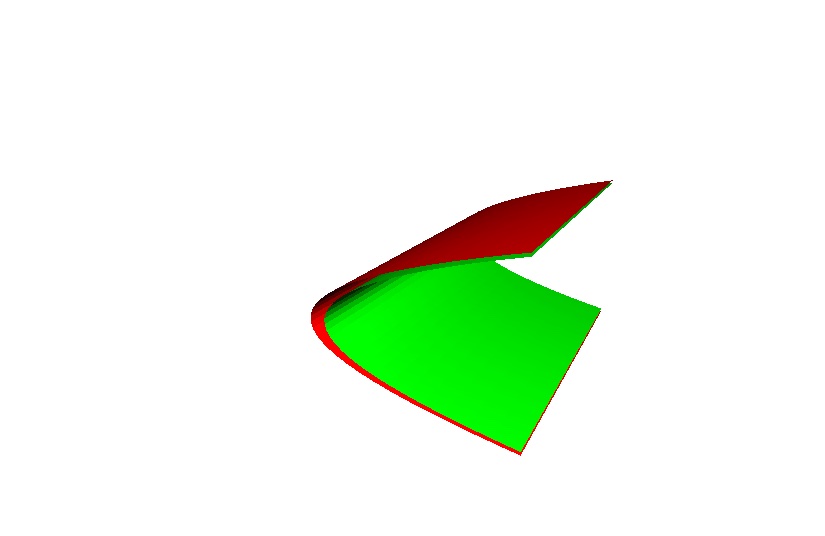}
    \caption{Non-reduced degree four spectral covers.}
    \label{fig:rednonred4}
\end{figure}

There are two cases that we are going to consider in this subsection. See Figure \ref{fig:rednonred4}. The spectral cover and the spectral sheaf in each case can be described generally as
\begin{eqnarray}
\begin{tikzcd}[ampersand replacement=\&, column sep=small]
S_4 = Z^2(a_2X Z+a_0Z^3), \& 0\arrow[r]\& \mathcal{U}_1 \arrow[r] \& \mathcal{L}_4 \arrow[r]\& \mathcal{L}_3\arrow[r]\& 0, \\
S_4 = (a_2X +a_0Z^2)^2,\& 0\arrow[r]\& \mathcal{L}_2 \arrow[r] \& \mathcal{L}_4 \arrow[r]\& \mathcal{L}'_2\arrow[r]\& 0,
\end{tikzcd}
\end{eqnarray}
We consider each case in turn. 
\begin{itemize}
    \item $\mathcal{L}_4$ is supported on $S_4= Z^2 (a_2X+a_0Z^2)$. 
\end{itemize}
We continue by iteration, and extend the result of \ref{sec312}
\begin{eqnarray}\label{V4L1L1L2}
\begin{tikzcd}[ampersand replacement=\&, column sep=small]
0\arrow[r] \& \mathcal{U}_1 \arrow[r] \& \mathcal{L}_4\arrow[r]\& \mathcal{L}_3 \arrow[r]\& 0,
\end{tikzcd}
\end{eqnarray}
where $\mathcal{U}_1$ is supported on $\sigma$, and $\mathcal{L}_3$ is given by \eref{redSU(3)sheaf}. 
First let us study the bundle $V$ itself and then continue to $\Lambda^2V$. The extension group can be derived by using dualising sheaf \cite{BBRH}
\begin{eqnarray}\label{z2s2EXT1}
RHom_{D(X)}(\mathcal{L}_3,\mathcal{U}_1)&=&RHom_{D(B)}(Li^*_{\sigma}\mathcal{L}_3,\mathcal{U}_1)\nonumber\\
&=& RHom_{D(B)}(\mathcal{L}_1\otimes K_B^{-1}[+1] \oplus \mathcal{F},\mathcal{U}_1), 
\end{eqnarray}
where $\mathcal{F}$ is given in \eref{V3pushforward},
\begin{eqnarray}
\begin{tikzcd}[ampersand replacement=\&, column sep=small]
0\arrow[r]\& \mathcal{L}_1 \arrow[r]\& \mathcal{F} \arrow[r] \& i_{\lbrace a_2=0 \rbrace*} \mathcal{L}_2 \arrow[r]\& 0.\nonumber
\end{tikzcd}
\end{eqnarray}
The equation gives a spectral sequence for $Ext^1$ group of the defining short exact sequence
\begin{eqnarray}
\begin{tikzcd}[ampersand replacement=\&]
0\arrow[r]\& Ext^1(\mathcal{F},\mathcal{U}_1)\arrow[r]\& Ext^1(\mathcal{L}_3,\mathcal{U}_1)\arrow[r]\& Hom(\mathcal{L}_1\otimes K_B^{-1},\mathcal{U}_1)\arrow[r]\& Ext^2(\mathcal{L}_3,\mathcal{U}_1)
\end{tikzcd}
\end{eqnarray}
The first term on the left contains the gluing of the line bundle $\mathcal{U}_1$ and the line bundle over $S_2$ i.e. $S_2$. It also contains a term ($Ext^1(\mathcal{L}_1,\mathcal{U}_1)$) that corresponds to making a rank 2 vector bundle over $\sigma$. This later part, if it contributes to $Ext^1(\mathcal{L}_3,\mathcal{U}_1)$, should be kept zero. However the term $Hom(\mathcal{L}_1\otimes K_B^{-1},\mathcal{U}_1)$, again if it contributes to $Ext^1(\mathcal{L}_3,\mathcal{U}_1)$, corresponds to a ``line bundle" over a degree two non-reduced surface, ad we need to keep in on generically (we denote a generic element of this group by $\mathcal{E}xt4$). The next, task as usual, is computing the derived pushforward of $V$,
\begin{eqnarray}
\begin{tikzcd}[ampersand replacement=\&, column sep=small, row sep=small]
0\arrow[r]\& \mathcal{U}_1\otimes K_B^{-1} \arrow[r]\& \pi_*V \arrow[r]\arrow[d, phantom, ""{coordinate, name=Z}]\& \mathcal{L}_1\otimes K_B^{-1}[+1]  \arrow[dll,to path={ -- ([xshift=2ex]\tikztostart.east)
|- (Z) [near end]\tikztonodes
-| ([xshift=-2ex]\tikztotarget.west)
-- (\tikztotarget)}]\& \\
\& \mathcal{U}_1 \arrow[r]\& R^1\pi_*V\arrow[r]\& \mathcal{F}\arrow[r]\& 0.
\end{tikzcd}
\end{eqnarray}
The coboundary may is given by the extension, therefore
\begin{eqnarray}
R\pi_*V = \mathcal{U}_1\otimes K_B^{-1}\oplus \bar{\mathcal{F}} [-1], \\
\begin{tikzcd}[ampersand replacement=\&, column sep=small]
0\arrow[r]\& \mathcal{U}_1|_{\mathcal{E}xt4=0} \arrow[r]\& \bar{\mathcal{F}} \arrow[r]\& \mathcal{F}\arrow[r]\& 0.
\end{tikzcd}
\end{eqnarray}
As usual, one needs the antisymmetrized Pontrjagin product 
\begin{eqnarray}
\begin{tikzcd}[ampersand replacement=\&, column sep=small]
\& \& \&0\arrow[d]\& \\
\& \& \&\pi^*\pi_*\mathcal{L}_1\otimes \mathcal{L}_2 \arrow[d]\& \\
0\arrow[r]\& \pi^*\pi_* \mathcal{U}_1 \otimes \mathcal{L}_3 \arrow[r]\& \mathcal{L}_4\star_A\mathcal{L}_4\arrow[r]\& \mathcal{L}_3\star_A\mathcal{L}_3\arrow[r]\arrow[d]\& 0 \\
\& \& \&i_{\sigma_*}Det \pi_* \mathcal{L}_2\arrow[d]\& \\
\& \& \&0 \&
\end{tikzcd}
\end{eqnarray}
The extension of this sequence is non-zero generally. Before continuing, note that we can read the spectral cover of $\Lambda^2V$ from this sequence,
\begin{eqnarray}
S_{\Lambda^2V} =Z^2 (a_2X+a_0Z^2)^2.
\end{eqnarray}
To compute $R^1\pi_*\Lambda^2V$, restrict $\mathcal{L}_4\star_A\mathcal{L}_4$ over $\sigma$
\begin{eqnarray}
\begin{tikzcd}[ampersand replacement=\&, column sep=small, row sep=small]
0\arrow[r]\& \mathcal{L}_1\otimes\mathcal{U}_1\otimes K_B^{-1} \arrow[r]\& L^{-1}i_{\sigma}^* \mathcal{L}_4\star_A\mathcal{L}_4\arrow[r] \arrow[d, phantom, ""{coordinate, name=Z}]\& L^{-1}i_{\sigma}^* \mathcal{L}_3\star_A\mathcal{L}_3 \arrow[dll, to path={ -- ([xshift=2ex]\tikztostart.east)
|- (Z) [near end]\tikztonodes
-| ([xshift=-2ex]\tikztotarget.west)
-- (\tikztotarget)}]\&\\
\& \mathcal{F}\otimes \mathcal{U}_1\arrow[r]\&  L^{0}i_{\sigma}^* \mathcal{L}_4\star_A\mathcal{L}_4\arrow[r] \& i_{\sigma}^* \mathcal{L}_3\star_A\mathcal{L}_3 \arrow[r]\&0.
\end{tikzcd}
\end{eqnarray}
The coboundary map of this long exact sequence is induced by the extension of the $\mathcal{L}_4\star_A\mathcal{L}_4$ sequence. As n the previous sections, there is a \textbf{10-16-16} candidate  
\begin{eqnarray}
\begin{tikzcd}[ampersand replacement=\&, column sep=small]
 H^1(\mathcal{U}_1\otimes K_B^{-1}) \otimes H^0(\mathcal{F}) \arrow[r] \& H^1(\mathcal{U}_1\otimes \mathcal{F}\otimes K_B^{-1}). 
\end{tikzcd}
\end{eqnarray}
As usual, one should check whether these cohomology groups contribute to $H^1(V)$ and $H^1(\Lambda^2V)$ or not. 

We should also mention that since $\mathcal{L}_4$ is defined by extending $\mathcal{L}_3$ with $\mathcal{U}_1$, one may expect the Yukawa couplings of the $\mathcal{L}_3$ should somehow contribute to the Yukawas for the $SU(4)$ bundle we constructed. However, note that the Yukawa couplings of the $SU(3)$ bundle depends on the zero modes $H^1(\mathcal{L}_1\otimes K_B^{-1})$. But these modes are killed by the extension morphism of the line bundles on the two copies of the zero section (the $Z^2$ part of $S_4$) i.e. $Hom(\mathcal{L}_1\otimes K_B^{-1},\mathcal{U}_1)$. So if one restricts to non-trivial spectral sheaf with a non-reduced component $Z^2$, the Yukawa couplings of $\mathcal{L}_3$ will not contribute.

The F-theory dual is quite similar to the $SU(3)$ bundle of \ref{sec312}.  The only difference is that the flux that lives inside the 7-brane bulk is $SU(2)$. So we ignore the detailed analysis.
\begin{itemize}
    \item $S_4=S_2^2=(a_2 X+a_0 Z^2)^2$
\end{itemize}
The defining sequence is given by
\begin{eqnarray}\label{S2^2}
\begin{tikzcd}[ampersand replacement=\&, column sep=small]
0\arrow[r]\& \mathcal{L}_2\arrow[r]\& \mathcal{L}_4\arrow[r]\& \mathcal{L}'_2 \arrow[r]\& 0,
\end{tikzcd}
\end{eqnarray}
where $\mathcal{L}_2$ and $\mathcal{L}_2'$ are two different line bundles supported over the same double cover $S_2=a_2X+a_0Z^2$. The extension of this short exact sequence, should be chosen such that $\mathcal{L}_4$ correspond to the a ``line bundle" (or numerically rank one to be precise) over the non-reduced surface $S_4$. The extension can be derived as usual by using the dualising sequence
\begin{eqnarray}
RHom_X(\mathcal{L}'_2,\mathcal{L}_2) = RHom_{S_2}(\mathcal{L}'_2,\mathcal{L}_2\otimes(\mathcal{O}\oplus \mathcal{O}_{S_2}(S_2)[-1])).
\end{eqnarray}
Therefore one gets a spectral sequence corresponding to this
\begin{eqnarray}
\begin{tikzcd}[ampersand replacement=\&, column sep=small]
0\arrow[r]\& Ext^1_{S_2}(\mathcal{L}_2',\mathcal{L}_2)\arrow[r]\& Ext^1_{X}(\mathcal{L}_2',\mathcal{L}_2)\arrow[r]\& Hom_{S_2}(\mathcal{L}_2',\mathcal{L}_2\otimes\mathcal{O}(S_2))\arrow[r,"\alpha"]\& Ext^2_{S_2}(\mathcal{L}_2',\mathcal{L}_2)\dots
\end{tikzcd}
\end{eqnarray}
So the first term on the left give a rank 2 bundle over $S_2$ and this is not a spectral sheaf. However, the third term, if the are in the kernel of $\alpha$, is a numerically rank one sheaf over the non reduced surface $S_4$.  For finding $H^*(V)$ we need the derived pushforward. It since $S_4$ intersects transversely with the zero section, only the first derived pushforward is non-zero
\begin{eqnarray}
R^1\pi_*V=i_{\sigma}^* \mathcal{L}_4 \Rightarrow H^i(V)=H^{i-1}(i_{\sigma}^* \mathcal{L}_4),
\end{eqnarray}
\begin{eqnarray}
\begin{tikzcd}[ampersand replacement=\&, column sep=small]
0\arrow[r]\& \mathcal{L}_2|_{c}\arrow[r]\& i_{\sigma}^* \mathcal{L}_4\arrow[r]\& \mathcal{L}'_2|_{c}\arrow[r]\& 0.
\end{tikzcd}
\end{eqnarray}
Note that the extension morphism of this sequence is induced by the extension group of the original sequence. The antisymmetrized Pontrjagin product can be computed by the following diagram
\begin{eqnarray}\label{L4L22}
\begin{tikzcd}[ampersand replacement=\&, column sep=small]
 \&0\arrow[d] \& \& \& \\
 \&\mathcal{L}_2\star_A \mathcal{L}_2\arrow[d]  \& \& \& \\
0\arrow[r] \& Q\arrow[d]\arrow[r] \& \mathcal{L}_4\star_A\mathcal{L}_4\arrow[r] \&\mathcal{L}'_2 \star_A \mathcal{L}'_2 \arrow[r] \& 0 \\
 \&\mathcal{L}_2\star\mathcal{L}'_2\arrow[d] \& \& \& \\
 \&0 \& \& \& \\
\end{tikzcd}
\end{eqnarray}
As in the double cover calculations, $\mathcal{L}_2\star_A\mathcal{L}_2=i_{\sigma*}Det(\pi_*\mathcal{L}_2)$ and $\mathcal{L}_2'\star_A\mathcal{L}_2'=i_{\sigma*}Det(\pi_*\mathcal{L}_2')$, so they are supported over the zero section. On the other hand, since the support of $\mathcal{L}_2$ and $\mathcal{L}_2'$ are the same, the support of $\mathcal{L}_2\star_A\mathcal{L}_2'$ should be reducible as $Z^2 S_2'$. So the spectral cover of $\Lambda^2V$ can be computed as \eref{S6V4S2S2'}
\begin{eqnarray}
S_{\Lambda^2V}=Z^4(4c_{tor} a_2X+a_0Z^2),
\end{eqnarray}
where $c_{tor}$ is the part of the branching locus of the double cover $S_2$ which corresponds to the torsion points of the elliptic fibration. To find $R^1\pi_*\Lambda^2V$, we restrict $\mathcal{L}_4\star_A\mathcal{L}_4$ to the zero section,
\begin{eqnarray}
Li_{\sigma}^* \mathcal{L}_4\star_A\mathcal{L}_4 =&& Det\pi_*\mathcal{L}'_2 \oplus Det \pi_* \mathcal{L}_2 \oplus \Gamma(\mathcal{L}_2,\mathcal{L}'_2) \oplus\nonumber \\ 
&&\left(Det\pi_*\mathcal{L}'_2  \oplus Det \pi_* \mathcal{L}_2 \right) \otimes K_B^{-1} [+1] \oplus  \overline{\Gamma}(\mathcal{L}_2,\mathcal{L}'_2)[+1],
\end{eqnarray}
where $\Gamma(\mathcal{L}_2,\mathcal{L}'_2)$ and $\overline{\Gamma}(\mathcal{L}_2,\mathcal{L}'_2)$ are a rank two bundles on $B$, derived from $\mathcal{L}_2\star_A \mathcal{L}'_2$ part. We will derive them in the examples. The details of $\Gamma$ and $\overline{\Gamma}$ are necessary to find the Yukawa couplings. 

In case the line bundles $\mathcal{L}_2$ and $\mathcal{L}_2'$ can be written as $\mathcal{O}_{S_2}(\sigma+D_1)$ and $\mathcal{O}_{S_2}(\sigma+D_2)$ respectively, there are Yukawa couplings from the bulk and localised zero modes interactions. We leave the details for the examples (the $SU(5)$ bundle example). First remember that
\begin{eqnarray}
c_{1}(\pi_* \mathcal{L}_4) &=& c_1(\pi_* \mathcal{L}_2) + c_1(\pi_* \mathcal{L}_2') = 4 c_1(K_B), \nonumber \\
&\Rightarrow& -\eta +D_1+D_2= c_1(K_B),
\end{eqnarray}
where the divisor class of the double cover is $[S_2]=2\sigma+\eta$. Also we can compute the derived pushforwards of $\Lambda^2V$,
\begin{eqnarray}
R\pi_*\Lambda^2V &=& \left[\mathcal{O}\oplus \mathcal{O}(K_B-2D_1-2D_2)\oplus \mathcal{O}(-\eta+2D_1) \oplus \mathcal{O}(-\eta +2D_2) \right] \\
&\oplus& \left[K_B\oplus \mathcal{O}(2D_1+2D_2)\oplus \mathcal{O}(K_B-\eta+2D_1) \oplus \mathcal{O}(K_B-\eta +2D_2) \right] [-1].
\end{eqnarray}
With these information (also remember $K_c= K_B+[c]=3K_B+\eta$) it is possible find the possible Yukawa couplings (modulo the coboundary maps as usual)
\begin{eqnarray}
\begin{tikzcd}[ampersand replacement=\&]
H^0(\mathcal{O}_c(K_B+D_1))\otimes H^0(\mathcal{O}_c(K_B+D_1))\otimes H^1(\mathcal{O}(-\eta+2D_2))\longrightarrow H^1(\mathcal{O}_c(K_c))\simeq \mathbb{C}, \\
H^0(\mathcal{O}_c(K_B+D_2))\otimes H^0(\mathcal{O}_c(K_B+D_2))\otimes H^1(\mathcal{O}(-\eta+2D_1))\longrightarrow H^1(\mathcal{O}_c(K_c))\simeq \mathbb{C}, \\
H^0(\mathcal{O}_c(K_B+D_1))\otimes H^0(\mathcal{O}_c(K_B+D_2))\otimes H^1(\mathcal{O})\longrightarrow H^1(\mathcal{O}_c(K_c))\simeq \mathbb{C}.
\end{tikzcd}
\end{eqnarray}
F-theory description of this is very similar to the reducible (but reduced) spectral cover $S_4=S_2S_2'$. So we comment on the F-theory dual briefly. The F-theory geometry has a split singularity $I_2^{*s}$ over the GUT 7-brane, which corresponds a $SO(12)$ effective gauge group. This would be the case if the extension morphism of the defining short exact sequence were chosen to be zero. The first difference relative to $S_4=S_2S_2'$ case is that over the curve $c$ the singularity enhances to $E_8$. We can see the adjoint of the of $E_8$ decomposes as
\begin{eqnarray}
E_8 &\longrightarrow& SU(2)\times SU(2)\times SO(12), \\
248 &\longrightarrow& (\textbf{3,1,1}) \oplus (\textbf{1,3,1})\oplus (\textbf{1,2,32'})\oplus (\textbf{1,1,66}) \oplus (\textbf{2,2,12}) \oplus (\textbf{2,1,32}). 
\end{eqnarray}
The seoncd difference is that the \textbf{12}s that are responsible for Higgsin $SO(12)$ to $SO(10)$ (or equivalently the extension maps) are non-localised inside the 7-brane. Geometrically they correspond to the deformations of $S_2$ (equivalently deformations of the 7-branes). So after Higgsing the remaining Yukawa couplings \textbf{12-32-32'} (\textbf{12}s from $H^1(\Lambda^2V)$) decompose to \textbf{10-16-16}. The third differece is that in this example the \textbf{16}s  are localise over the 7-brane, and \textbf{10}s live in the bulk. 

\subsection{$SU(5)$ Model}\label{sec33}
We conclude this section with $SU(5)$ models, which can be constructed by putting a $SU(5)$ vector bundle $V$ over the Calabi-Yau $X$. The strategy is the same as before. Simply rewrite the spectral data as extension of a line bundles on different components, and use the results have been derived for lower degree spectral covers in the previous sections. 

The new feature of $SU(5)$ models is that there are two possible Yukawa couplings, \textbf{10-10-5} and \textbf{10-$\overline{\textbf{5}}$-$\overline{\textbf{5}}$}, where the first one corresponds to the usual map,
\begin{eqnarray}
H^1(V) \otimes H^1(V) \otimes H^1(\Lambda^3V)\longrightarrow \mathbb{C} \quad \Longrightarrow\quad H^1(V)\otimes H^1(V) \longrightarrow H^2(\Lambda^2V) ,
\end{eqnarray}
while the second Yukawa coupling corresponds to
\begin{eqnarray}
H^1(V)\otimes H^1(\Lambda^2 V)\otimes H^1(\Lambda^2V)\longrightarrow \mathbb{C}.
\end{eqnarray}

\subsubsection{Smooth Spectral Cover}\label{331}
\begin{figure}
    \centering
    \includegraphics[width=0.5\textwidth]{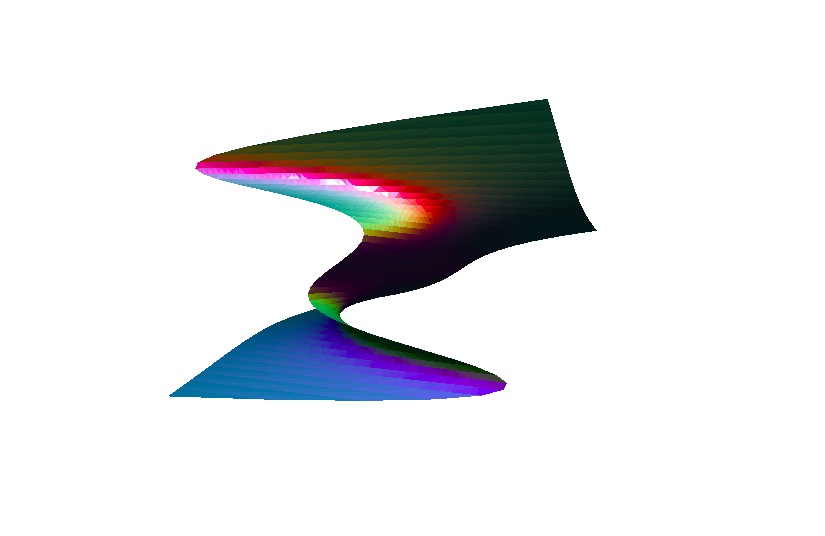}
    \caption{Smooth degree five spectral cover}
    \label{fig:5pl smooth}
\end{figure}

As always the starting point is the generic spectral cover, Figure \ref{fig:5pl smooth}. We start with computing the pushforward $R^1 \pi_* \Lambda^2 V$ for smooth generic spectral cover of $SU(5)$ bundle,
\begin{eqnarray}
S_5= a_5 X Y+a_4 X^2Z+a_3YZ^2+a_2XZ^3+a_0Z^5.
\end{eqnarray}
Consider a generic fiber, and the spectral cover intersect with it at five points $P_1,\dots,P_5$ such that $P_1\boxplus \dots \boxplus P_5=0$. So $S_{\Lambda^2V}$ intersect with the same fiber at the following ten points,
\begin{eqnarray}
P_1\boxplus P_2,\quad P_1\boxplus P_3, \dots,\quad P_4\boxplus P_5.
\end{eqnarray}
So the contribution to the $R^1\pi_*\Lambda^2V$ coming from a double cover. Similar to $SU(3)$ and $SU(4)$ bundles, if for example $P_1\boxplus P_2=0$ then $P_3\boxplus P_4\boxplus P_5=0$. So we should look for a locus where the spectrak cover decomposes into a degree two and a degree three cover. In other words, the non-zero contribution is coming from $S_5 \cap \tau(S_5)$, 
\begin{eqnarray}\label{smoothS5dec}
a_5 X+a_3Z^2=0,\quad a_4 X^2+a_2 X Z^2++a_0 Z^4=0,
\end{eqnarray}
by eliminating $X$ in these equations we get a double cover $\tilde{c}$ over curve on base defined by
\begin{eqnarray}
\tilde{c}:\quad a_5 X+a_3Z^2=0,\quad a_4 a_3^2 -a_2a_3a_5+a_0a_5^2=0.
\end{eqnarray}
So $R^1\pi_*\Lambda^2V$ is given by the formula \eref{Final}, but the double cover $\tilde{c}$ defined as above. 

Similar to the previous Yukawa couplings for smooth spectral covers. There are pointlike couplings (both \textbf{10-10-5} and \textbf{$\overline{\textbf{5}}$-$\overline{\textbf{5}}$-10}) coming from the intersection of the matter curve (or the \textbf{10} curve $c$) $a_5=0$ and the \textbf{5} curve (or $\overline{\textbf{5}}$ curve) $a_4 a_3^2 -a_2a_3a_5+a_0a_5^2=0$ ($c'$). The F-theory dual is quite simple. Inside the $SU(5)$ GUT 7-brane the singularity enhances to $SO(10)$ over $c$. This is not surprising since either from the heterotic view of local F-theory Higgs bundle, the spectral cover reduces to $S_4$. Over this curve the \textbf{10} and $\overline{\textbf{10}}$ hypermultiplets live, as can be seen by the decomposition of \textbf{45} of $SO(10)$,
\begin{eqnarray}
SO(10) &\longrightarrow& SU(5)\times U(1), \\
\textbf{45} &\longrightarrow& \textbf{1}_0 \oplus \textbf{10}_4\oplus \overline{\textbf{10}}_{-4} \oplus \textbf{24}_0.
\end{eqnarray}
Over the curve $c'$, the singularity enhance by $1$, i.e. the order of the discriminant jumps from 5 to 6. However, the spectral cover becomes reducible \eref{smoothS5dec}, so the singularity of the F-theory geometry also will be of split type, i.e. $I_6^s$. So the singularity enhances to $SU(6)$ over $c'$. Again this is not unexpected from heterotic side since the bundle (ignoring the extension morphisms) becomes $S(U(2)\times U(3))$ over $c'$, and its commutant in $E_8$ is $SU(6)\times U(1)$ (the extra $U(1)$ is anomalous). By decomposing the \textbf{35} of $SU(6)$ one can figure out that the \textbf{5} and $\overline{\textbf{5}}$ hypermultiplets live over $c'$
\begin{eqnarray}
SU(6) &\longrightarrow& SU(5)\times U(1), \\ 
\textbf{35} &\longrightarrow& \textbf{1}_0\oplus \textbf{5}_6\oplus \overline{\textbf{5}}_{-6} \oplus\textbf{24}.
\end{eqnarray}
Hence we get the advertised Yukawa couplings overs the intersections of $c$ and $c'$.
\subsubsection{Reducible Reduced Spectral Covers}\label{sec332}
\begin{figure}
    \centering
    \includegraphics[width=0.33\textwidth]{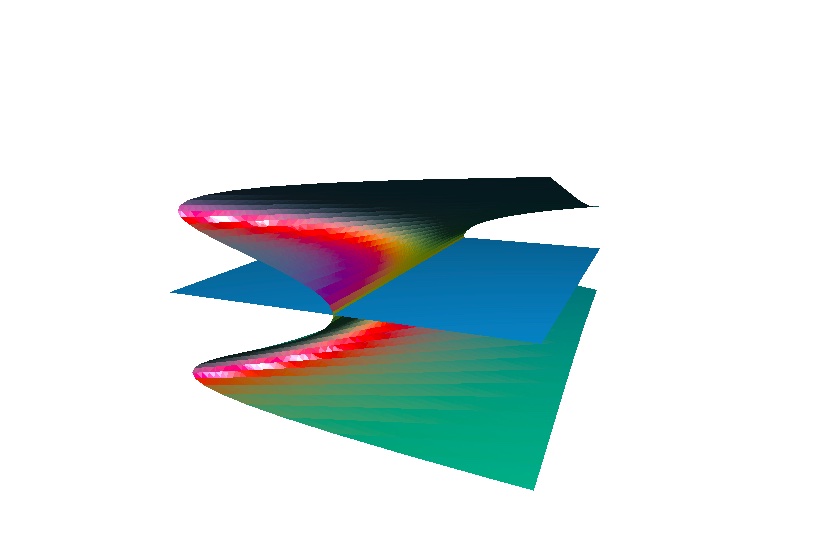}\includegraphics[width=0.33\textwidth]{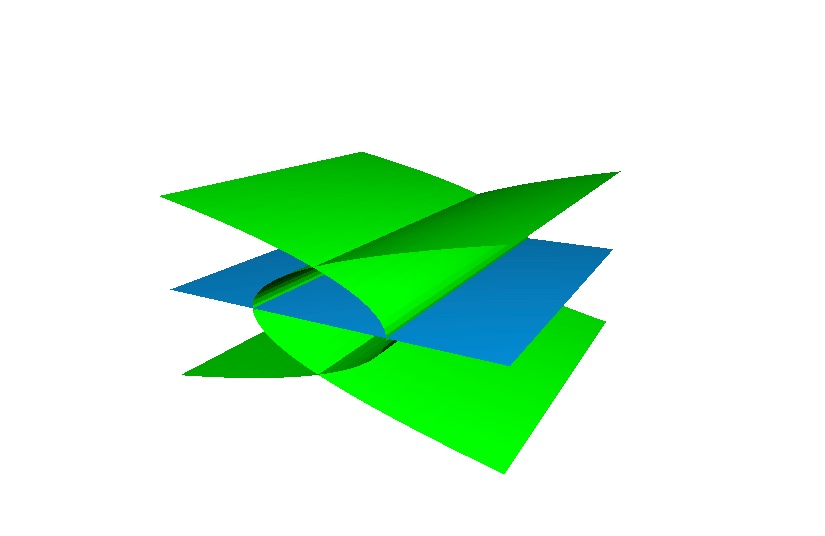}\includegraphics[width=0.33\textwidth]{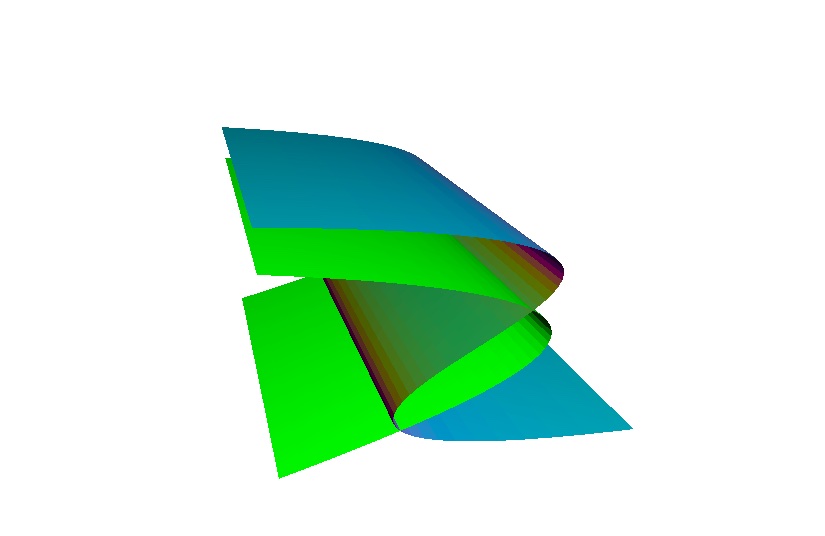}
    \caption{Reducible degree five spectral covers with smooth components.}
    \label{fig:redV5}
\end{figure}

In this case there are three possibilities (see Figure \ref{fig:redV5}),
\begin{eqnarray}
\begin{tikzcd}[ampersand replacement=\&, column sep=small]
S_5=Z(a_4 X^2 + a_3 Y Z + a_2 X Z^2 +a_0 Z^4),\& 0\arrow[r]\& \mathcal{W}_1 \arrow[r] \& \mathcal{L}_5 \arrow[r]\& \mathcal{L}_4|_{S_4} \arrow[r] \& 0, \\
S_5=Z(a_2 X + a_0Z^2)(b_2 X+b_0 Z^2),\& 0\arrow[r]\& \mathcal{W}_1 \arrow[r] \& \mathcal{L}_5 \arrow[r]\& \mathcal{L}_4|_{S_2 \tilde{S}_2} \arrow[r] \& 0, \\
S_5=S_2 S_3\& 0\arrow[r]\& \mathcal{L}_2|_{S_2} \arrow[r] \& \mathcal{L}_5 \arrow[r]\& \mathcal{L}_3|_{S_3} \arrow[r] \& 0.
\end{tikzcd}
\end{eqnarray}
We study each case, by iteration as before. 
\begin{itemize}
    \item $S_5=Z(a_4 X^2 + a_3 Y Z + a_2 X Z^2 +a_0 Z^4)$
\end{itemize}
This case, i.e. a reducible spectral cover with one component being zero section and the other component being generic, should be easy by now. The dfining short exact sequence is 
\begin{eqnarray}
\begin{tikzcd}[ampersand replacement=\&, column sep=small]
0\arrow[r]\& \mathcal{W}_1 \arrow[r] \& \mathcal{L}_5 \arrow[r]\& \mathcal{L}_4|_{S_4} \arrow[r] \& 0.
\end{tikzcd}
\end{eqnarray}
The extension morphisms live inside the group
\begin{eqnarray}
Ext^1(\mathcal{L}_4,\mathcal{W}_1) = H^0(c,\mathcal{W}_1\otimes\mathcal{L}_4^* \mathcal{O}_c(c)),
\end{eqnarray}
where $c=S_4\cap \sigma= \lbrace a_4=0 \rbrace$. To find $H^1(V)$ we need the derived pushforwards of $V$
\begin{eqnarray}
&R\pi_*V = \mathcal{W}_1 \otimes K_B^{-1} \oplus \mathcal{F}[-1],& \nonumber \\
&\begin{tikzcd}[ampersand replacement=\&, column sep=small]
0\arrow[r] \& \mathcal{W}_1 \arrow[r]\& \mathcal{F} \arrow[r]\& \mathcal{L}_4|_{a_4=0} \arrow[r]\& 0.
\end{tikzcd}&
\end{eqnarray}
For $R\pi_* \Lambda^2 V$, again we use the antisymmetrized sequence,
\begin{eqnarray}
\begin{tikzcd}[ampersand replacement=\&, column sep=small]
0\arrow[r] \& \pi_1^* \mathcal{W}_1 \otimes \pi_2^* \mathcal{L}_4|_{S_4} \arrow[r]\& \pi_1^* \mathcal{L}_5 \otimes_A \pi_2^*\mathcal{L}_5 \arrow[r] \& \pi_1^* \mathcal{L}_4 \otimes_A \pi_4^*\mathcal{L}_2 \arrow[r] \& 0.
\end{tikzcd}
\end{eqnarray}
Hence,
\begin{eqnarray}
\begin{tikzcd}[ampersand replacement=\&, column sep=small]
0\arrow[r] \& R^0m_* \pi_1^* \mathcal{W}_1 \otimes \pi_2^* \mathcal{L}_4|_{S_4} \arrow[r]\& R^0m_* \pi_1^* \mathcal{L}_5 \otimes_A \pi_2^*\mathcal{L}_5 \arrow[r] \& R^0m_*  \pi_1^* \mathcal{L}_4 \otimes_A \pi_2^*\mathcal{L}_4 \arrow[r] \& 0.
\end{tikzcd}
\end{eqnarray}
Acting on this with $Li^*_{\sigma}$,
\begin{eqnarray}
\begin{tikzcd}[ampersand replacement=\&, column sep=small]
0\arrow[r] \& (\mathcal{W}_1 \otimes \mathcal{L}_4)|_{a_4=0} \arrow[r] \& i_{\sigma}^*\mathcal{L}_5 \star_A\mathcal{L}_5 \arrow[r] \& P_* \mathcal{L}_4|_{a_3^2=0}\arrow[r] \& 0,
\end{tikzcd}
\end{eqnarray}
where $P$ is the morphism defined in \ref{sec321}. As a byproduct, it is possible to read the spectral cover of $\Lambda^2V$ from this sequence
\begin{eqnarray}
S_{\Lambda^2V}=(-a_3^2 X^3+(a_2^2-4a_0a_4) X^2 Z^2 + a_0^2 Z^6)(a_4 X^2 + a_3 Y Z + a_2 X Z^2 +a_0 Z^4).
\end{eqnarray}
So similar to the previous section there is a candidate for non-vanishing Yukawa coupling \textbf{10-10-5} when the following cohomologies contribute in $H^1(V)$ and $H^2(\Lambda^2V)$
\begin{eqnarray}
H^1(B,\mathcal{W}_1\otimes K_B^{-1}) \otimes H^0(\mathcal{L}_4|_{a_4=0}) \longrightarrow H^1((\mathcal{W}_1 \otimes \mathcal{L}_4)|_{a_4=0}\otimes K_B^{-1}).
\end{eqnarray}
There are also several possibilities for the \textbf{10-$\overline{\textbf{5}}$-$\overline{\textbf{5}}$} couplings, which of course have to be checked whether the following cohomologies inject into the $H^1(V)$ and $H^1(\Lambda^2V)$,
\begin{eqnarray}
H^0(\mathcal{W}_1)&\otimes& H^0((\mathcal{W}_1 \otimes \mathcal{L}_4)|_{a_4=0})\otimes H^0(P_* \mathcal{L}_4|_{a_3^2=0}) \longrightarrow \mathbb{C}, \\
H^0(\mathcal{L}_4|_{a_4=0})&\otimes& H^0((\mathcal{W}_1 \otimes \mathcal{L}_4)|_{a_4=0})\otimes H^0(P_* \mathcal{L}_4|_{a_3^2=0})\longrightarrow \mathbb{C}, \\
H^0(\mathcal{L}_4|_{a_4=0})&\otimes& H^0(P_* \mathcal{L}_4|_{a_3^2=0})\quad\quad \otimes H^0(P_* \mathcal{L}_4|_{a_3^2=0})\longrightarrow \mathbb{C}.
\end{eqnarray}
F-theory dual again is given in terms of T-branes like \ref{sec312} . From the spectral data it is clear that some of the \textbf{10}s are living in the bulk of the $SO(10)$ 7-brane, while some are in the $a_4=0$ curve. However, all of the \textbf{5} and $\overline{\textbf{5}}$ live in either $a_4=0$ or $a_4=0$. 

To see this in F-theory, first note that the Calabi-Yau's singularity is $SO(10)$, this means there is a $U(1)$ gauge group inside the 7-brane (T-brane). In principle this $U(1)$ and the associated zero modes in the 7-brane bulk can be read from the spectral data or equivalently from the Higgs bundle hypercohomology \cite{Donagi:2011jy}. However this outside of the scope of this paper. The singularity structure of the F-theory geometry is basically the same as smooth $S_4$. In other words there are enhancements to $SO(11)$ and $E_6$ over $a_3=0$ and $a_4=0$ respectively. Purely from geometry one might say the gauge group in the effective theory is $SO(10)$, and there are \textbf{10}s of $SO(10)$ in $a_3=0$ and \textbf{16}s and $\overline{\textbf{16}}$s on $s_4=0$. In addition there is a gauge field in the bulk in the \textbf{45} (adjoint) of $SO(10)$. Now consider the effect of the $U(1)$ living in the bulk of the 7-brane. First of all the \textbf{45} breaks as
\begin{eqnarray}
\textbf{45}\longrightarrow \textbf{1}_0\oplus \textbf{10}_4\oplus \overline{\textbf{10}}_{-4}\oplus \textbf{24}_0.
\end{eqnarray}
So one gets the bulk \textbf{10}s and $\overline{\textbf{10}}$s of the $SU(5)$, identified with $H^1(\mathcal{W}_1\otimes K_B^{-1})$. The \textbf{16} that lives in $a_4=0$ breaks as
\begin{eqnarray}
\textbf{16} \longrightarrow 1_{-5}\oplus \overline{\textbf{5}}_3\oplus \textbf{10}_{-1}.
\end{eqnarray}
Similar decomposition for $\overline{\textbf{16}}$s. So this is the source of the \textbf{5}s and \textbf{10}s (and their conjugates) of $SU(5)$ over $a_4=0$. Similarly the \textbf{10} of $SO(10)$ over $a_3=0$ break into \textbf{5} and $\overline{\textbf{5}}$ of $SU(5)$. So we see there are no \textbf{10}s or $\overline{\textbf{10}}$s of $SU(5)$ on $a_3=0$. In agreement with the heterotic calculations.
\begin{itemize}
    \item $S_5=Z(a_2 X + a_0Z^2)(b_2 X+b_0 Z^2)$
\end{itemize}
In this case there is \textbf{10-10-5} from $H^1(V)\otimes H^1(V)\otimes H^1(\Lambda^3V)$ in the usual way. There are \textbf{$\overline{\textbf{5}}$-$\overline{\textbf{5}}$-10} candidates from
\begin{eqnarray}
H^0\left(c_{\Lambda^2V_4},\pi_*(\mathcal{L}_2\otimes \tau^* \mathcal{L}'_2)\right) \otimes H^0(c, \mathcal{W}_1 \otimes \mathcal{L}_2|_{\sigma}) \otimes H^0(c',\mathcal{L}'_2)\longrightarrow \mathbb{C},
\end{eqnarray}
where the first two terms can contribute in $H^1(\Lambda^2V)$ and the third one contribute in $H^1(V)$. The curves $c_{\Lambda^2V_4}$, $c$ and $c'$ are given by equations $a_2 b_0 -a_0b_2=0$, $a_2=0$ and $b_2=0$ respectively. One can also swap the $\mathcal{L}_2$ and $\mathcal{L}'_2$ in the last two factors. To be complete let us add the details in this case too, take the defining sequence
\begin{eqnarray}\label{V5S2S2P}
\begin{tikzcd}[ampersand replacement=\&, column sep=small]
0\arrow[r]\& \mathcal{W}_1\arrow[r]\& \mathcal{L}_5\arrow[r]\& \mathcal{L}_4\arrow[r]\& 0,
\end{tikzcd}
\end{eqnarray}
the atisymmetrized product can be found as usual,
\begin{eqnarray}
\begin{tikzcd}[ampersand replacement=\&, column sep=small]
0\arrow[r]\& \pi^*\pi_*\mathcal{W}_1\otimes \mathcal{L}_4\arrow[r]\& \mathcal{L}_5\star_A\mathcal{L}_5 \arrow[r]\& \mathcal{L}_4\star_A\mathcal{L}_4 \arrow[r]\& 0,
\end{tikzcd}
\end{eqnarray}
where $\mathcal{L}_4\star_A\mathcal{L}_4$ is given in \eref{L4wL2L2}. Therefore the spectral cover is
\begin{eqnarray}
S_{\Lambda^2V}= Z^2 \left((a_2 b_0 -a_0b_2)^2X^2+2a_0b_0(a_2 b_0 +a_0b_2)X Z^2+a_0^2b_0^2 Z^4\right)(a_2 X + a_0Z^2)(b_2 X+b_0 Z^2).\nonumber \\
\quad
\end{eqnarray}
Therefore the \textbf{10-10-5} coupling is related to the following cohmologies,
\begin{eqnarray}
H^0(B, \mathcal{W}_1\otimes K_B^{-1})\otimes H^0(c\cup c', \mathcal{L}_4|_{a_2b_2=0})\longrightarrow H^1(\pi^*\pi_*\mathcal{W}_1\otimes \mathcal{L}_4|_{a_2b_2=0}\otimes K_B^{-1}),
\end{eqnarray}
which is coming from the bulk and localised zero modes.

F-theory view is also similar to the previous case. Adding a copy of the zero section in the spectral cover is equivalent to a $5\times 5$ Higgs filed with rank four inside the 7-brane, hence there will be q bulk gauge field living inside the surface the 7-brane wraps. This $U(1)$ flux breaks the $SO(10)$ further to $SU(5)$ compatible with the heterotic view. Consequently there will be \textbf{10} hypermultiplets of $SU(5)$ identified with bulk zero modes $H^1(\mathcal{W}_1\otimes K_B^{-1})$. The \textbf{16}s and $\overline{\textbf{16}}$s of the $SO(10)$ on $c$ and $c'$ break into the \textbf{10} and \textbf{5} (and conjugates) of $SU(5)$. Similarly the \textbf{10} over $c_{\Lambda^2V_4}$ will break into \textbf{5} and $\overline{\textbf{5}}$.  
\begin{itemize}
    \item $S_5=S_2S_3 = (a_2X+a_0Z^2)(b_3 Y+b_2XZ+b_0 Z^3)$
\end{itemize}
Start with the defining short exact sequence,
\begin{eqnarray}
\begin{tikzcd}[ampersand replacement=\&, column sep=small]
0\arrow[r]\& \mathcal{L}_2\arrow[r]\& \mathcal{L}_5\arrow[r]\& \mathcal{L}_3 \arrow[r]\& 0.
\end{tikzcd}
\end{eqnarray}
The extension morphism of this sequence lives in the following group,
\begin{eqnarray}
RHom_{D(X)}(\mathcal{L}_3,\mathcal{L}_2) &=& RHom_{D(S_3)} (\mathcal{L}_3,\mathcal{L}_2|_{S_2\cdot S_3}\otimes \mathcal{O}(S_3)[-1])\nonumber \\
&\Downarrow& \\
Ext_X^1(\mathcal{L}_3,\mathcal{L}_2) &=& H^0\left(S_2\cdot S_3,\mathcal{L}_3^*\otimes \mathcal{L}_2\otimes \mathcal{O}(S_3)\right).\nonumber
\end{eqnarray}
Since $S_5$ intersects with zero section transversely, $\pi_*V=0$. $H^1(V)$ can be computed easily as,
\begin{eqnarray}
&H^1(V)= H^0 (R^1\pi_*V),& \\
&\begin{tikzcd}[ampersand replacement=\&, column sep=small]
0\arrow[r]\& \mathcal{L}_2|_{c_2}\arrow[r]\& R^1\pi_*V\arrow[r]\& \mathcal{L}_3|_{c_3} \arrow[r]\& 0,
\end{tikzcd}&
\end{eqnarray}
where $c_2$ and $c_3$ are the curves defined by $a_2=0$ and $b_3=0$. The extension of the sequence above is induced by the extension of the original sequence. The next part as usual is to compute the $R\pi_*\Lambda^2V$ via antisymmetrized Pontrjagin product 
\begin{eqnarray}
\begin{tikzcd}[ampersand replacement=\&, column sep=small]
 \&0\arrow[d] \& \& \& \\
 \&\mathcal{L}_2\star_A \mathcal{L}_2\arrow[d]  \& \& \& \\
0\arrow[r] \& Q\arrow[d]\arrow[r] \& \mathcal{L}_5\star_A\mathcal{L}_5\arrow[r] \&\mathcal{L}_3 \star_A \mathcal{L}_3 \arrow[r] \& 0 \\
 \&\mathcal{L}_2\star \mathcal{L}_3\arrow[d] \& \& \& \\
 \&0 \& \& \& \\
\end{tikzcd}
\end{eqnarray}
We need to restrict this on the zero section as usual. Remember $\mathcal{L}_2\star_A\mathcal{L}_2 = i_{\sigma*} Det(\pi_*\mathcal{L}_2)$, and $Li_{\sigma}^*\mathcal{L}_3\star_A\mathcal{L}_3=i_{c_3*}Det(\mathcal{L}_3|_{\tilde{c}})$ .\footnote{$\tilde{c}$ is the double cover of $c_3$ define in \ref{sec311}.} So we only need to find the restriction of $\mathcal{L})2\star \mathcal{L}_3$ on the zero section. Then suppose $S_2$ intersect a generic fiber at two points $P_1$ and $P_2$ ($P_1\boxplus P_2=0$), and $S_3$ intersect at $Q_1$, $Q_2$ and $Q_3$. Then $\mathcal{L})2\star \mathcal{L}_3$ is supported on the following points
\begin{eqnarray}
P_1&\boxplus& Q_1,\quad P_1\boxplus Q_2,\quad P_1\boxplus Q_3 \nonumber\\
P_2&\boxplus& Q_1,\quad P_2\boxplus Q_2,\quad P_2\boxplus Q_3.
\end{eqnarray}
So the zero section intersect when one of these points say $P_1\boxplus Q_1$ becomes zero, which means $P_2=\mathcal{Q}_1$. We conclude that the curve $\tau^*S_2 \cdot S_3$ maps to the ``matter curve" of $\mathcal{L})2\star \mathcal{L}_3$ (which is a $2:1$ map as will be mentioned shortly). Therefore the derived pushforwards of $\Lambda^2V$ is given by
\begin{eqnarray}
R\pi_*\Lambda^2V=Det(\pi_*\mathcal{L}_2)\otimes K_B^{-2} \oplus \mathcal{H}\otimes K_B^{-1} [-1], 
\end{eqnarray}
where $\mathcal{H}$ is
\begin{eqnarray}
\begin{tikzcd}[ampersand replacement=\&, column sep=small]
 \&0\arrow[d] \& \& \& \\
 \&Det(\pi_*\mathcal{L}_2)\arrow[d]  \& \& \& \\
0\arrow[r] \& Q|_{\sigma}\arrow[d]\arrow[r] \& \mathcal{H}\arrow[r]\&Det(\pi_*\mathcal{L}_3|_{\tilde{c}}) \arrow[r] \& 0 \\
 \&\pi_*\mathcal{L}_2\otimes\mathcal{L}_3|_{S_2\cdot S_3}\arrow[d] \& \& \& \\
 \&0 \& \& \& \\
\end{tikzcd}
\end{eqnarray}
where $\tilde{c}$ is the double cover of $b_3=0$ defined previously for $E_6$ models, and $S_2\cdot S_3$ is a double cover of \footnote{$F$ and $G$ are the defining polynomials in the Weierstrass model.}
\begin{eqnarray}
c_{\Lambda^2V_5}:=a_0^3b_3^2+a_0^2a_2b_2^2-2a_0a_2^2b_0b_2+a_2^3b_0^2+a2^2b_3^2(a_0 F-a_2G).
\end{eqnarray}
This model has point like \textbf{10-$\overline{\textbf{5}}$-$\overline{\textbf{5}}$} candidates on the intersection locus of the matter curves.
\begin{eqnarray}
H^0(c_2,\mathcal{L}_2|_{c_2}) \otimes H^0(c_3,\mathcal{L}_3\star_A\mathcal{L}_3|_{c_3}) \otimes H^0(\Lambda^2V_5,\mathcal{L}_2\star \mathcal{L}_3|_{c_{\Lambda^2V_5}}) \longrightarrow \mathbb{C}.
\end{eqnarray}
The intuition is similar to the $SU(4)$ bundle with reducible spectral cover mentioned before. However one should be careful whether these cohomologies contribute in $H^1(\Lambda^2V)$. In particular the extension of the sequence $Q$ can be non-zero. 

There are also pointlike \textbf{10-10-5} in this case
\begin{eqnarray}
H^0(\mathcal{L}_2|_{c_2})\otimes H^0(\mathcal{L}_3|_{c_3})\otimes H^0(\mathcal{H})\rightarrow \mathbb{C},
\end{eqnarray}
where the first two factors are $H^1(V)$ subgroups (possibly zero) and the third one is the $H^1(\Lambda^3V)$ contribution. We have used the relation $\Lambda^2V^*\simeq \Lambda^3V$ and the fact that $\Phi(V^*)=\tau^* \Phi(V)$ to compute the relevant part of $H^1(\Lambda^3V)$.

The F-theory analysis is very similar to $S_4=S_2S_2'$ case in \ref{sec322}. The F-theory Calabi-Yau has a $SU(6)$ singularity corresponding to the $S(U(2)\times U(3))$ on the heterotic (the extra $U(1)$ is anomalous) when the extension in the defining short exact sequence were zero. Before turning on the extension, there are \textbf{20} hypermultiplets of $SU(6)$ in $a_2=0$, \textbf{15} and $\overline{\textbf{15}}$ in $b_3=0$, and \textbf{6} and $\overline{\textbf{6}}$ in $c_{\lambda^2V_5}$.\footnote{In heterotic picture, the \textbf{6} and $\overline{\textbf{6}}$ hypermultiplets live in either the deformations of the vector bundle (in particular the extension morphisms) and in $H^1(\Lambda^2V)$.} So at the intersections there are \textbf{20-15-6} and \textbf{20-$\overline{\textbf{15}}$-$\overline{\textbf{6}}$} Yukawa couplings. Turning on the Higgs field, corresponds to turning on the extension of the defining short exact sequence, and it break $SU(6)$ to $U(1)$. The Yukawa couplings at the intersections decompose into the \textbf{10-10-5} and \textbf{10-$\overline{\textbf{5}}$-$\overline{\textbf{5}}$} couplings, in agreement with our heterotic calculations.\footnote{As mentioned before one should check whether the hypermultiplets in the Yukawa couplings remain massless after Higgsing or not, this is exactly what we mean by checking whether the zero modes are killed by the extension morphisms.} We ignore the details at this point.
\subsubsection{Spectral Cover With Non-Reduced Components}
\begin{figure}
    \centering
    \includegraphics[width=0.33\textwidth]{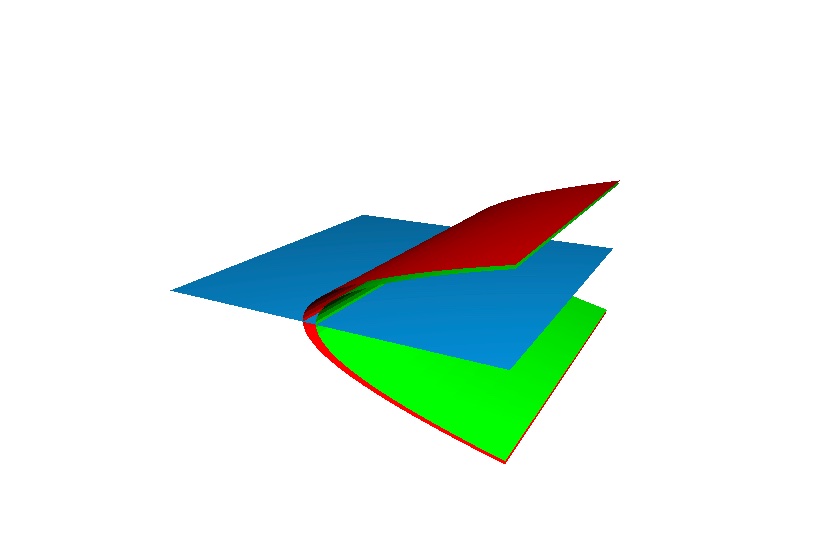}\includegraphics[width=0.33\textwidth]{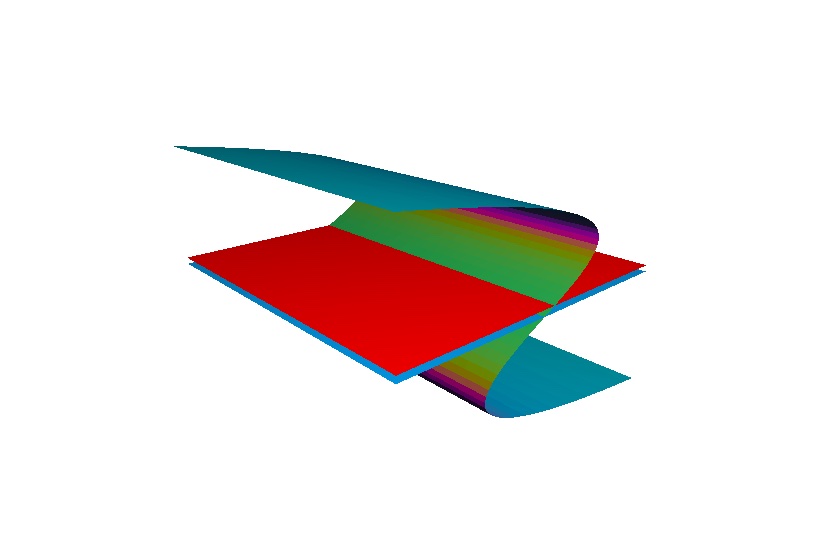}\includegraphics[width=0.33\textwidth]{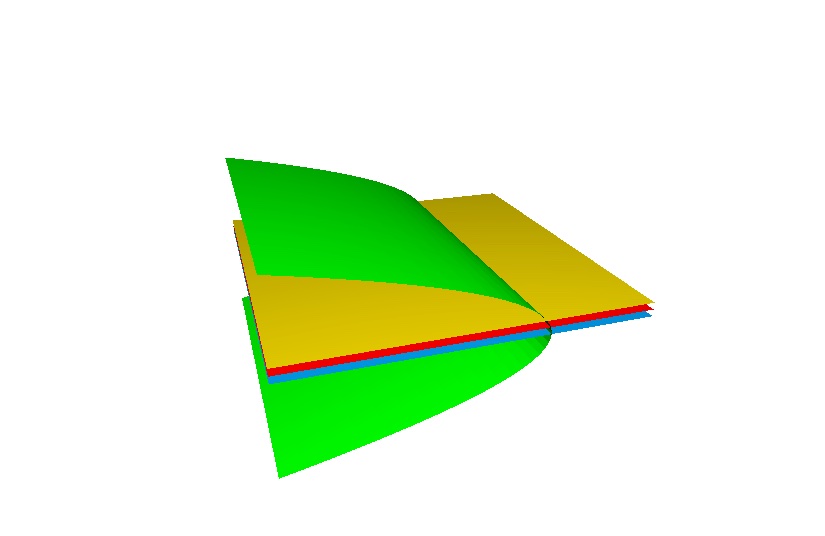}
    \caption{Degree five spectral covers with non-reduced components.}
    \label{fig:v5nonred}
\end{figure}

In this final subsection we study the degree five spectral cover with non-reduced components, Figure \ref{fig:v5nonred}. These cases correspond to the following spectra sheaves
\begin{eqnarray}
\begin{tikzcd}[ampersand replacement=\&, column sep=small]
S_5=Z(a_2 X +a_0 Z^2)^2,\& 0\arrow[r]\& \mathcal{W}_1 \arrow[r] \& \mathcal{L}_5 \arrow[r]\& \mathcal{L}_4|_{(S_2)^2} \arrow[r] \& 0, \\
S_5=Z^2(a_3Y+a_2 X Z+ a_0Z^3),\& 0\arrow[r]\& \mathcal{W}_1 \arrow[r] \& \mathcal{L}_5 \arrow[r]\& \mathcal{L}_4|_{ZS_3} \arrow[r] \& 0, \\
S_5=Z^3(a_2X+a_0Z^2)\& 0\arrow[r]\& \mathcal{W}_1 \arrow[r] \& \mathcal{L}_5 \arrow[r]\& \mathcal{L}_4|_{Z^2S_2} \arrow[r] \& 0.
\end{tikzcd}
\end{eqnarray}
The F-theory picture of these $SU(5)$ heterotic models, roughly speaking, can be explained by adding an extra $U(1)$ flux to the corresponding ``$U(4)$" model inside the 7-brane and consider its interaction with the bulk/localised zero modes that are already there. Since we have briefly done a similar calculations of this type the previous sections, we ignore the F-theory dual in this subsection.
\begin{itemize}
    \item $S_5=Z (a_2X+a_0Z^2)^2$
\end{itemize}
This case corresponds to adding a zero section to the spectral cover of \eref{S2^2}. So the parameters of the defining sequence is (modulo the defining extensions of $\mathcal{L}_4$)
\begin{eqnarray}
RHom_{X}(\mathcal{L}_4,\mathcal{W}_1)&=& RHom_{c}(\mathcal{L}_4|_c,\mathcal{W}_1\otimes \mathcal{O}_c(c)[-1]) \nonumber \\
&\Downarrow &\\
Ext^1_X(\mathcal{L}_4,\mathcal{W}_1)&=&Hom_{c}(\mathcal{L}_4|_c,\mathcal{W}_1\otimes \mathcal{O}_c(c)). \nonumber
\end{eqnarray}
Next, to compute $H^1(V)$ the derived pushforwards $R\pi_*V$ can be used
\begin{eqnarray}\label{PiV5S2S2P}
\begin{tikzcd}[ampersand replacement=\&, column sep=small, row sep=small]
0\arrow[r]\& \mathcal{W}_1\otimes K_B^{-1} \arrow[r,"\simeq"]\& \pi_*V \arrow[r]\arrow[d, phantom, ""{coordinate, name=Z}]\& 0 \arrow[dll,to path={ -- ([xshift=2ex]\tikztostart.east)
|- (Z) [near end]\tikztonodes
-| ([xshift=-2ex]\tikztotarget.west)
-- (\tikztotarget)} ]\&\\
\& \mathcal{W}_1 \arrow[r]\& R^1\pi_*V \arrow[r]\& \mathcal{L}_4|_c\arrow[r]\& 0.
\end{tikzcd}
\end{eqnarray}
The cohomologies of $\Lambda^2V$ can be computed by using the derived pushforward of it. For that, the antisymmetrized Pontrjagin product of $\mathcal{L}_5$ have to be computed
\begin{eqnarray}\label{V5S2S2}
\begin{tikzcd}[ampersand replacement=\&, column sep=small]
0\arrow[r]\& \pi^*\pi_* \mathcal{W}_1\otimes \mathcal{L}_4 \arrow[r]\& \mathcal{L}_5\star_A \mathcal{L}_5 \arrow[r]\& \mathcal{L}_4\star_A \mathcal{L}_4 \arrow[r]\& 0.
\end{tikzcd}
\end{eqnarray}
As a by product that can help at some point, it is easy to read the spectral cover of $\Lambda^2V$ 
\begin{eqnarray}
S_{\Lambda^2V}= Z^4 (a_2X+a_0Z^2)^2(4C_{tor} a_2 X+a_0Z^2),
\end{eqnarray}
where $c_{tor}$ is defined above \eref{L4L22}. We've proved $R\pi_*\Lambda^2V=Li_{\sigma}^* \mathcal{L}_5\star_A \mathcal{L}_5\otimes K_B^{-1}$,
\begin{eqnarray}
\begin{tikzcd}[ampersand replacement=\&, column sep=small, row sep=small]
 \& 0\arrow[r]\& L^{-1}i_{\sigma}^* \mathcal{L}_5\star_A \mathcal{L}_5 \arrow[r]\arrow[d, phantom, ""{coordinate, name=Z}]\& L^{-1}i_{\sigma}^* \mathcal{L}_4\star_A\mathcal{L}_4  \arrow[dll,to path={ -- ([xshift=2ex]\tikztostart.east)
|- (Z) [near end]\tikztonodes
-| ([xshift=-2ex]\tikztotarget.west)
-- (\tikztotarget)}]\& \\
 \& (\mathcal{W}_1\otimes \mathcal{L}_4)|_c \arrow[r]\& i_{\sigma}^* \mathcal{L}_5\star_A \mathcal{L}_5 \arrow[r]\& i_{\sigma}^* \mathcal{L}_4\star_A\mathcal{L}_4 \arrow[r] \& 0
\end{tikzcd}
\end{eqnarray}
The only relevant terms for the Yukawa coupling \textbf{10-10-5} is the second term, which gives a localised Yukawa coupling on $c$ is
\begin{eqnarray}\label{YCV5S2S2}
H^1(\mathcal{W}_1\otimes K_B^{-1})\otimes H^0(\mathcal{L}_4|_c) \longrightarrow H^1((\mathcal{W}_1\otimes \mathcal{L}_4)|_c\otimes K_B^{-1}).
\end{eqnarray}
There are also possible \textbf{$\overline{\textbf{5}}$-$\overline{\textbf{5}}$-10} non localised Yukawa couplings. First notice that
\begin{eqnarray}
c_1(\pi_* \mathcal{L}_n)=5 K_B.
\end{eqnarray}
So, 
\begin{eqnarray}
\mathcal{W}_1 \otimes Det(\pi_* \mathcal{L}_2)\otimes Det(\pi_*\mathcal{L}_2') = K_B^5.
\end{eqnarray}
Therefore the are following morphisms, 
\begin{eqnarray}\label{YC2V5S2S2}
&&H^1(\mathcal{W}_1  \otimes K_B^{-1}) \otimes H^1(Det\pi_*\mathcal{L}_2 \otimes K_B^{-2}) \otimes H^0(Det\pi_*\mathcal{L}'_2 \otimes K_B^{-1}) \longrightarrow \mathbb{C},\nonumber \\
&&
\end{eqnarray}
or,
\begin{eqnarray}\label{YC3V5S2S2}
&&H^1(\mathcal{W}_1  \otimes K_B^{-1}) \otimes H^0(Det\pi_*\mathcal{L}_2 \otimes K_B^{-1}) \otimes H^1(Det\pi_*\mathcal{L}'_2 \otimes K_B^{-2})  \longrightarrow \mathbb{C},\nonumber \\
&&
\end{eqnarray}
or,\footnote{The third possibility must be zero for when $K_B^{-1}$ is effective. This because when $H^0(\mathcal{W}_1)$ is non-zero the $H^(\mathcal{W}_1\otimes K_B^{-1})$ can be non-zero too, which means $H^0(V)$ is non-zero, then $V$ is unstable.}
\begin{eqnarray}\label{YC4V5S2S2}
&&H^0(\mathcal{W}_1) \otimes H^1(Det\pi_*\mathcal{L}_2 \otimes K_B^{-2}) \otimes H^1(Det\pi_*\mathcal{L}'_2 \otimes K_B^{-2})  \longrightarrow \mathbb{C}.\nonumber \\
&&
\end{eqnarray}
In these cases we should check whether $H^{i} (Det\pi_*\mathcal{L}_2 \otimes K_B^{-1-i})$ and $H^{i} (Det\pi_*\mathcal{L}'_2 \otimes K_B^{-1-i})$ contribute in $H^1(\Lambda^2V)$. From the analysis of the degree four spectral cover it seems the coboundary morphisms don't kill the elements in the corresponding cohomologies and therefore all of the Yukawa couplings above are non-vanishing. A detailed example of this type will be given later in this paper.

\begin{itemize}
    \item $S_5=Z^2(a_3Y+a_2XZ+a_0Z^3)$
\end{itemize}
This is the same as adding a zero section to the spectral cover $S_4$ in \eref{V4L1L3}. The extension morphism of the spectral sheaf (modulo the parts that corresponds to the deformations of $\mathcal{L}_4$) live the following group
\begin{eqnarray}
RHom_{X}(\mathcal{L}_4,\mathcal{W}_1) &=& RHom_B(\mathcal{F}\oplus \mathcal{L}_1\otimes K_B^{-1}[+1], \mathcal{W}_1),
\end{eqnarray}
which leads to the following long exact seuence
\begin{eqnarray}
\begin{tikzcd}[ampersand replacement=\&, column sep=small]
0\arrow[r]\&  Ext^1_B(\mathcal{F},\mathcal{W}_1)\arrow[r]\& Ext^1_X(\mathcal{L}_4,\mathcal{W}_1)  \arrow[r]\& Ext^0 (\mathcal{L}_1\otimes K_B^{-1}, \mathcal{W}_1) \arrow[r]\& Ext^2_B(\mathcal{F},\mathcal{W}_1)
\end{tikzcd}
\end{eqnarray}
where $\mathcal{F}$ is defined before,
\begin{eqnarray}
\begin{tikzcd}[ampersand replacement=\&, column sep=small]
0\arrow[r]\& \mathcal{L}_1 \arrow[r]\& \mathcal{F}\arrow[r]\& i_{c*} \mathcal{L}_3\arrow[r]\& 0.\nonumber
\end{tikzcd}
\end{eqnarray}
The cohomology of $V$ can be computed by the derived pushforward $R\pi_*V=Li_{\sigma}^* \mathcal{L}_5[-1]$, 
\begin{eqnarray}
\begin{tikzcd}[ampersand replacement=\&, column sep=small, row sep=small]
0\arrow[r]\& \mathcal{W}_1\otimes K_B^{-1}\arrow[r]\& L^{-1}i_{\sigma}^* \mathcal{L}_5\arrow[r]\arrow[d, phantom, ""{coordinate, name=Z}]\& \mathcal{L}_1\otimes K_B^{-1} \arrow[dll,to path={ -- ([xshift=2ex]\tikztostart.east)
|- (Z) [near end]\tikztonodes
-| ([xshift=-2ex]\tikztotarget.west)
-- (\tikztotarget)}]\& \\
\& \mathcal{W}_1 \arrow[r]\& L^0i_{\sigma}^* \mathcal{L}_5 \arrow[r]\& \mathcal{F}\arrow[r]\& 0.
\end{tikzcd}
\end{eqnarray}
Therefore,
\begin{eqnarray}
Li_{\sigma}^* \mathcal{L}_5=\Bar{\mathcal{N}} \oplus \mathcal{W}_1\otimes K_B^{-1}[+1].\\
\begin{tikzcd}[ampersand replacement=\&, column sep=small]
0\arrow[r]\& \mathcal{W}_1|_{\lbrace\mathcal{E}xt=0\rbrace} \arrow[r]\& \Bar{\mathcal{N}} \arrow[r]\& \mathcal{F}\arrow[r]\& 0.
\end{tikzcd}
\end{eqnarray}
As always the next step is computing $R\pi_*\Lambda^2V=Li_{\sigma}^*\mathcal{L}_5\star_A\mathcal{L}_5\otimes K_B^{-1}[-1]$. So let us compute $\mathcal{L}_5\star_A\mathcal{L}_5$ 
\begin{eqnarray}
\begin{tikzcd}[ampersand replacement=\&, column sep=small]
\& \& \&0\arrow[d] \& \\
\& \& \&\pi^* \pi_* \mathcal{L}_1 \otimes \mathcal{L}_3\arrow[d] \& \\
0\arrow[r]\& \pi^*\pi_* \mathcal{W}_1\otimes \mathcal{L}_4 \arrow[r]\& \mathcal{L}_5\star_A \mathcal{L}_5 \arrow[r]\& \mathcal{L}_4\star_A \mathcal{L}_4 \arrow[r]\arrow[d]\& 0\\
\& \& \&\mathcal{L}_3\star_A\mathcal{L}_3 \arrow[d] \& \\
\& \& \&0 \& \\
\end{tikzcd}
\end{eqnarray}
One can read the spectral cover of the associated bundle easily
\begin{eqnarray}
S_{\Lambda^2V}=Z (a_3Y+a_2XZ+a_0Z^3)^2(-a_3Y+a_2XZ+a_0Z^3).
\end{eqnarray}
Anyways, the restriction over the zero section is given by the following formula
\begin{eqnarray}
Li_{\sigma}^* \mathcal{L}_5\star_A\mathcal{L}_5 = \mathcal{W}_1\otimes\mathcal{L}_1\otimes K_B^{-1}[+1]\oplus \mathcal{M} \\
\begin{tikzcd}[ampersand replacement=\&, column sep=small]
\& \& \&0\arrow[d] \& \\
\& \& \&i_{c*}(\mathcal{L}_1 \otimes \mathcal{L}_3)\arrow[d] \& \\
0\arrow[r]\& \mathcal{W}_1\otimes \mathcal{F} \arrow[r]\& \mathcal{M} \arrow[r]\& \mathcal{L}_4\star_A \mathcal{L}_4|_{\sigma} \arrow[r]\arrow[d]\& 0\\
\& \& \&i_{c*}\left(Det \pi_* \mathcal{L}_3|_{\tilde{c}}\right) \arrow[d] \& \\
\& \& \&0 \& \\
\end{tikzcd}
\end{eqnarray}
There is a ``candidate" for Yukawa coupling \textbf{10-10-5}
\begin{eqnarray}
H^1(\mathcal{W}_1\otimes K_B^{-1}) \otimes H^0 (\mathcal{F}) \longrightarrow H^1(\mathcal{W}_1\otimes \mathcal{F}\otimes K_B^{-1})
\end{eqnarray}
The coboundary map $H^0(i_{c*} (\mathcal{L}_1\otimes \mathcal{L}_3)\otimes K_B^{-1}) \longrightarrow H^1(\mathcal{W}_1\otimes \mathcal{F}\otimes K_B^{-1})$ is induced by the extension of $\mathcal{W}_1$ and $\mathcal{L}_3$ in the original sequence. So it is non-zero generically, and to see whether \textbf{10-10-5} localised, one needs to check whether the corresponding terms are not killed by this coboundary map. 

For \textbf{$\overline{\textbf{5}}$-$\overline{\textbf{5}}$-10} There are also candidates on $c$. For this case we should use the relation between the line bundles $\mathcal{W}_1$, $\mathcal{L}_1$, $Det(\pi_*\mathcal{L}_3)$. The relation is,
\begin{eqnarray}
\mathcal{W}_1|_c \otimes\mathcal{L}_1|_c\otimes\mathcal{L}_3|_c\otimes Det(\pi_* \mathcal{L}_3|_{\tilde{c}})=K_c\otimes K_B^{+3},
\end{eqnarray}
where $K_c$ is the canonical bundle of the curve $c$. This relation can be derived as follows. First start with the, by now, well known relation,
\begin{eqnarray}
c_1(\pi_*\mathcal{L}_5)=5c_1(K_B)=c_1(\mathcal{W}_1)+c_1(\mathcal{L}_1)+c_1(\pi_*\mathcal{L}_3).
\end{eqnarray}
Note that the restriction of $S_3$ over $pi^{-1}c$ is reducible, therefore $i_{\pi^{-1}c}^*\mathcal{L}_3$ can be represented by a short exact sequence,
\begin{eqnarray}
\begin{tikzcd}[ampersand replacement=\&, column sep=small]
0\arrow[r]\& \mathcal{L}_3|_c \otimes \mathcal{O}_c(-\tilde{c}\cdot c)\arrow[r]\& i_{\pi^{-1}c}^*\mathcal{L}_3 \arrow[r]\&  \mathcal{L}_3|_{\tilde{c}}\arrow[r]\& 0.
\end{tikzcd}
\end{eqnarray}
To compute $\mathcal{O}_c(-\tilde{c}\cdot c)$, note that the intersection locus of $c$ and $\tilde{c}$ is in one-to-one correspondence to the intersection locus of $a_3=a_2=0$ in the base. So one gets
\begin{eqnarray}
\mathcal{O}_c(-\tilde{c}\cdot c)=K_c^{-1}\otimes K_B^2,
\end{eqnarray}
where we used the adjunction formula for the curve $c$. Consequently we get the a simle relation for $c_1(\pi_*\mathcal{L}_3)$ when it is restricted over $\pi^{-1}c$,
\begin{eqnarray}
i_{\pi^{-1}c}^*c_1(\pi_*\mathcal{L}_3)=c_1(i_{\pi^{-1}c}^*\pi_*\mathcal{L}_3) = 2c_1(K_B)-K_c+c_1(\mathcal{L}_3|_c)+c_1(\pi_*\mathcal{L}_3|_{\tilde{c}}).
\end{eqnarray}
Putting everything together one gets the advertised relation. Then it is easy to see that there are actually possible \textbf{$\overline{\textbf{5}}$-$\overline{\textbf{5}}$-10} from the interaction of the bulk and localised zero modes,
\begin{eqnarray}
H^0(\mathcal{L}_c|_c)\otimes H^1(\mathcal{W}_1\otimes \mathcal{L}_1\otimes K_B^{-2})\otimes H^0(Det(\pi_*\mathcal{L}_3|_{\tilde{c}})) \longrightarrow H^1(K_c) \simeq H^0(\mathcal{O}_c) =\mathbb{C}. 
\end{eqnarray}

\begin{itemize}
    \item $S_5=Z^3(a_2X+a_0Z^2)$
\end{itemize}
This corresponds to adding a zero section to the degree four spectral cover of \eref{V4L1L1L2}. Start with the extension group as usual
\begin{eqnarray}
RHom_{D(X)} (\mathcal{L}_4,\mathcal{W}_1)&=&RHom_{D(B)} (Li_{\sigma}^*\mathcal{L}_4,\mathcal{W}_1)\\
&=& RHom_{D(B)}(\mathcal{U}_1\otimes K_B^{-1}[+1]\oplus \bar{\mathcal{F}},\mathcal{W}_1).
\end{eqnarray}
This leads to a long exact sequence
\begin{eqnarray}
\begin{tikzcd}[ampersand replacement=\&,column sep=small]
0\arrow[r]\& Ext^1_B(\bar{\mathcal{F}},\mathcal{W}_1)\arrow[r]\& Ext^1_{D(X)} (\mathcal{L}_4,\mathcal{W}_1) \arrow[r]\& Ext^0_B(\mathcal{U}_1\otimes K_B^{-1},\mathcal{W}_1) \arrow[r]\& Ext^2_B(\bar{\mathcal{F}},\mathcal{W}_1)
\end{tikzcd}
\end{eqnarray}
Again the first term on the left contains elements that corresponds to gluing the line bundles to a rank two vector bundle on the zero section. We keep these terms off. It also contains the gluing between $\mathcal{W}_1$ and the line bundle over $S_2$. The third term corresponds to a sheaf with numeric rank 1 on the non reduced surface $Z^2=0$. This term is generically on to construct generic spectral sheaves. Nest let us compute $R\pi_*V=Li_{\sigma}^*\mathcal{L}_5[-1]$
\begin{eqnarray}
\begin{tikzcd}[ampersand replacement=\&, column sep=small, row sep=small]
0\arrow[r]\& \mathcal{W}_1\otimes K_B^{-1}\arrow[r]\& L^{-1}i_{\sigma}^* \mathcal{L}_5 \arrow[r]\arrow[d, phantom, ""{coordinate, name=Z}]\& \mathcal{U}_1 \otimes K_B^{-1} \arrow[dll,to path={ -- ([xshift=2ex]\tikztostart.east)
|- (Z) [near end]\tikztonodes
-| ([xshift=-2ex]\tikztotarget.west)
-- (\tikztotarget)}]\& \\
\& \mathcal{W}_1 \arrow[r]\& L^{0}i_{\sigma}^* \mathcal{L}_5 \arrow[r]\& \bar{\mathcal{F}}\arrow[r]\& 0.
\end{tikzcd}
\end{eqnarray}
Therefore,
\begin{eqnarray}
Li_{\sigma}^* \mathcal{L}_5=\mathcal{W}_1\otimes K_B^{-1}[+1]\oplus \mathcal{J}, \\
\begin{tikzcd}[ampersand replacement=\&, column sep=small]
0\arrow[r]\&\mathcal{W}_1|_{\mathcal{E}xt_5=0} \arrow[r]\& \mathcal{J}\arrow[r]\& \bar{\mathcal{F}}\arrow[r]\& 0.
\end{tikzcd}
\end{eqnarray}
The next part is computing the $\mathcal{L}_5\star_A\mathcal{L}_5$
\begin{eqnarray}
\begin{tikzcd}[ampersand replacement=\&,column sep=small]
0\arrow[r]\& \pi^*\pi_* \mathcal{W}_1\otimes\mathcal{L}_4 \arrow[r]\& \mathcal{L}_5\star_A\mathcal{L}_5 \arrow[r]\& \mathcal{L}_4\star_A\mathcal{L}_4\arrow[r] \& 0.
\end{tikzcd}
\end{eqnarray}
We can also read the spectral cover of $\Lambda^2V$
\begin{eqnarray}
S_{\Lambda^2V} = Z^4 (a_2 X +a_0 Z^2)^3.
\end{eqnarray}
Restrict to the zero section to get $R\pi_*\Lambda^2V=Li_{\sigma}^*\mathcal{L}_5\star_A\mathcal{L}_5\otimes K_B^{-1}[-1]$
\begin{eqnarray}
\begin{tikzcd}[ampersand replacement=\&, column sep=small,row sep=small]
0\arrow[r]\&\mathcal{W}_1\otimes \mathcal{U}_1\otimes K_B^{-1}\arrow[r]\& L^{-1}i_{\sigma}^* \mathcal{L}_5\star_A \mathcal{L}_5\arrow[r]\arrow[d, phantom, ""{coordinate, name=Z}]\& L^{-1}i_{\sigma}^* \mathcal{L}_4\star_A \mathcal{L}_4\arrow[dll,to path={ -- ([xshift=2ex]\tikztostart.east)
|- (Z) [near end]\tikztonodes
-| ([xshift=-2ex]\tikztotarget.west)
-- (\tikztotarget)}]\& \\
\& \mathcal{W}_1 \otimes \bar{\mathcal{F}}\arrow[r]\& L^{0}i_{\sigma}^* \mathcal{L}_5\star_A \mathcal{L}_5 \arrow[r]\& L^{0}i_{\sigma}^* \mathcal{L}_4\star_A \mathcal{L}_4\arrow[r]\& 0
\end{tikzcd}
\end{eqnarray}
The coboundary map can receive non-zero contributions from $Hom_{D(B)}(\mathcal{U}_1\otimes K_B^{-1},\mathcal{W}_1)$. Therefore if the following terms
\begin{eqnarray}
H^1(\mathcal{W}_1\otimes K_B^{-1})\otimes H^0(\bar{\mathcal{F}})\longrightarrow H^1(\mathcal{W}_1\otimes \bar{\mathcal{F}}\otimes K_B^{-1})
\end{eqnarray}
have non-zero contribution in $H^1(L^{0}i_{\sigma}^* \mathcal{L}_5\star_A \mathcal{L}_5)$, then we can have Yukawa coupling \textbf{10-10-5}. Note that theses couplings are not coming from localised modes.

There are three possibilities for the \textbf{$\overline{\textbf{5}}$-$\overline{\textbf{5}}$-10} coupling due to the following relation,
\begin{eqnarray}
\mathcal{W}_1\otimes \mathcal{U}_1\otimes \mathcal{L}_1\otimes Det(\pi_* \mathcal{L}_2) = K_{B}^5.
\end{eqnarray}
The cohomologies are,
\begin{eqnarray}
H^1(\mathcal{W}_1\otimes \mathcal{U}_1\otimes K_B^{-2}) \otimes H^1(Det\pi_*\mathcal{L}_2\otimes K_B^{-2}) \otimes H^0(\mathcal{L}_1) \longrightarrow \mathbb{C}, \\
H^1(\mathcal{L}_1\otimes \mathcal{U}_1\otimes K_B^{-2}) \otimes H^0(Det\pi_*\mathcal{L}_2\otimes K_B^{-1}) \otimes H^1(\mathcal{W}_1\otimes K_B^{-1}) \longrightarrow \mathbb{C},\\
H^0(\mathcal{L}_1\otimes \mathcal{U}_1\otimes K_B^{-1}) \otimes H^1(Det\pi_*\mathcal{L}_2\otimes K_B^{-2}) \otimes H^1(\mathcal{W}_1\otimes K_B^{-1}) \longrightarrow \mathbb{C},
\end{eqnarray}
Of course one needs to check whether the elements of these cohomologies contribute in $H^1(\Lambda^2V)\otimes H^1(\Lambda^2V)\otimes H^1(V)$.



\section{Vertical Component}
The cases with a vertical component is somewhat subtle. This is because of the existence of singular fibers. In case the ambiguity is not too harsh, as in the completely degenerate spectral cover, we can still use the spectral data. However it is more satisfying if we can give an alternative way to compute the cohomologies of $\Lambda^2V$ that works in any case that spectral cover contains vertical components. As in the previous cases we first consider a degree two spectral cover as a warm up example, and then study the effect of these vertical components for the $SU(3)$ , $SU(4)$ and $SU(5)$ examples in details.

Consider a spectral cover of the form $\alpha(b_2X+b_0Z^2)$ which is double cover with a vertical component,
\begin{eqnarray}
S_2= \alpha (b_2 X + b_0 Z^2),
\end{eqnarray}
where $\alpha$, $b_0$ and $b_2$ are suitable polynomials in the base. $b_0$ and $b_2$ should be generic without common factors, and $\alpha$ can be reducible but shouldn't have factors with multiplicities more than one. First assume that $\alpha$ is a generic irreducible polynomial. Then the spectral cover is now a union of a smooth vertical component and a smooth double cover. The spectral sheaf, without loss of generality can be written as,
\begin{eqnarray}
\begin{tikzcd}[ampersand replacement=\&, column sep=small]
0\arrow[r]\& \mathcal{L}1_{V} \arrow[r]\& \mathcal{L}_2 \arrow[r]\& \mathcal{L}'_2 \arrow[r]\& 0,
\end{tikzcd}
\end{eqnarray}
where $\mathcal{L}_2$ is the spectral sheaf as usual.\footnote{We don't write the inclusion pushforward explicitly from now on, unless it causes confusion. For example instead of $i_{S_2*}\mathcal{L}_2$ we simply write $\mathcal{L}_2$.} $\mathcal{L}1_V$ is a rank one coherent sheaf over the vertical components $\alpha=0$ (call it $S_V$), and $\mathcal{L}'_2$ is rank one coherent sheaf supported over $b_2X+b_0Z^2=0$. Since the inverse Fourier transform of the spectral sheaf must be a vector bundle, the coherent sheaves $\mathcal{L}1_V$ and $\mathcal{L}'_2$ can have singularity only at the intersection locus of the components, however as we will argue later, these singularities do not contribute in the Yukawa couplings, therefore from now on we will assume that the restriction of the spectral sheaf over each component is smooth. Therefore $\mathcal{L}1_V$ and $\mathcal{L}'_2$ are line bundles. The extension group of this sequence is also important to know,
\begin{eqnarray}
Ext^1(\mathcal{L}'_2,\mathcal{L}1_V) = H^0(S_2' \cap S_V, \mathcal{L}1_V\otimes \mathcal{L}^{'*}_2 \otimes \mathcal{O}(+S'_2)).
\end{eqnarray}
The pushforwards of $V$ can be computed easily as before,
\begin{eqnarray}
&R\pi_*V= i_{\sigma}^* \mathcal{L}_2,& \\
&\begin{tikzcd}[ampersand replacement=\&, column sep=small]
0\arrow[r]\& \mathcal{L}1_V|_{\sigma} \arrow[r]\& i_{\sigma}^*\mathcal{L}_2 \arrow[r]\& \mathcal{L}'_2|_{\sigma\cap S_2'}\arrow[r]\& 0, 
\end{tikzcd}&
\end{eqnarray}
where the first sheaf is supported on the curve $\alpha=0$ (call it $c^m_1$) and the third sheaf is supported on $b_2=0$ (call it $c^m_2=0$).

Before continuing to the Pontrjagin product some comments about this spectral sheaf should be added. The first comments is that as usual the vertical components show up in the spectral data when the corresponding bundle is unstable on that locus. By similar arguments mentioned in \cite{Anderson:2019axt} one can show since the vertical component is reduced and $\mathcal{L}1_V$ is a line bundle over this component, the dual vector bundle must become a direct sum of line bundles on $S_V$ one with relative degree $-1$ and the other one with relative degree $+1$.\footnote{More clearly $V|_{S_V}=\mathcal{O}(\sigma+A)\oplus \mathcal{O}(-\sigma +B)$ where $A$ and $B$ are divisors in the base.} The summand with positive degree do not contribute in the spectral sheaf (due to the flatness of the projection maps and smoothness of the bundle $V$). Then a well known Fourier-Mukai property (\cite{BBRH} 6.34) shows that the restriction of $\mathcal{L}_2$ on $S_V$ must have relative degree $+1$. Since $\mathcal{L}'_2$ is supported over a double cover, the relative degree of $\mathcal{L}1_V$ must be $-1$ (because 2-1=1).\footnote{This is good because for the specific Fourier-Mukai that we are using here (defined with the Poincare sheaf as the kernel) the pushforward of $\mathcal{L}_2$ to the base must correspond to the restriction of $V$ to zero section $\sigma$, and hence it should be smooth. However if the relative degree of $\mathcal{L}1_V$ was positive, the pushforward $\pi_*\mathcal{L}_2$ would be singular.} The second comment is that as before $c_1(\pi_*\mathcal{L}_2) = 2 c_1(K_B)$. This should put constraint on $\mathcal{L}'_2$ which will be useful later.
\begin{eqnarray}
\begin{tikzcd}[ampersand replacement=\&, column sep=small]
0\arrow[r]\& \pi_*\mathcal{L}_2 \arrow[r]\& \pi_*\mathcal{L}'_2 \arrow[r]\& i_{c^m_1} R^1\pi_* \mathcal{L}1_V \arrow[r]\& 0.
\end{tikzcd}
\end{eqnarray}
So one can show,
\begin{eqnarray}\label{c1RedRedV2}
c_1(\pi_*\mathcal{L}'_2) = 2c_1(K_B)+[c^m_1].
\end{eqnarray}
The next task is to compute $\mathcal{L}_2\star_A \mathcal{L}_2$. But, as mentioned, dealing with Pontrjagin product that involves the $\mathcal{L}1_V$ is somewhat ambiguous, and it is better to avoid this. The idea is that we know how to deal with the $\mathcal{L}'_2$ part. So at this point we do an inverse Fourier transform and try to compute $\Lambda^2V$ in terms directly. More clearly, since the degree of $\mathcal{L}1_V$ is (-1) over the fibers, the inverse Fourier transform gives,
\begin{eqnarray}\label{defVVertical}
\begin{tikzcd}[ampersand replacement=\&, column sep=small]
0\arrow[r]\& V\arrow[r]\& \bar{V} \arrow[r]\& i_{c_1^m*} \mathcal{F} \arrow[r]\& 0,
\end{tikzcd}
\end{eqnarray}
where $\mathcal{F}$ is a line bundle with relative degree (+1) over the fibers. Since we can compute $\mathcal{L}'_2 \star_A \mathcal{L}'_2$ we can compute $\Lambda^2\bar{V}$. To use the above sequence to compute $\Lambda^2V$ we have to dualise the sequence above,

\begin{eqnarray}
\begin{tikzcd}[ampersand replacement=\&, column sep=small]
0\arrow[r]\& \bar{V}^*\arrow[r]\& V^* \arrow[r]\& i_{c_1^m*} \mathcal{F}^* \otimes \mathcal{O}(c_1^m) \arrow[r]\& 0.
\end{tikzcd}
\end{eqnarray}
Now, as explained in the \ref{Appendix}, one can derive the following sequence from this,
\begin{eqnarray}\label{defLVVertical}
\begin{tikzcd}[ampersand replacement=\&, column sep=small]
 \& \& 0\arrow[d] \& \& \\
 \& \& \Lambda^2 \bar{V}^*\arrow[d] \& \& \\
0\arrow[r]\& (\Lambda^2 i_{c_1^m*} \mathcal{F}^* \otimes \mathcal{O}(c_1^m))^{-1} \arrow[r]\& Q \arrow[d]\arrow[r] \& \Lambda^2V^* \arrow[r]\& 0 \\
 \& \& \bar{V}^* \otimes i_{c_1^m*} \mathcal{F}^* \otimes \mathcal{O}(c_1^m)\arrow[d] \& \& \\
 \& \& 0 \& \& 
\end{tikzcd}
\end{eqnarray}
From \eref{c1RedRedV2} we know that $\Lambda^2\bar{V}^* = \mathcal{O}_X(-c^m_1)$. In addition \begin{eqnarray}
\Lambda^2 i_{c_1^m*} \mathcal{F}^* \otimes \mathcal{O}(c_1^m) = i_{c_1^m*}\mathcal{F}^{*\otimes 2}\otimes \mathcal{O}(c_1^m)[+1].
\end{eqnarray}
Consequently, the cokernel of $\lbrace (\Lambda^2 i_{c_1^m*} \mathcal{F}^* \otimes \mathcal{O}(c_1^m))^{-1} \longrightarrow Q \rbrace$ will be,
\begin{eqnarray}
\begin{tikzcd}[ampersand replacement=\&, column sep=small]
0\arrow[d] \\
\mathcal{O}_X(-c^m_1) \arrow[d] \\
\Lambda^2V^* \arrow[d] \\
i_{c^m_1*}\mathcal{O} \arrow[d]\\ 
0
\end{tikzcd}
\end{eqnarray}
So clearly $\Lambda^2 V^* =\mathcal{O}_X$ as expected. In this case the result was already known, but we will repeat the same techniques for other higher rank bundles in the following. The idea is basically add a vertical fiber the cases that we could compute the $\Lambda^2 V$ from spectral covers, and check how these vertical components affect the Yukawa couplings.

\subsection{General Results for Yukawa Couplings}
Let us see how this vertical components can possibly contribute the Yukawa couplings. In general we are looking for a map 
\begin{eqnarray}
H^1(V)\otimes H^1(V)\longrightarrow H^2(\Lambda^2V)=H^1(\Lambda^2V^*).
\end{eqnarray}
From the way we defined $V$ in \eref{defVVertical} and the resulting diagram \eref{defLVVertical}, one can study the effect of the vertical components. Very roughly speaking the coholomogy group $H^1(V)$ is made of the following groups
\begin{eqnarray}
H^1(\bar{V}),\quad H^0(i_{c_1^m*}\mathcal{F}),
\end{eqnarray}
and $H^2(\Lambda^2V)=H^1(\Lambda^2V^*)$ are given by,
\begin{eqnarray}
H^1(\Lambda^2\bar{V}^*),\quad H^1(\bar{V}^*\otimes i_{c_1^m*}\mathcal{F}^*\otimes \mathcal{O}(c_1^m)),\quad H^2(i_{c_1^m*}\mathcal{F}^{\star \otimes 2}\otimes \mathcal{O}(c_1^m)).
\end{eqnarray}
Of course one should check the long exact sequences to see whether these groups actually have non-zero contribution into $H^1(V)$ and $H^2(\Lambda^2V)$. It is quit possible that one cannot decompose the cohomologies into constituent smaller pieces. For example the   First of all, there are possible Yukawa couplings from $\bar{V}$,
\begin{eqnarray}
H^1(\bar{V})\otimes H^1(\bar{V}) \longrightarrow H^1(\Lambda^2 \bar{V}^*) = H^2(\Lambda^2\bar{V}).
\end{eqnarray}
Such couplings can be found from the results of the previous sections. The new possibilities are,
\begin{eqnarray}\label{E6VC1}
H^1(\bar{V})\otimes H^0(i_{c_1^m*}\mathcal{F}) \longrightarrow H^1(\bar{V}\otimes i_{c_1^m*}\mathcal{F}) \simeq H^1(\bar{V}^*\otimes i_{c_1^m*}\mathcal{F}^*\otimes \mathcal{O}(c_1^m)),
\end{eqnarray}
and
\begin{eqnarray}\label{E6VC2}
H^0(i_{c_1^m*}\mathcal{F})\otimes H^0(i_{c_1^m*}\mathcal{F})\longrightarrow H^0(i_{c_1^m*} \mathcal{F}^{\otimes 2})\simeq H^2(i_{c_1^m*} \mathcal{F}^{*\otimes 2}\otimes \mathcal{O}(c_1^m)).
\end{eqnarray}
We should add a couple of comments here. Firs of all the second row is always vanishing. This is related to the smoothness of the bundle $V$. In other words, the cohomology group $H^2(i_{c_1^m*} \mathcal{F}^{*\otimes 2}\otimes \mathcal{O}(c_1^m))$ does not contribute in $H^1(\Lambda^2V^*)$. This is because we demand the bundle $V$ and therefore $\Lambda^2V$ to be stable and smooth. In this case, the Fourier transform of the bundle should always be WIT-1 (this is a necessary condition for the smoothness plus stability). Also note that for $SU(5)$ models, there are no new \textbf{$\overline{\textbf{5}}$-$\overline{\textbf{5}}$-10} Yukawa couplings from the vertical components. 

So we have the tools for computing the Yukawa couplings for the cases that the spectral cover has a vertical component. In the following these tools are applied for the $E_6$, $SO(10)$ and $SU(5)$ models.
\subsection{Bundles With Degenerate Spectral Covers}
In this situation the spectral cover is a degree three non-redced cover with vertical components. We assume that the vertical component is smooth. So the algebraic eaquation for the spectral cover is,
\begin{eqnarray}
S_3=a_0 Z^3.
\end{eqnarray}
Therefore the spectral sheaf is can be written generally as,
\begin{eqnarray}
\begin{tikzcd}[ampersand replacement=\&, column sep=small]
0\arrow[r]\& \mathcal{L}_V \arrow[r]\& \mathcal{L}_3\arrow[r]\& \mathcal{L}'_3\arrow[r]\& 0.
\end{tikzcd}
\end{eqnarray}
First sheaf is localised over the non-reduced cover $Z^3=0$. It can be constructed iteratively as 
\begin{eqnarray}
\begin{tikzcd}[ampersand replacement=\&, column sep=small, row sep=small]
0\arrow[r]\& \mathcal{B}_1\arrow[r]\& \mathcal{L}'_3\arrow[r]\& \mathcal{L}_2\arrow[r]\& 0, \\
0\arrow[r]\& \mathcal{A}_1\arrow[r]\& \mathcal{L}_2\arrow[r]\& \mathcal{L}_1\arrow[r]\& 0,
\end{tikzcd}
\end{eqnarray}
where $\mathcal{A}_1$, $\mathcal{B}_1$, $\mathcal{L}_1$ are line bundles over the zero section $\sigma$, and the extensions of these two sequences are given by,
\begin{eqnarray}
\begin{tikzcd}[ampersand replacement=\&, column sep=small, row sep=small]
0\arrow[r]\& H^1(B,\mathcal{L}^{\star}_1\otimes\mathcal{A}_1)\arrow[r]\& Ext^1(\mathcal{L}_1,\mathcal{A}_1) \arrow[r]\&  H^0(B,\mathcal{L}^{\star}_1\otimes\mathcal{A}_1\otimes K_B)\arrow[r]\& \dots
\end{tikzcd}\\
\begin{tikzcd}[ampersand replacement=\&, column sep=small]
 \&0\arrow[d] \& \&0\arrow[d] \& \\
 \& H^1(B,\mathcal{L}^{\star}_1\otimes\mathcal{B}_1)\arrow[d]\& \&H^1(B,\mathcal{A}^{\star}_1\otimes\mathcal{B}_1)\arrow[d]\&  \\
\dots\arrow[r]\& Ext^1(\mathcal{L}_1,\mathcal{B}_1) \arrow[r]\arrow[d] \&EXt^1(\mathcal{L}_2,\mathcal{B}_1)\arrow[r]\& Ext^1(\mathcal{A}_1,\mathcal{B}_1) \arrow[r]\arrow[d]\& \dots \\
 \& H^0(B,\mathcal{L}^{\star}_1\otimes\mathcal{B}_1\otimes K_B)\arrow[d]\& \&H^0(B,\mathcal{A}^{\star}_1\otimes\mathcal{B}_1\otimes K_B)\arrow[d]\&  \\
 \&\vdots \& \&\vdots \&
\end{tikzcd}
\end{eqnarray}
We assume the only terms that are turned on in the extension groups are $H^0(B,\mathcal{L}^{\star}_1\otimes\mathcal{A}_1\otimes K_B)$, $ H^0(B,\mathcal{L}^{\star}_1\otimes\mathcal{B}_1\otimes K_B)$ and $H^0(B,\mathcal{A}^{\star}_1\otimes\mathcal{B}_1\otimes K_B)$. Numerical rank of the spectral sheaf is one.\footnote{By numerical rank one we mean that the Hilbert polynomial of $\mathcal{L}'_3$ is similar to the Hilbert polynomial of a sine bundle over a smooth triple cover.} The other groups corresponds to vector bundles (or rank two or three) living over the zero section. 

The corresponding vector bundle can be found easily by the inverse Fourier transform,
\begin{eqnarray}
\begin{tikzcd}[ampersand replacement=\&, column sep=small]
0\arrow[r]\& V_3\arrow[r]\& \bar{V}_3 \arrow[r]\& i_{c_V*}\mathcal{F}\arrow[r]\& 0,
\end{tikzcd}\\
\begin{tikzcd}[ampersand replacement=\&, column sep=small, row sep=small]
0\arrow[r]\& \mathcal{B}_1\otimes K_B^{-1}\arrow[r]\& \bar{V}_3\arrow[r]\& V_2 \arrow[r]\& 0, \\
0\arrow[r]\& \mathcal{A}_1\otimes K_B^{-1}\arrow[r]\& V_2\arrow[r]\& \mathcal{L}_1\otimes K_B^{-1} \arrow[r]\& 0.
\end{tikzcd}
\end{eqnarray}
First let us take a look at the Yukawa coupling candidates from $\bar{V}_3$. We can easily construct $\Lambda^2\bar{V}_3$,
\begin{eqnarray}
\begin{tikzcd}[ampersand replacement=\&, column sep=small]
 \&0\arrow[d] \& \& \& \\
 \&\mathcal{A}_1\otimes\mathcal{B}_1\otimes K_B^{-2} \arrow[d] \& \& \& \\
0\arrow[r] \&\mathcal{B}_1\otimes V_2\otimes K_B^{-1}\arrow[d]\arrow[r] \& \Lambda^2\bar{V}_3\arrow[r] \& \mathcal{A}_1\otimes\mathcal{L}_1\otimes K_B^{-2} \arrow[r] \& 0 \\
 \&\mathcal{B}_1\otimes\mathcal{L}_1\otimes K_B^{-2} \arrow[d] \& \& \& \\
 \&0 \& \& \&
\end{tikzcd}
\end{eqnarray}
Clearly there are three possible Yukawa \textbf{27-27-27} couplings from $H^1(\bar{V}_3)\otimes H^1(\bar{V}_3)\longrightarrow H^2(\Lambda^2\bar{V}_3)$,
\begin{eqnarray}\label{degE6Coup}
\begin{tikzcd}[ampersand replacement=\&, column sep=small]
 H^1(\mathcal{B}_1\otimes K_B^{-1})\otimes H^1(\mathcal{L}_1\otimes K_B^{-1})\arrow[r]\& H^2(\mathcal{B}_1\otimes\mathcal{L}_1\otimes K_B^{-2}) \\
 H^1(\mathcal{B}_1\otimes K_B^{-1})\otimes H^1(\mathcal{A}_1\otimes K_B^{-1})\arrow[r]\& H^2(\mathcal{A}_1\otimes\mathcal{B}_1\otimes K_B^{-2}) \\
 H^1(\mathcal{A}_1\otimes K_B^{-1})\otimes H^1(\mathcal{L}_1\otimes K_B^{-1})\arrow[r]\& H^2(\mathcal{A}_1\otimes\mathcal{L}_1\otimes K_B^{-2}) 
\end{tikzcd}
\end{eqnarray}
Naively it seems all of the zero modes involving the Yukawa couplings are non-localised in the base manifold $B$. However, note that the extension groups that mentioned above, can kill these zero modes or localise them over curves in $B$. This will be illustrated in the examples. In addition to these couplings there are other candidates for example corresponding the maps in the first row, there are other versions,
\begin{eqnarray}
\begin{tikzcd}[ampersand replacement=\&, column sep=small]
H^0(\mathcal{B}_1)\otimes H^1(\mathcal{L}_1\otimes K_B^{-1})\arrow[r]\& H^1(\mathcal{B}_1\otimes\mathcal{L}_1\otimes K_B^{-1}) \\ 
H^1(\mathcal{B}_1\otimes K_B^{-1})\otimes H^1(\mathcal{L}_1)\arrow[r]\& H^1(\mathcal{B}_1\otimes\mathcal{L}_1\otimes K_B^{-1})
\end{tikzcd}
\end{eqnarray}

There are similar stories for other vector bundles. Since the vertical components didn't contribute in the \textbf{$\overline{\textbf{5}}$-$\overline{\textbf{5}}$-10}, it is interesting to check to see whether when spectral cover completely degenerates it is possible to have this kind of coupling or not. The analysis is the same as before. Consider only $\bar{V}_5$ part in \eref{defVVertical}, and it is simply defined iteratively as the extension of line bundles,
\begin{eqnarray}
\begin{tikzcd}[ampersand replacement=\&, column sep=small]
0\arrow[r]\& \mathcal{W}_1\arrow[r]\& \bar{V}_5\arrow[r]\& V_4\arrow[r]\& 0, \\
0\arrow[r]\& \mathcal{U}_1\arrow[r]\& V_4\arrow[r]\& V_3\arrow[r]\& 0, \\
0\arrow[r]\& \mathcal{B}_1\arrow[r]\& V_3\arrow[r]\& V_2\arrow[r]\& 0, \\
0\arrow[r]\& \mathcal{A}_1\arrow[r]\& V_2\arrow[r]\& \mathcal{L}_1\arrow[r]\& 0,
\end{tikzcd}
\end{eqnarray}
and the corresponding sequences for $\Lambda^2\bar{V}$ are
\begin{eqnarray}
\begin{tikzcd}[ampersand replacement=\&, column sep=small]
0\arrow[r]\& \mathcal{W}_1\otimes V_4\arrow[r]\& \Lambda^2\bar{V}_5 \arrow[r]\& \Lambda^2V_4\arrow[r]\& 0,\\
0\arrow[r]\& \mathcal{U}_1\otimes V_3\arrow[r]\& \Lambda^2 V_4 \arrow[r]\& \Lambda^2V_3\arrow[r]\& 0,\\
0\arrow[r]\& \mathcal{B}_1\otimes V_2\arrow[r]\& \Lambda^2 V_3 \arrow[r]\& \mathcal{A}_1\otimes \mathcal{B}_1 \arrow[r]\& 0.
\end{tikzcd}
\end{eqnarray}
There is also a relation due to the $c_1(V)=0$,
\begin{eqnarray}
\mathcal{W}_1\otimes\mathcal{U}_1\otimes\mathcal{B}_1\otimes\mathcal{A}_1\otimes\mathcal{L}_1 = K_B^5\otimes\mathcal{O}(c_V).
\end{eqnarray}
One can prove there are candidates for \textbf{$\overline{\textbf{5}}$-$\overline{\textbf{5}}$-10} using the relation above between the line bundles. We are looking for morphisms $H^1(\bar{V})\otimes H^1(\Lambda^2\bar{V})\longrightarrow H^2(\Lambda^3 \bar{V})\simeq H^1(\Lambda^2\bar{V}^* \otimes \mathcal{O}(c_V))$. Where in the last isomorphism we used $\Lambda^5\bar{V}=\mathcal{O}(c_V)$. Putting everything together one can show for example,
\begin{eqnarray}
\begin{tikzcd}[ampersand replacement=\&, column sep=small]
H^1(\bar{V})\otimes H^1(\Lambda^2\bar{V})\arrow[r]\&  H^2(\Lambda^2\bar{V}^* \otimes \mathcal{O}(c_V)) \\
H^1(B,\mathcal{W}_1)\otimes H^1(B,\mathcal{U}_1\otimes \mathcal{B}_1)\arrow[u, hook']\arrow[r]\& H^2(B, \mathcal{A}_1^*\otimes \mathcal{L}_1^*\otimes \mathcal{O}(c_V))\arrow[u, dashed]
\end{tikzcd}
\end{eqnarray}
Where the dashed arrow emphasizes that the relation between these to spaces is given by the Leray spectral sequence, and it is not an injection necessarily. There are 15 Yukawa coupling candidates like this one which, as mentioned in the $E_6$ model, seems to come from the bulk zero modes inside the 7-branes. However one need to check, case by case, whether these zero modes are killed or localised over some curve (by coboundary morphisms or extension maps of the short exact sequences) or not. There are also 30 other candidates. For example corresponding to the diagram above there are two other diagrams,
\begin{eqnarray}
\begin{tikzcd}[ampersand replacement=\&, column sep=small]
H^1(\bar{V})\otimes H^1(\Lambda^2\bar{V})\arrow[r]\&  H^2(\Lambda^2\bar{V}^* \otimes \mathcal{O}(c_V)) \\
H^0(B,\mathcal{W}_1\otimes K_B)\otimes H^1(B,\mathcal{U}_1\otimes \mathcal{B}_1)\arrow[u, dashed]\arrow[r]\& H^1(B, \mathcal{A}_1^*\otimes \mathcal{L}_1^*\otimes \mathcal{O}(c_V)\otimes K_B)\arrow[u, dashed]
\end{tikzcd}
\end{eqnarray}
and
\begin{eqnarray}
\begin{tikzcd}[ampersand replacement=\&, column sep=small]
H^1(\bar{V})\otimes H^1(\Lambda^2\bar{V})\arrow[r]\&  H^2(\Lambda^2\bar{V}^* \otimes \mathcal{O}(c_V)) \\
H^1(B,\mathcal{W}_1)\otimes H^0(B,\mathcal{U}_1\otimes \mathcal{B}_1\otimes K_B)\arrow[u, dashed]\arrow[r]\& H^1(B, \mathcal{A}_1^*\otimes \mathcal{L}_1^*\otimes \mathcal{O}(c_V)\otimes K_B)\arrow[u, dashed]
\end{tikzcd}
\end{eqnarray}
\section{Examples}
In this section we illustrate the ideas of the previous sections by means of simple examples. In this paper we don't try to find models that have correct phenomenological properties such as correct number of generations, moduli stabilization etc.
\subsection{$E_6$ Model}
For this subsection lets consider a rank three bundle defined as
\begin{eqnarray}
\begin{tikzcd}[ampersand replacement=\&, column sep=small]
 \& \& \& 0\arrow[d] \& \\
 \& \& \& \mathcal{O}(-\sigma+D_2)\arrow[d] \& \\
0\arrow[r]\& \mathcal{O}(D_1)\arrow[r] \& V_3 \arrow[r]\& V_2\arrow[r]\arrow[d]\& 0 \\
 \& \& \& \mathcal{O}(\sigma+D_3)\arrow[d] \& \\
  \& \& \& 0 \&
\end{tikzcd}
\end{eqnarray}
Where $D_1,D_2,D_3$ are divisors in the base that satisfy
\begin{eqnarray}
D_1+D_2+D_3=0.
\end{eqnarray}
The Fourier transform of this bundle us quite easy to find,
\begin{eqnarray}
\begin{tikzcd}[ampersand replacement=\&, column sep=small]
  \& \& \& 0\arrow[d] \& \\
  \& \& \& \mathcal{O}(-\sigma+D_3+K_B)\arrow[d,"\mathcal{S}_2"] \& \\
  \& \& \& \mathcal{O}(\sigma+D_2)\arrow[d] \& \\
0\arrow[r]\& \mathcal{O}_{\sigma}(D_1+K_B)\arrow[r] \& \Phi^1(V_3) \arrow[r]\& \mathcal{O}_{S_2}(\sigma+D_2)\arrow[r]\arrow[d]\& 0 \\
 \& \& \& 0 \& 
\end{tikzcd}
\end{eqnarray}
Therefore the spectral cover in this case is
\begin{eqnarray}
S_3=Z \mathcal{S}_2 = Z(a_2 X+a_0 Z^2),\quad [\mathcal{S}_2]=2\sigma-D_3+D_2-K_B
\end{eqnarray}
This is our starting point. The purpose is to compute the Yukawa couplings of this bundle, then consider various deformations of this bundle and see how they affect the Yukawa couplings. The Parameters of this bundle are the polynomials $a_2$ and $a_0$ which determine the double cover, and the extension group of $V_2$ and $\mathcal{O}(D_1)$,
\begin{eqnarray}
Ext^1(\mathcal{O}_{S_2}(\sigma+D_2),\mathcal{O}_{\sigma}(D_1+K_B))&=&H^0(c,\mathcal{O}(D_1-D_2)\otimes \mathcal{O}(c))\nonumber \\
&=& H^0\left(c,\mathcal{O}(K_B+D_1-D_3)\right),
\end{eqnarray}
where $c$ is the ``matter curve" $c=\sigma\cap \mathcal{S}_2=K_B+D_2-D_3$.

The first task will be computing the cohomologies and we can just use the results of \ref{sec312} with $\mathcal{L}_1=\mathcal{O}(D_1+K_B)$ and $\mathcal{L}_2=\mathcal{O}_{S_2}(\sigma+D_2)$. We have seen that there are possible Yukawa coupling in this case from the map
\begin{eqnarray}
H^1(\mathcal{L}_1\otimes K_B^{-1}) \otimes H^0(i_{c*} \mathcal{L}_2) \longrightarrow H^1(i_{c*} (\mathcal{L}_1\otimes \mathcal{L}_2\otimes K_B^{-1})).\nonumber
\end{eqnarray}
We only need to check whether they actually contribute in the cohmologies $H^1(V)$ and $H^2(\Lambda^2V)$. Let us start with the former,
\begin{eqnarray}
&\pi_*V=\mathcal{O}(D_1)&\\
&\begin{tikzcd}[ampersand replacement=\&]
  0\arrow[r]\& \mathcal{O}(D_1+K_B)\arrow[r]\& R^1\pi_*V\arrow[r]\& \mathcal{O}_c(D_2+K_B)\arrow[r] \& 0.
\end{tikzcd}&
\end{eqnarray}
Clearly $H^1(\mathcal{O}(D_1))$ injects into $H^1(V)$. But whether any element in  $H^0(\mathcal{O}_c(D_2+K_B)$ can be inside $H^1(V)$ or not, clearly depends on the choices of the divisors. For example when $B=\mathbb{F}_0$ and $D_1=(2,-4)$, $D_2=(2,9)$ ($[c]=(4,12)$) then $h^*(\mathcal{O}_c(D_2+K_B))=(8,12)$ and it is indeed part of the $H^1(V)$ because $H^2(\pi_*V)=0$ in \eref{LerayE6Red}. In addition the extension of the defining equation is non-zero : $h^0(c,\mathcal{O}(K_B+D_1-D_3))=12$.\footnote{It is easy to check that with this choice of divisors there is always a region in Kahler cone such that the slope of $\mathcal{O}(D_1)$ is negative, so the bundle is not semistable.} 

We also need to check whether $H^1(i_{c*} (\mathcal{L}_1\otimes \mathcal{L}_2\otimes K_B^{-1}))= H^1(c,\mathcal{O}(K_B+D_1+D_2))$ also maps into $H^1(R^1\pi_*\Lambda^2V)$. It turns out that this is not the case. So there are no \textbf{27-27-27} couplings in this case and they are killed by the extension of the defining equation. To see this in more detail remember the the long exact sequence that gave us the derived pushforwards of $\Lambda^2V$ in section \ref{sec312},
\begin{eqnarray}
\begin{tikzcd}[ampersand replacement=\&, row sep=small]
 \& 0\arrow[r] \& \pi_* \Lambda^2V\arrow[r]\arrow[d, phantom, ""{coordinate, name=Z}]\& \mathcal{O}(-D_1)\arrow[dll,
to path={ -- ([xshift=2ex]\tikztostart.east)
|- (Z) [near end]\tikztonodes
-| ([xshift=-2ex]\tikztotarget.west)
-- (\tikztotarget)}]\& \\
 \& \mathcal{O}_c(K_B+D_1+D_2) \arrow[r]\& R^1\pi_* \Lambda^2V\arrow[r]\& \mathcal{O}(-D_1+K_B)\arrow[r]\& 0.
\end{tikzcd}\nonumber
\end{eqnarray}
Note that the coboundary map is exactly the same as the extension of the original defining sequence. So the cokernel of this map are torsion sheaves supported over the points where the extension vanishes i.e. the points where a global section (that one chooses as the extension ``parameter") of $\mathcal{O}_c(K_B+D_1-D_3)$ vanishes. So we can conclude
\begin{eqnarray}
\pi_* \Lambda^2V = \mathcal{O}(-K_B+2D_3), \quad  R^1\pi_* \Lambda^2V = T\oplus \mathcal{O}(-D_1+K_B),
\end{eqnarray}
where the $T$ is the torsion sheaf mentioned before, and with our choice of the divisors, $T$ is supported over 44 generic points over the curve $c$. So we see that elements of  $H^1(i_{c*} (\mathcal{L}_1\otimes \mathcal{L}_2\otimes K_B^{-1}))$ cannot inject into $H^2(\Lambda^2V)$ and these modes are removed by the coboundary map (or extension). So when the extension is chosen to be zero there can be \textbf{27-27-27} couplings from localised over $c$.  

For \textbf{$\overline{\textbf{27}}$-$\overline{\textbf{27}}$-$\overline{\textbf{27}}$} couplings we can use the same techniques as before and we have all information that we need,
\begin{eqnarray}
\begin{tikzcd}[ampersand replacement=\&]
H^1(V^*)\otimes H^1(V^*) \arrow[r]\& H^2(\Lambda^2V^*) \\
H^2(V)\otimes H^2(V) \arrow[r]\& H^2(V)
\end{tikzcd}
\end{eqnarray}
The sequences above are equivalent. So in this case all we need is just the spectral data of $V$. It is not too hard to show that the relevant cohomologies for this particular coupling are
\begin{eqnarray}
\begin{tikzcd}[ampersand replacement=\&]
H^1(\mathcal{O}(D_1+K_B))\otimes H^1(\mathcal{O}_c(D_2+K_B)) \simeq H^1(\mathcal{O}(-D_1))\otimes H^0(\mathcal{O}_c(K_B-D_3))\arrow[r] \& H^1(\mathcal{O}_c(D_2+K_B), 
\end{tikzcd}
\end{eqnarray}
where we have used $D_1+D_2+D_3=0$. The only thing wee need to show is that whether the elements of $H^1(\mathcal{O}(D_1+K_B))$ enter into $H^1(R^1\pi_*V)$,
\begin{eqnarray}
\begin{tikzcd}[ampersand replacement=\&]
\dots\arrow[r]\& H^0(\mathcal{O}_c(D_2+K_B)) \arrow[r]\& H^1(\mathcal{O}(D_1+K_B)) \arrow[r]\& H^1(R^1\pi_* V)\arrow[r]\& \dots,
\end{tikzcd}
\end{eqnarray}
where the second map is simply the extension of the short exact sequence. With the choices of $D_1$, $D_2$ and $D_3$ in this example, one should find the cokernel of $H^0(\mathcal{O}_c(0,7))\longrightarrow H^1(\mathcal{O}(0,-6))$ where the maps is an element of $H^1(\mathcal{O}_c(4,-1))$. The basis of these cohomology groups and the generic form of the homomorphism can be represented by (rational) monomials as follows
\begin{eqnarray}
&X_1^7,\quad X_1^6 X_2,\quad X_1^5 X_2^2,\quad X_1^4 X_2^3,\quad X_1^3 X_2^4,\quad X_1^2 X_2^5,\quad X_1 x_2^6,\quad X_2^7,& \\
&\frac{1}{X_1 X_2 X_1^4},\quad \frac{1}{X_1 X_2 X_1^3 X_2},\quad \frac{1}{X_1 X_2 X_1^2 X_2^2},\quad \frac{1}{X_1 X_2 X_1 X_2^3},\quad \frac{1}{X_1 X_2 X_2^4},& \\
&\frac{1}{X_1 X_2 f_{11}(X_1,X_2)},&
\end{eqnarray}
where the variables $X_1$ and $X_2$ are homogeneous coordinates of the $\mathbb{P}^1$ base of $\mathbb{F}_0$, the monomials in the first row are the bases of $H^0(\mathcal{O}_c(0,7))$, the rational monomials of the second row are the bases of $H^1(\mathcal{O}(0,-6))$, and the third row is the rational monomial that corresponds to the extension of the defining short exact sequence (with $f_{11}$ being a generic degree 11 polynomial). It is now very easy to show that this map is surjective, in other words $H^1(\mathcal{O}(D_1+K_B))$ do not inject into $H^1(R^1\pi_*V)$ and hence no contribution in $H^2(V)$. So there is no  \textbf{$\overline{\textbf{27}}$-$\overline{\textbf{27}}$-$\overline{\textbf{27}}$} coupling in this model. Here we clearly see one of the main points of this paper. Instead of computing the Yukawa couplings by directly computing the cohomology groups of $V$, one can dissect this problem to smaller pieces using the spectral data, reduce the problem to simple calculations.

Before considering the deformations, it is somewhat important to mention that there is another Yukawa coupling, namely \textbf{1-27-$\overline{\textbf{27}}$}, which can be non-zero. This is given by the Yoneda pairing,
\begin{eqnarray}
\begin{tikzcd}[ampersand replacement=\&]
Ext^1(\mathcal{O},V)\otimes Ext^1(V,V)\otimes Ext^1(V,\mathcal{O})\arrow[r]\& Ext^3(\mathcal{O},\mathcal{O})=H^3(\mathcal{O})=\mathbb{C}.
\end{tikzcd}
\end{eqnarray}
The middle term in this pairing counts the infinitesimal deformations of the bundle. In the current example they correspond to the deformations of $\mathcal{S}_2$ and the extension group elements. When the bundle moduli are stabilized to some fixed values, these couplings give mass to the \textbf{27-$\overline{\textbf{27}}$} pairs but the net number of generations remains unchanged. Anyways, in this example these couplings are non-zero and are localised over the curve $c$ and they are counted by the following map 
\begin{eqnarray}
\begin{tikzcd}[ampersand replacement=\&]
H^0(\mathcal{O}_c(K_B+D_2))\otimes H^0(\mathcal{O}_c(K_B+D_1-D_3))\otimes H^1(\mathcal{O}(-D_1)) \arrow[r]\& H^1(c,K_c) \simeq \mathbb{C}.
\end{tikzcd}
\end{eqnarray}
With the choices we have made all of the cohomologies above are non-zero. But the previous calculations shows that $H^1(\mathcal{O}(-D_1))$ doesn't contribute to $H^2(\mathcal{O}(-D_1))$. Therefore this coupling is also vanishing.

In the next part we deform the spectral data to a completely degenerate one by setting the polynomial $a_2$ to zero. The spectral sheaf can be rewritten in the form of the previous subsection,
\begin{eqnarray}
\begin{tikzcd}[ampersand replacement=\&]
0\arrow[r]\&\mathcal{L}_{c_V}\arrow[r]\& \mathcal{L}_3\arrow[r]\& \bar{\mathcal{L}}_3\arrow[r]\& 0.
\end{tikzcd}
\end{eqnarray}
It is possible to find $\mathcal{L}_V$ and $\bar{\mathcal{L}}_3$ by computing $Li_{c_V}^*\mathcal{L}_3$ and $Li_{\sigma}^*\mathcal{L}_3$ using two different representations of $\mathcal{L}_3$, and compare them. The final result is,
\begin{eqnarray}
&\mathcal{L}_{c_V}= \mathcal{O}_{c_V}(-2\sigma+D_2),& \\
&\begin{tikzcd}[ampersand replacement=\&, row sep=small]
0\arrow[r]\& \mathcal{O}_{\sigma}(-2D_3)\arrow[r]\& \bar{\mathcal{L}}_3\arrow[r]\& \bar{\mathcal{L}}_2 \arrow[r]\& 0, \\
0\arrow[r]\& \mathcal{O}_{\sigma}(D_2)\arrow[r]\& \bar{\mathcal{L}}_2\arrow[r]\& \mathcal{O}_{\sigma}(D_2+K_B) \arrow[r]\& 0.
\end{tikzcd}&
\end{eqnarray}
All of the extension morphisms are non-zero and generic. As usual, instead of working with the vertical components directly, one can perform an inverse Fourier transform,
\begin{eqnarray}
&\begin{tikzcd}[ampersand replacement=\&, row sep=small]
0\arrow[r]\& V_3\arrow[r]\& \bar{V}_3\arrow[r]\& i_{c_V*}\mathcal{F}\arrow[r]\& 0,
\end{tikzcd}&\\
&\begin{tikzcd}[ampersand replacement=\&, row sep=small]
0\arrow[r]\& \mathcal{O}_{c_V}(-4\sigma+D_2+K_B)\arrow[r]\& (K_B\oplus K_B^{-1}\oplus K_B^{-2})\otimes \mathcal{O}_{c_V}(-\sigma+D_2)\arrow[r]\& i_{c_V*}\mathcal{F}\arrow[r]\& 0.
\end{tikzcd}&\nonumber \\
\quad
\end{eqnarray}
To do concrete calculations one needs to choose the divisors $D_1$, $D_2$ and $D_3$.\footnote{With the previous choices the bundle becomes unstable as in the limit where the spectral cover is degenerate.} We choose $D_1=(2,-4)$ and $D_2(2,3)$. Then $D_3=(-4,1)$ and $c_V=(8,4)$. In this case one can show the cohomologies of $i_{c_V*}\mathcal{F}=(60,0,0,0)$. Pushforwads are also easy to find,
\begin{eqnarray}
\begin{tikzcd}[ampersand replacement=\&, row sep=small]
0\arrow[r]\& \mathcal{O}(-2D_3-K_B)\arrow[r]\& \pi_*V\arrow[r]\& \mathcal{O}(D_2-K_B)\arrow[r]\& 0, \\
0\arrow[r]\& \mathcal{O}(-2D_3)\arrow[r]\& \pi_*V\arrow[r]\& \mathcal{O}(D_2+K_B)\arrow[r]\& 0.
\end{tikzcd}
\end{eqnarray}
In principle between the first row and the second raw is non-zero, but in this example it is vanishing. By the analysis of the previous section, especially the relations \eref{degE6Coup}, there are only three ``types" of candidates for the Yukawa couplings from the degenerate part of the spectral cover
\begin{eqnarray}
\begin{tikzcd}[ampersand replacement=\&, row sep=small]
H^1(\mathcal{B}_1\otimes K_B^{-1})\otimes H^1(\mathcal{A}_1\otimes K_B^{-1})\arrow[r]\& H^2(\mathcal{B}_1\otimes\mathcal{A}_1\otimes K_B^{-2}), \\
H^0(\mathcal{B}_1)\otimes H^1(\mathcal{A}_1\otimes K_B^{-1})\arrow[r]\& H^1(\mathcal{B}_1\otimes\mathcal{A}_1\otimes K_B^{-1}), \\
H^1(\mathcal{B}_1\otimes K_B^{-1})\otimes H^0(\mathcal{L}_1)\arrow[r]\& H^1(\mathcal{B}_1\otimes\mathcal{A}_1\otimes K_B^{-1}),
\end{tikzcd}
\end{eqnarray}
where $\mathcal{B}_1=\mathcal{O}(-2D_3)$, $\mathcal{A}_1=\mathcal{O}(D_2)$ and $\mathcal{L}_1=\mathcal{O}(D_2+K_B)$. One needs to check whether these cohomologies can contribute into $H^1(V)$ and $H^2(\Lambda^2 V)$. Since the cohomologies of $\mathcal{F}$ are $(60,0,0)$, any element of $H^1(\bar{V})$ can be lifted to an element in $H^1(V)$. However, all of the relevant cohomologies of the line bundles $\mathcal{A}_1$ $\mathcal{B}_1$ and $\mathcal{L}_1$, that can feed non-zero elements into $H^1(\bar{V})$, and at the same time give non-zero Yukawa couplings, are vanishing. Therefore there is no \textbf{27-27-27} couplings of this type.

The final part is checking for the couplings induced by the vertical component of the spectral cover. The candidates are given in the relations \eref{E6VC1} and \eref{E6VC2}. The cohomologies of $i_{c_V*}\mathcal{F}^{*\otimes 2}\otimes \mathcal{O}(c_V)$ are $(0,824,0)$. Therefore the couplings corresponding to \eref{E6VC2} are zero, consistent with our expectation. For the other option one gets $h^1(\bar{V})=2$ which is lifted from $h^1(\mathcal{O}(D_2))$, and $h^0(i_{c_V*}\mathcal{F})=60$. However one can check easily $h^1(i_{c_V*}\mathcal{F}\otimes \bar{V})=0$. So nothing left and all of the \textbf{27-27-27} couplings are zero.   

\subsection{SU(5) Model}
We consider a vector bundle defined by \eref{V5S2S2}, 
\begin{eqnarray}
\begin{tikzcd}[ampersand replacement=\&, column sep=small]
\& \& \&0\arrow[d] \& \\
\& \& \&\mathcal{O}_{S_2}(\sigma+D_2)\arrow[d] \& \\
0\arrow[r]\&i_{\sigma*}\mathcal{O}(D_1)\arrow[r] \& \mathcal{L}_5\arrow[r] \&\mathcal{L}_4\arrow[r]\arrow[d] \& 0 \\
\& \& \&\mathcal{O}_{S_2}(\sigma+D_3)\arrow[d] \& \\
\& \& \&0 \& 
\end{tikzcd}
\end{eqnarray}
where $S_2$ is a double cover with divisor class $[S_2]=2\sigma+\eta$. We have showed that there is a \textbf{10-10-5} coupling candidate \eref{YCV5S2S2}, and three candidates for \textbf{$\overline{\textbf{5}}$-$\overline{\textbf{5}}$-10} \eref{YC2V5S2S2},\eref{YC3V5S2S2},\eref{YC4V5S2S2}. We only need to check whether in a certain model the cohomologies in the couplings are non-zero, they contribute to the relevant cohomologies of $V$ and $\Lambda^2V$. We need a bundle $V$ which is stable and degree zero. For the later condition, $Det(\pi_*\mathcal{L}_5)=K_B^5$. This puts a constraint between $\eta$, $D_1$, $D_2$ and $D_3$. To find this one should compute the pushforward of $\pi_*\mathcal{O}_{S_2}(\sigma)$ first,
\begin{eqnarray}
\begin{tikzcd}[ampersand replacement=\&, column sep=small]
0\arrow[r]\& \mathcal{O}\arrow[r]\& \pi_*\mathcal{O}_{S_2}(\sigma)\arrow[r]\& \mathcal{O}(K_B-\eta)\arrow[r]\& 0.
\end{tikzcd}
\end{eqnarray}
Therefore,
\begin{eqnarray}
Det(\pi_* \mathcal{O}_{S_2}(\sigma+D_2)) = \mathcal{O}(K_B-\eta+2D_3), \quad Det(\pi_* \mathcal{O}_{S_2}(\sigma+D_2)) = \mathcal{O}(K_B-\eta+2D_3).
\end{eqnarray}
Deriving the final result is now easy,
\begin{eqnarray}
D_1+2D_2+2D_3-2\eta=3K_B.
\end{eqnarray}
On the other hand, the stability requires $h^0(V)=h^3(V)=0$. So combining \eref{PiV5S2S2} with the Leray spectral sequence, one gets 
\begin{eqnarray}
H^0(V)=H^0(\pi_*V),\quad H^3(V)=H^2(R^1\pi_*V).
\end{eqnarray}
The following choice satisfies the these conditions,
\begin{eqnarray}
B=\mathbb{F}_1,\quad D_1=(-4,3),\quad D_2=(3,3),\quad D_3=(1,2) \Rightarrow \eta=(5,11).
\end{eqnarray}
With this choice the extension of the vertical column can be computed as usual,\footnote{$\Gamma$ is the global section functor.}
\begin{eqnarray}
&RHom(i_{S_2*}\mathcal{O}(\sigma+D_3),i_{S_2*}\mathcal{O}(\sigma+D_2)) = R\Gamma(\mathcal{O}_{S_2}(D_2-D_3)\otimes (\mathcal{O}\oplus \mathcal{O}(S_2)[-1]))&\nonumber\\
&\Downarrow& \nonumber\\
&\begin{tikzcd}[ampersand replacement=\&, column sep=small]
0\arrow[r]\& Ext^1_X(\mathcal{O}_{S_2}(\sigma+D_3),\mathcal{O}_{S_2}(\sigma+D_2))\arrow[r]\& H^0(\mathcal{O}_{S_2}(S_2+D_2-D_3)) \arrow[r]\& H^2(\mathcal{O}_{S_2}(D_2-D_3))\arrow[r]\& \dots
\end{tikzcd}&\nonumber\\
&\quad&
\end{eqnarray}
It is not hard to show $h^0(\mathcal{O}_{S_2}(S_2+D_2-D_3))=95$ and $h^2(\mathcal{O}_{S_2}(D_2-D_3))=39$. So there are at least 56 independent extension morphisms that can give a ``line bundle" over the degree four cover $S_2^2$ (as opposed to a rank two bundle over $S_2$).   

To have non-zero Yukawa couplings $H^1(B,\mathcal{O}(D_1-K_B))$ but be non-empty. The cohmologies of this line bundle are $(0,8,0)$. Together with the Leray spectral sequence,
\begin{eqnarray}
\begin{tikzcd}[ampersand replacement=\&, column sep=small]
0\arrow[r]\& H^1(\mathcal{O}(D_1)\otimes K_B^{-1})\arrow[r]\& H^1(V)\arrow[r]\& H^0(R^1\pi_*V)\arrow[r]\& 0,
\end{tikzcd}
\end{eqnarray}
it is clear that the elements of $H^1(B,\mathcal{O}(D_1-K_B))$ injects into $H^1(V)$. What about $H^0(R^1\pi_*V)$? For Yukawa couplings one needs to check whether at least some elements of $\mathcal{L}_4|_c$ (where $c=S_2\cdot \sigma$) can lift into some non-zero elements in $H^0(R^1\pi_*V)$. Recall the following diagram,
\begin{eqnarray}
\begin{tikzcd}[ampersand replacement=\&, column sep=small]
\& \& \&0\arrow[d] \& \\
\& \& \&\mathcal{O}_c(K_B+D_2)\arrow[d]\arrow[dll,dotted,"extension=zero"description, bend right=35] \& \\
0\arrow[r]\& \mathcal{O}(D_1)\arrow[r]\& R^1\pi_*V\arrow[r]\&\mathcal{L}_4|_c\arrow[r]\arrow[d]\& 0 \\
\& \& \&\mathcal{O}_c(K_B+D_3)\arrow[d]\arrow[ull,dotted,"extension=non-zero"description,bend left=35] \& \\
\& \& \&0 \& 
\end{tikzcd}
\end{eqnarray}
The extension of the sequence that defines $R^1\pi_*V$ is non zero, and in principle it comes from two pieces one piece $\mathcal{O}_c(K_B+D_2)$ and one from $\mathcal{O}_c(K_B+D_3)$. With this information we can show that $h^*(R^1\pi_*V)=(5,22,0)$, and all of the five independent elements of $h^0(R^1\pi_*V)$ are coming from $\mathcal{L}_4|_c$. More precisely,
\begin{eqnarray}
\begin{tikzcd}[ampersand replacement=\&, column sep=small]
h^*(\mathcal{O}(D_1)) \& h^*(R^1\pi_*V) \& h^*(\mathcal{L}_4|_c) \\
0 \& 5 \& 5 \leftrightarrow h^0(\mathcal{O}_c(K_B+D_2)) \arrow[dll,dotted,"extension=zero"description] \\
18 \& 22 \& 4 \leftrightarrow h^1(\mathcal{O}_c(K_B+D_3))\\
0 \& 0 \& 0
\end{tikzcd}
\end{eqnarray}
So far we have identified 40 possible \textbf{10-10-5} couplings from $H^1(B,\mathcal{O}(D_1-K_B))\otimes H^0(c,\mathcal{O}_c(D_2+K_B))$. To make sure they are actually non-zero the cohomologies of $\Lambda^2V$ is required. The generalities are explained before in previous sections, here we only add some of the details. Remeber we had the following short exact sequence,
\begin{eqnarray}
\begin{tikzcd}[ampersand replacement=\&, column sep=small]
0\arrow[r]\& \mathcal{O}(D_1)\otimes \mathcal{L}_4 \arrow[r]\& \mathcal{L}_5\star_A\mathcal{L}_5\arrow[r]\& \mathcal{L}_4\star_A\mathcal{L}_4\arrow[r]\& 0.
\end{tikzcd}
\end{eqnarray}
The first term on the right is hard to compute. But it is already done before,
\begin{eqnarray}
\mathcal{L}_4\star_A\mathcal{L}_4 &=& i_{\sigma*}Det(\pi_*\mathcal{O}_{S_2}(\sigma)+D_2)\oplus  i_{\sigma*}Det(\pi_*\mathcal{O}_{S_2}(\sigma)+D_3)\oplus \mathcal{L}_2\star\mathcal{L}_2'\nonumber \\
&=& \mathcal{O}_\sigma (K_B-\eta+2D_2)\oplus \mathcal{O}_\sigma (K_B-\eta+2D_3)\oplus \mathcal{L}_2\star\mathcal{L}_2'.
\end{eqnarray}
We only need to compute the last term,
\begin{eqnarray}
\mathcal{L}_2\star\mathcal{L}_2'=\mathcal{O}(D_2+D_3)\otimes \mathcal{O}_{S_2}(\sigma)\star\mathcal{O}_{S_2}(\sigma).
\end{eqnarray}
Here we give a rather heuristic, but direct way, and in the appendix we prove this formula indirectly by using inverse Fourier transform. It has been shown that the antisymmetric part of this product is simply the determinant of the pushforward. For general product we know the product should be supported over a degree four but reducible cover, which one component of that is a non-reduced degree two copy of the zero section. The definition of the Pontrjagin product helps to derive a formula for the restriction of $\mathcal{O}_{S_2}(\sigma)\star\mathcal{O}_{S_2}(\sigma)$ on the zero section,
\begin{eqnarray}
r:\mathcal{O}_{S_2}(\sigma)\star\mathcal{O}_{S_2}(\sigma) \longrightarrow \pi_{*} \left(\mathcal{O}_{S_2}(\sigma)\otimes \tau^*\mathcal{O}_{S_2}(\sigma)\right),
\end{eqnarray}
where $\tau$ is the involution of the elliptic fibers ($y\rightarrow -y$). Since any double cover and also the divisor $\sigma$ are invariant under the involution then $\tau^*\mathcal{O}_{S_2}(\sigma) = \mathcal{O}_{S_2}(\sigma)$. Therefore we can compute the restriction over zero section,
\begin{eqnarray}
r(\mathcal{O}_{S_2}(\sigma)\star\mathcal{O}_{S_2}(\sigma)) = \pi_*\mathcal{O}_{S_2}(2\sigma),
\end{eqnarray}
\begin{eqnarray}
\begin{tikzcd}[ampersand replacement=\&, column sep=small]
0\arrow[r]\& \mathcal{O}(2K_B+\eta)\arrow[r]\& \pi_*\mathcal{O}_{S_2}(2\sigma) \arrow[r]\& \mathcal{O}(K_B-\eta)\arrow[r]\& 0.
\end{tikzcd}
\end{eqnarray}
Note that the last term on the right is just the antisymmetric part of the Pontrjagin product, i.e. $Det(\pi_*\mathcal{O}_{S_2}(\sigma))=\mathcal{O}_{S_2}(\sigma)\star_A \mathcal{O}_{S_2}(\sigma)$. So the kernel should be the restriction of the symmetric product restricted over the fiber.\footnote{This can also be seen by noting that the two terms differ by $\frac{Br}{2}$ where $Br=4\eta+2K_B$ is the branching divisor of $S_2$ \cite{Friedman}} So we expect $\mathcal{O}_{S_2}(\sigma)\star\mathcal{O}_{S_2}(\sigma)$ can be represented by the following diagram,
\begin{eqnarray}
\begin{tikzcd}[ampersand replacement=\&, column sep=small]
\&0\arrow[d] \& \& \& \\
\&\mathcal{T}_1\arrow[d] \& \& \& \\
0\arrow[r]\&\mathcal{O}_{S_2}(\sigma)\star_S\mathcal{O}_{S_2}(\sigma)\arrow[r]\arrow[d] \& \mathcal{O}_{S_2}(\sigma)\star\mathcal{O}_{S_2}(\sigma)\arrow[r] \& \mathcal{O}_{S_2}(\sigma)\star_A\mathcal{O}_{S_2}(\sigma) \& 0, \\
\&\mathcal{F}_2 \arrow[d] \& \& \& \\
\&0 \& \& \& \\
\end{tikzcd}
\end{eqnarray}
where $\mathcal{T}_1$ is a line bundle over the zero section to be determined soon, $\mathcal{F}_2$ is a line bundle over the double cover mentioned in sec ??. As emphasized before, we expect that $i_{\sigma}^*\mathcal{O}_{S_2}(\sigma)\star_S\mathcal{O}_{S_2}(\sigma)=\mathcal{O}(2K_B+\eta)$.\footnote{Of course $i_{\sigma}^*\mathcal{O}_{S_2}(\sigma)\star_S\mathcal{O}_{S_2}(\sigma)=\mathcal{O}(2K_B+\eta)=\mathcal{O}(K_B-\eta)$.} To determine $\mathcal{T}_1$ note that the zero section and the support of $\mathcal{F}_2$ intersect on the branching divisor of $S_2$. Therefore we expect that 
 \begin{eqnarray}
 \mathcal{T}_1=\mathcal{O}(2K_B+\eta)\otimes\mathcal{O}(-Br)=\mathcal{O}(-3\eta).
 \end{eqnarray}
So the final result about $Li_{\sigma}^*\mathcal{L}_2\star\mathcal{L}_2'$ can be derived easily,
\begin{eqnarray}
\begin{tikzcd}[ampersand replacement=\&, column sep=small,row sep=small]
0\arrow[r]\& \mathcal{O}(-3\eta+D_2+D_3-K_B)\arrow[r]\& L^{-1}i_{\sigma}^*\mathcal{L}_2\star\mathcal{L}_2' \arrow[r]\arrow[d, phantom, ""{coordinate, name=Z}]\& \mathcal{O}(-\eta+D_2+D_3)\arrow[dll,to path={ -- ([xshift=1ex]\tikztostart.east)
|- (Z) [near end]\tikztonodes
-| ([xshift=-2ex]\tikztotarget.west)
-- (\tikztotarget)}]\&\&  \\
\& \mathcal{O}(\eta+D_2+D_3+2K_B)\arrow[r]\& i_{\sigma}^*\mathcal{L}_2\star\mathcal{L}_2' \arrow[r]\& \mathcal{O}(-\eta+D_2+D_3+K_B)\arrow[r]\& 0.
\end{tikzcd}
\end{eqnarray}
A comment on the coboundary maps is useful at this point. Note that the coboundary map of the sequence cannot be induced by the Extension groups of the original sequence that defined $\mathcal{L}_4$  for us. Similarly, same thing happens for the maps between $Det\pi_*\mathcal{L}_2$ and $\mathcal{L}_2'$ and other components of $\mathcal{L}_4\star_A\mathcal{L}_4$. 

Now everything for computing $Li_{\sigma}^*\mathcal{L}_5\star_A\mathcal{L}_5$ is ready,
\begin{eqnarray}
\begin{tikzcd}[ampersand replacement=\&, column sep=small]
 \&0\arrow[r] \& \pi_*\Lambda^2V\arrow[r]\arrow[d, phantom, ""{coordinate, name=Z}] \&\begin{array}{c}
     \mathcal{O}(-K_B-\eta+2D_2) \oplus \mathcal{O}(-K_B-\eta+2D_3)  \\
      \oplus \mathcal{O}(-2K_B-3\eta+D_2+D_3)\oplus \mathcal{O}(-K_B-\eta+D_2+D_3)
 \end{array}\arrow[dll,to path={ -- ([xshift=1ex]\tikztostart.east)
|- (Z) [near end]\tikztonodes
-| ([xshift=-2ex]\tikztotarget.west)
-- (\tikztotarget)}]\& \\
  \&  \mathcal{O}(D_1)\otimes \mathcal{L}_4|_c\arrow[r] \& R^1\pi_*\Lambda^2V\arrow[r] \& \begin{array}{c}
     \mathcal{O}(-\eta+2D_2) \oplus \mathcal{O}(-\eta+2D_3)  \\
      \oplus \mathcal{O}(\eta+D_2+D_3+K_B)\oplus \mathcal{O}(-\eta+D_2+D_3)
 \end{array}\arrow[r]\& 0
\end{tikzcd}
\nonumber\\
\quad
\end{eqnarray}
Even though the cokernel of this long exact sequence is non-zero, with the current choices of $D_1$, $D_2$ and $D_3$ all of the cohomologies of $\mathcal{W}_1\otimes \mathcal{L}_4|_c\otimes K_B^{-1} = \mathcal{O}_c(D_1-K_B)\otimes \mathcal{L}_4$ actually inject into the cohomologies of $R^1\pi_*\Lambda^2V$. In particular, for the \textbf{10-10-5}
 coupling only $H^1(R^1\pi_*\Lambda^2V)$ is needed \eref{YCV5S2S2}, and by Leray sequence all the elements of this cohomology group lift into $H^2(\Lambda^2V)$. However, the relevant cohomologies of $Det(\pi_*\mathcal{L}_2)$ and $Det(\pi_*\mathcal{L}'_2)$ in \eref{YC2V5S2S2} and \eref{YC3V5S2S2} are zero. So there are no \textbf{$\overline{\textbf{5}}$-$\overline{\textbf{5}}$-10} couplings in this theory. To check whether the \textbf{10-10-5} couplings actually exist in the effective theory, one needs to check whether the elements of $H^1(\mathcal{O}(D_1-K_B))$ and $H^0(\mathcal{L}_4|_c)$ can ``contract" with each other and give an element of $H^1(\mathcal{O}_c(D_1-K_B)\otimes \mathcal{L}_4)$. Luckily in this example we can say that without any calculations. Recall that $H^0(\mathcal{L}_4|_c)$ has five independent elements which are injected from $H^0(\mathcal{O}_c(K_B+D_2))$. So to check whether the map in \eref{YCV5S2S2} is non-zero, we should examine whether the following map is non-zero
 \begin{eqnarray}
 \begin{tikzcd}[ampersand replacement=\&,column sep=small]
 H^1(\mathcal{O}(D_1-K_B))\otimes H^0(\mathcal{O}_c(D_2+K_B)) \arrow[r, "?" ]\& H^1(\mathcal{O}_c(D_1+D2)) \arrow[r,dashed]\& H^1(\mathcal{O}_c(D_1-K_B)\otimes \mathcal{L}_4).
 \end{tikzcd}\nonumber\\
 \quad
 \end{eqnarray}
It tuns out $h^1(\mathcal{O}_c(D_1+D_2))=0$. So there is no \textbf{10-10-5} couplings in this effective theory either.

\section{Elliptic Fibrations Without Holomorphic Section}
So far, we assumed the Calabi-Yau threefold is given by the Weierstrass model. As long as the elliptic fibration has a holomorphic (in other words, the section is isomorphic to the base manifold), it is easy to generalize these results. Because to define the Poincare sheaf, the section of the fibration should hit to \textit{every} fiber at exactly one point. However, most of the elliptically fibered Calabi-Yau threefolds do not have any holomorphic sections. In these cases if someone defines an integral transform with kernel chosen to be the ``naive" Poincare sheaf $\mathcal{I}_{\Delta}\otimes \pi_1^*\mathcal{O}(\sigma)\otimes\pi_2^*\mathcal{O}(\sigma)$, then that integral transform is not a Fourier-Mukai transform anymore.

It is very important for phenomenological applications to generalize the approach presented in the previous sections such that they can be used for the non-holomorphic section too. We only need to find the right kernel such that when the transform is restricted to any elliptic fiber (reducible or not)  is reduced to the ordinary Fourier-Mukai transform. 

Here we only briefly propose a solution for case the reducible fibers are $I_2$. This approach can be generalized for $I_n$ fibers. However the for other types of reducible elliptic fibers, one needs more sophisticated approaches. 

First note that for a single elliptic curve, reducible or not, the Poincare sheaf always defines a Fourier-Mukai transform. The reason is one can always choose the zero section to be a point on the curve. For smooth elliptic curves it is well known, and for $I_2$ fibers it is proved by Caldararu \cite{Cldraru2000DerivedCO}. The problem appears when one consider an elliptic fibration where both types of fibers can appear in one family. When the section of the family is rational, the restriction of $\mathcal{I}_{\Delta}\otimes \pi_1^*\mathcal{O}(\sigma)\otimes\pi_2^*\mathcal{O}(\sigma)$ over smooth fibers is the Poincare sheaf, but its restriction over $I_2$ fibers (if the section wraps around one component, say $c_1$) is not. More precisely, suppose $i_p : E_p \hookrightarrow X$ be the inclusion morphism of the fiber over the point $p\in B$. If $E_p$ is $I_2$ ans $\sigma$ is the rational section that wraps around $c_1$, then the restriction over $E_p$ is given by,
\begin{eqnarray}
\begin{tikzcd}[ampersand replacement=\&]
0\arrow[r]\& i_p^*\mathcal{O}_X(\sigma) \arrow[r]\& \mathcal{O}_{c_1}(-1)\oplus \mathcal{O}_{c_2}(+2)\arrow[r]\& T\arrow[r]\& 0,
\end{tikzcd}
\end{eqnarray}
where $T$ is a torsion sheaf supported over the intersection points of $c_1$ and $c_2$. However, a true section that hits a point in $c_1$ (denoted by $\sigma_0$) can be represented as
\begin{eqnarray}
\begin{tikzcd}[ampersand replacement=\&]
0\arrow[r]\& \mathcal{O}_{E_p}(\sigma_0) \arrow[r]\& \mathcal{O}_{c_1}(+1)\oplus \mathcal{O}_{c_2}\arrow[r]\& T\arrow[r]\& 0.
\end{tikzcd}
\end{eqnarray}
Ideally, if one can modify $\mathcal{O}(\sigma)$ such that over smooth fibers it is just the ordinary section that hits a single point, i.e. the modification has no effect over smooth fibers, but over the $I_2$ fibers it modifies $i_p^*\mathcal{O}_X(\sigma)$ to $\mathcal{O}_{E_p}(\sigma_0)$ then we reach our purpose. The way we have been able to modify $\mathcal{O}(\sigma)$, turns the restriction on the reducible elliptic fiber into $\mathcal{O}(-\sigma_0)$. The corresponding integral transform is still a Fourier-Mukai.\footnote{The kernel of the integral transform is basically the same as the Poincare sheaf twisted by a $\mathcal{O}(+2)$} However, this special transform does not map the semistable degree zero bundles over the elliptic fiber into spectral cover. But it is in principle possible to easily study the Fourier transform of semistable sheaves over such isolated elliptic fibers. On the other hand, over generic irreducible fibers, the Fourier-Mukai is just the same as the usual transform that we used so far. So globally, the Fourier-Mukai transform of a stable degree zero bundle will be a spectral sheaf plus some additional data over the irreducible fibers. We will explain this briefly in the following.

Define the sheaf $\Sigma$ as
\begin{eqnarray}
\begin{tikzcd}[ampersand replacement=\&, column sep=small]
0\arrow[r]\& \Sigma \arrow[r]\& \mathcal{I}_{c_1}\otimes \mathcal{O}_X(2\sigma)\arrow[r]\& i_{\sigma*} \mathcal{O}_{\sigma} (c_1)\arrow[r]\& 0.
\end{tikzcd}
\end{eqnarray}
Clearly the restriction over smooth fibers is the zero section. Let us compute the restriction over the components of the $I_2$ fiber
\begin{eqnarray}
Li_{c_1}^* \Sigma= \mathcal{O}_{c_1}(-1),\quad  Li_{c_2}^* \Sigma =\mathcal{O}_{c_2}.
\end{eqnarray}
So the restriction of $\Sigma$ on the reducible fiber is $\mathcal{O}(-\sigma_0)$
\begin{eqnarray}
\begin{tikzcd}[ampersand replacement=\&,column sep=small]
0\arrow[r]\& \Sigma\arrow[r]\& \mathcal{O}_{c_1}(-1)\oplus \mathcal{O}_{c_2}\arrow[r]\& T\arrow[r]\& 0.
\end{tikzcd}
\end{eqnarray}
Then define the modified Poincare sheaf as
\begin{eqnarray}
\overline{\mathcal{P}} := \mathcal{I}_{\Delta}\otimes \pi_1^* \Sigma \otimes \pi_2^* \Sigma \otimes K_B^{-1}.
\end{eqnarray}
As mentioned before with this kernel the new or modified Fourier-Mukai is almost the same as the usual one for Weierstrass models, but over the reducible fibers, it maps the degree zero line bundles to rank two bundles. For example
\begin{eqnarray}
R\pi_{2*}(\pi_1^* \mathcal{O}_{E_p}(q-\sigma_0)) = \mathcal{F},
\end{eqnarray}
where $q$ and $\sigma_0$ are points in the same component $c_1$, and $\mathcal{F}$ is 
\begin{eqnarray}
\begin{tikzcd}[ampersand replacement=\&,column sep=small]
0\arrow[r]\& \Sigma^{\otimes 2}\otimes \mathcal{O}_{E_p} (p-\sigma_0) \arrow[r]\& \mathcal{F}\arrow[r]\& \Sigma\arrow[r]\& 0.
\end{tikzcd}
\end{eqnarray}
So the usual ``spectral data" is modified by adding some exceptional data on non-generic fiber. More clearly, if the vector bundle over the reducible fiber is a direct sum on line bundles $\bigoplus_i \mathcal{O}(q_i-\sigma_0)$ then the Fourier-Mukai transform of that is $\bigoplus_i \mathcal{F}_i$. We ignore the calculations, but it is basically similar to the standard case that can be found in the literature.   

It is now possible to continue the classification of the ``spectral data" for non-Weierstrass Calabi-Yau threefolds. However this is a very formidable work to do. Because there hundreds of millions of Calabi-Yau threefolds and each one may have many rational zero sections that wrap around many rational curves. In addition the fibers doesn't necessarily degenerate in a simple form like the $I_2$ that we considered here, and for each case maybe another ``zero section" $\Sigma$ needs to be defined. On the other hand, the F-theory dual of such heterotic models (it they exist) is not very well known at least to the author. So we leave the complete classification the Yukawa couplings to the future.
\section{Conclusion} \label{sec7}
In this paper, we tried to introduce a new approach for computing the Yukawa couplings of the effective $4D$ $\mathcal{N}=1$ GUT theories, derived from heterotic string compactifications over Calabi-Yau threefold manifolds. The key inside was that the Fourier-Mukai transform (with the kernel being the Poincare sheaf) interchanges the tensor product and the Pontrjagin product, at least over a smooth elliptic curve. We argued that for computing the cohomologies of a product of vector bundles, we can still use the Pontjagin product in most cases. In other words, when the spectral cover has no vertical component, it is still possible to use the antisymmetrized Pontrjagin product for computing the cohomologies of $\Lambda^2V$. On the other hand, when there are vertical components, we used indirect computations, even though it seems it is still possible to directly use the spectral data. 

Assuming the Calabi-Yau manifold is defined by the Weierstrass equation, one can classify possible spectral data that can appear for a smooth degree zero vector bundle and use the Pontrjagin product to find the relevant ``pieces" of the spectral data that can give specific Yukawa couplings. This has two advantages. First, relative to the standard approaches, the calculations become much simpler, or at least the problem dissects into smaller and easy steps. Second, the relation between the intersecting brane models and T-brane becomes very clear. We didn't study the F-theory implications of these calculations thoroughly. However, in principle, these spectral data can be used to define the Higgs bundle one the 7-branes. Hence, we can see whether the specific couplings in the effective heterotic theory result from the triple intersection of branes or the interaction of the zero modes localized inside the 7-brane and curves.

Finally, since most of the elliptically Calabi-Yau manifolds constructed so far has only rational sections, it is interesting to see whether this idea can be extended to the Calabi-Yau manifolds without holomorphic sections. We gave the idea of how this can be done, but the full classification of the spectral data over such manifolds is much harder, plus there are many manifolds that need to be checked. In addition, the F-theory dual of such heterotic models is not very well known. So this generalization is outside of the scope of this paper, and we leave it for the future.

\section*{Acknowledgments}
MK is grateful for the insightful discussions with Tony Pantev, James Gray, Juntao Wang, Wei Cui, Seung-Joo Lee, Xin Gao, and Paul Oehlmann. Also, special thanks to Elham Mahmoudian for being supportive during the writing and preparing this paper.

\appendix

\section{Useful Formulas/Terminologies} \label{Appendix}
In this appendix we briefly explain the terminologies, and some of the formulas that we used in the bulk of this paper. 
\begin{itemize}
    \item \textit{Integral Transform}
\end{itemize}
Integral transform is defined as 
\begin{eqnarray}
\Phi: D(X)&\longrightarrow& D(Y)\\
\Phi(\bullet) &:=& R\pi_{2*} (\pi_1^*\bullet \otimes K^{\bullet}),
\end{eqnarray}
where $K^{\bullet}$ is an object of $D(X\times Y)$, and $\pi_1$ and $\pi_2$ are projection morphisms
\begin{eqnarray}
\begin{tikzcd}[ampersand replacement=\&]
 \& X\times Y \arrow[dl,"\pi_1"']\arrow[dr,"\pi_2"] \& \\
 X \& \& Y
\end{tikzcd}
\end{eqnarray}
An integral transform is called a Fourier-Mukai when it is an equivalence of the derived categories. For this to be true, the kernel of the integral transform should satisfy an ``orthogonality condition", see \cite{BBRH} Theorem \textbf{1.27} 
\begin{itemize}
    \item WIT-i 
\end{itemize}
As mentioned before $\Phi$ ``maps" objects $D(X)$, i.e. complexes of coherent sheaves in $X$, to objects of $D(Y)$. So one can view a coherent sheaf $\mathcal{F}$ as an object in derived category
\begin{eqnarray}
\begin{tikzcd}[ampersand replacement=\&]
 \& \& \& \& 0-th \arrow[d]\& \& \\
\mathcal{F}^{\bullet} \simeq \& \& \dots \arrow[r]\& 0\arrow[r]\& \mathcal{F}\arrow[r]\& 0\arrow[r]\& \dots 
\end{tikzcd}
\end{eqnarray}
We call a sheaf $\mathcal{F}$ is WIT-i respect to the functor $\Phi$ if its integral transform $\mathcal{G}^{\bullet}=\Phi(\mathcal{F})$ is a coherent sheaf located on the i-th position 
\begin{eqnarray}
\begin{tikzcd}[ampersand replacement=\&]
 \& \& \& \& i-th \arrow[d]\& \&\\
\mathcal{G}^{\bullet} \simeq \& \& \dots \arrow[r]\& 0\arrow[r]\& \mathcal{G}\arrow[r]\& 0\arrow[r]\& \dots 
\end{tikzcd}
\end{eqnarray}
In this case we write
\begin{eqnarray}
\Phi(\mathcal{F})=\mathcal{G}[-i].
\end{eqnarray}
Note that $[-i]$ is a shift functor, and it simply shift the complex i places to the right.
\begin{itemize}
    \item If $V$ is a rank $n$ stable degree zero vector bundle over X (which is an elliptically fibered Calabi-Yau threefold), then $c_1(\pi_*\mathcal{L}_n)=n c_1(K_B)$.
\end{itemize}
This can be proved easily. First note that the Chern character of $\Phi(V)=i_{S_n*}\mathcal{L}_n[-1]$ can be computed in terms of the Chern classes of $V$ \cite{Anderson:2019agu}
\begin{eqnarray}
ch(i_{S_n*}\mathcal{L}_n) = [S_n]+[S_n]\cdot (-\frac{c_1(B)}{2}) + \frac{1}{2}c_2(V)f + \dots
\end{eqnarray}

where $[S_n]= n\sigma+\eta$ is the divisor class of the spectral cover, and $f$ is the class of the elliptic fiber. One can use this information, and the Grothendieck-Riemann-Roch theorem to find $ch(\pi_*i_{S_n*}\mathcal{L}_n)$
\begin{eqnarray}
ch(\pi_* i_{S_n*}\mathcal{L}_n) = \pi_*\left(ch(i_{S_n*}\mathcal{L}_n) \frac{td(X)}{td(B)} \right) = \pi_* \left( (n\sigma+\eta) -c_1(B)(n\sigma+\eta)-\frac{c_3(V)}{2}f +\dots\right).
\end{eqnarray}
Then we use the fact that $\pi_*\sigma = 1$, $\pi_*\eta = 0$ (when we right $\eta$ we really mean $\pi^*\eta$) and $\pi_* f=0$. Therefore
\begin{eqnarray}
ch(\pi_* i_{S_n*}\mathcal{L}_n) = n -n c_1(B)+\dots=n+ n c_1(K_B) +\dots.
\end{eqnarray}
\begin{itemize}
    \item Numeric Rank.
    By numeric rank of a sheaf $\mathcal{E}$ which is supported over a non-reduced subscheme we mean the Hilbert polynomial of that sheaf $P(s):\chi(X,\mathcal{E}\otimes\mathcal{O}(s))$, is the same as a the Hilbert polynomial of a coherent sheaf over a smooth subscheme with the same class.
    \end{itemize}

\begin{itemize}
    \item Proof of \eref{defLVVertical}
\end{itemize}
Consider a short exact sequence, 
\begin{eqnarray}
\begin{tikzcd}[ampersand replacement=\&]
 0\arrow[r]\& A \arrow[r,"f"]\& B \arrow[r]\& C \arrow[r]\& 0,
\end{tikzcd}
\end{eqnarray}
where $A$ and $B$ are locally free, but $C$ can be a general coherent sheaf. In derived category we can rewrite this as an isomorphism
\begin{eqnarray}
&C^{\bullet}& : \dots \longrightarrow 0 \longrightarrow C \longrightarrow 0 \longrightarrow \dots \\
&Cone(f)&: \dots\longrightarrow 0\longrightarrow A \longrightarrow B \longrightarrow 0 \longrightarrow \dots
\end{eqnarray}
\begin{eqnarray}
C^{\bullet} \simeq Cone(f).
\end{eqnarray}
Note that $Cone(f)^0=B$ and $Cone(f)^{-1}=A$ So there is a isomorphism after antisymmetrized product
\begin{eqnarray}\label{higherisom}
\Lambda^2 C^{\bullet} \simeq \Lambda^2 Cone(f).
\end{eqnarray}
When $C$ is a vector bundle
\begin{eqnarray}
&\Lambda^2 C^{\bullet}& = \Lambda^2C, \\
&\Lambda^2 Cone(f)&: 0\longrightarrow S^2A \longrightarrow A\otimes B \longrightarrow \Lambda^2B\longrightarrow 0.
\end{eqnarray}
Again please remember $\Lambda^2 Cone(f)^0=\Lambda^2B$, $\Lambda^2 Cone(f)^{-1}=A\otimes B$ and $\Lambda^2 Cone(f)^{-2}=S^2A$. So \eref{higherisom} implies the usual long exact sequence
\begin{eqnarray}
0\longrightarrow S^2A \longrightarrow A\otimes B \longrightarrow \Lambda^2B\longrightarrow \Lambda^2C\longrightarrow 0.
\end{eqnarray}
However, if $C$ is a torsion sheaf supported on a smooth divisor $D$
\begin{eqnarray}
C^{\bullet} = i_{D*}\mathcal{M},
\end{eqnarray}
where $\mathcal{M}$ is line bundle on $D$, then the antidymmetrized product will be in the $(-1)$-th position
\begin{eqnarray}
\Lambda^2C^{\bullet} = i_{D*}\mathcal{M} i_{D*}\mathcal{M} &=& i_{D*}\left(\mathcal{M}\otimes_{\Lambda} (\mathcal{M}\oplus\mathcal{M}\otimes \mathcal{N}^*_{D/X} [+1])\right) \nonumber\\
&=& i_{D*}  \mathcal{M}^{\otimes 2}\otimes  \mathcal{N}^*_{D/X} [+1],
\end{eqnarray}
$ \mathcal{N}_{D/X}$ is the normal bundle of $D$. So the isomorphism \eref{higherisom} implies
\begin{eqnarray}\label{finalcoh}
\mathcal{H}^{-1}(Cone(f)) \simeq i_{D*}  \mathcal{M}^{\otimes 2}\otimes  \mathcal{N}^*_{D/X}.
\end{eqnarray}
Namely, the complex $Cone(f)$ is not exact on the middle term anymore. Using the following short exact sequence for the bundle $A$
\begin{eqnarray}
0\longrightarrow S^2A \longrightarrow A\otimes A \longrightarrow \Lambda^2A \longrightarrow 0
\end{eqnarray}
we can rewrite \eref{finalcoh} as

\begin{eqnarray}
\begin{tikzcd}[ampersand replacement=\&, column sep=small]
 \& \& 0\arrow[d] \& \& \\
 \& \& \Lambda^2 A\arrow[d] \& \& \\
0\arrow[r]\& i_{D*}  \mathcal{M}^{\otimes 2}\otimes  \mathcal{N}^*_{D/X} \arrow[r]\& Q \arrow[d]\arrow[r] \& \Lambda^2B \arrow[r]\& 0 .\\
 \& \& B\otimes i_{D*}\mathcal{M} \arrow[d] \& \& \\
 \& \& 0 \& \& 
\end{tikzcd}
\end{eqnarray}

\begin{itemize}
    \item $Li_{\sigma}^*\mathcal{O}_{S_2}(\sigma)\star\mathcal{O}_{S_2}(\sigma)$
\end{itemize}
Here instead of computing the Pontrjagin product we use the converse transform and compute the tensor product of the corresponding bundles and then we take the Fourier-Mukai transform of that. First to fix the notation, assume $[S_2]=2\sigma+\eta$.

It is easy to find the inverse transform of $\mathcal{O}_{S_2}(\sigma)$, which is denoted by $V$
\begin{eqnarray}
\begin{tikzcd}[ampersand replacement=\&]
0\arrow[r]\& \mathcal{O}_X(-\sigma)\arrow[r]\& V \arrow[r]\& \mathcal{O}_X(\sigma-\eta-K_B)\arrow[r]\& 0.
\end{tikzcd}
\end{eqnarray}
Now all we need is the Fourier transform of $V\otimes V$ and use the identity 
\begin{eqnarray}
Li_{\sigma}^*\Phi^1(V\otimes V)=Li_{\sigma}^* \mathcal{O}_{S_2}(\sigma)\star\mathcal{O}_{S_2}(\sigma)\otimes K_B^{-1}[-1].
\end{eqnarray}
We ignore the detailed calculations,\footnote{All we need are the Fourier transforms of $\mathcal{O}_X$, $\mathcal{O}_X(2\sigma)$ and $\mathcal{O}_X(-2s)$. They can be computed by using the defining short exact sequence for $\mathcal{I}_{\Delta}$ \cite{BBRH,Anderson:2019agu}} and we just write the final result for $\Phi(V\otimes V)$
\begin{eqnarray}
\begin{tikzcd}[ampersand replacement=\&]
 0\arrow[r]\&\mathcal{F}_2\arrow[r]\&\mathcal{F}_1\arrow[r]\&\Phi^1(V\otimes V)\arrow[r]\&0,
\end{tikzcd}
\end{eqnarray}
where $\mathcal{F}_1$ is defined as
\begin{eqnarray}
\begin{tikzcd}[ampersand replacement=\&]
 \&0\arrow[d] \& \& \& \\
 \&K_B^{-1}\arrow[d]\& \& \& \\
0\arrow[r]\& \mathcal{B}\arrow[d]\arrow[r] \&\mathcal{F}_1\arrow[r]\& \mathcal{O}_{\sigma}(-\eta)\arrow[r]\&0 \\
 \&\mathcal{O}_X(\sigma)\arrow[d] \& \& \& \\
 \&0 \& \& \& \\
\end{tikzcd}
\end{eqnarray}
and $\mathcal{F}_2$,
\begin{eqnarray}
\begin{tikzcd}[ampersand replacement=\&]
\& \&0\arrow[d] \& \& \\
0\arrow[r]\&\mathcal{F}_2\arrow[r]\&\mathcal{A}\otimes\mathcal{O}(-2\eta-2K_B)\arrow[r]\arrow[d] \&\mathcal{O}_{\sigma}(-\eta)\arrow[r]\&0\\
\& \&(\mathcal{O}\oplus K_B^2\oplus K_B^3)\otimes \mathcal{O}(\sigma-2\eta-3K_B)\arrow[d] \& \& \\
\& \&\mathcal{O}(4\sigma-2\eta-3K_B)\arrow[d] \& \& \\
\& \&0 \& \& 
\end{tikzcd}
\end{eqnarray}
Even though these diagrams look very difficult, the restriction on the zero section is quite easy
\begin{eqnarray}
\begin{tikzcd}[ampersand replacement=\&]
\mathcal{M}^{\bullet}\simeq\& 0\arrow[r]\&\begin{array}{c}
    \mathcal{O}_B(-K_B-\eta)   \\
     \oplus \\
    \mathcal{O}_B(-3\eta-2K_B) 
 \end{array} \arrow[r]\& \begin{array}{c}
      \mathcal{O}_B(\eta+K_B) \\
      \oplus\\
      \mathcal{O}_B(-\eta)
 \end{array}\arrow[r]\& 0,
\end{tikzcd}\nonumber \\
\quad
\end{eqnarray}
where the sequence $\mathcal{M}^{\bullet}$ on the left is not exact, and it is isomorphic (in derived category $D(B)$) to the pullback
$Li_{\sigma}^*\Phi^1(V\otimes V)$. This mean the zeroth cohomology $\mathcal{H}^0(\mathcal{M}^{\bullet})$ of this sequence gives $i_{\sigma}^*\Phi^1(V\otimes V)$, and the ``(-1)-th" cohomology $\mathcal{H}^{-1}(\mathcal{M}^{\bullet})$ gives $L^{-1}i_{\sigma}^*\Phi^1(V\otimes V)$. In addition to this, there is a surjection
\begin{eqnarray}
V\otimes V\longrightarrow \Lambda^2V=\mathcal{O}_X(-\eta-K_B)\longleftrightarrow 0,
\end{eqnarray}
which induces a surjection
\begin{eqnarray}
\mathcal{O}_B(\eta+K_B)\longrightarrow i_{\sigma}^*\Phi^1(V\otimes V)\longrightarrow \mathcal{O}_{\sigma}(-\eta)\longrightarrow 0.
\end{eqnarray}
On the other hand, the map between the line bundles $\mathcal{O}_V(-\eta-K_B)$ and $\mathcal{O}_V(\eta+K_B)$ can be induced by the extension morphism of the original sequence.\footnote{The extension group decomposes into two subgroups $Ext^1(\mathcal{O}(\sigma-\eta-K_B),\mathcal{O}(-\sigma)) = H^0(\mathcal{O}(\eta))\oplus H^0(\mathcal{O}(\eta+2K_B))$. The choice of the extension morphism uniquely fixes the algebraic equation of the double cover $S_2= a_2 X+a_0 Z^2$ and the coefficients are identified with the elements of the subgroups mentioned.} They could be $a_0 a_2$. But between the line bundles $\mathcal{O}_V(-3\eta-2K_B)$ and $\mathcal{O}_V(\eta+K_B)$ cannot be induced by the extension morphism. Since the bundle $V\otimes V$ is associated to $V$, we expect that the morphisms between the constituents of $V\otimes V$ is also induced by the morphisms that defined $V$, i.e. the extension morphism. So We conclude that $\mathcal{O}(-3\eta-2K_B)$ should be injected intro $\mathcal{H}^{-1}$. But whether $\mathcal{O}(-\eta-K_B)$ is inside $\mathcal{H}^{-1}$ and $\mathcal{O}(\eta+K_B)$ is inside $\mathcal{H}^0$ depends on the the morphism mentioned above. Therefore the final result is
\begin{eqnarray}
\begin{tikzcd}[ampersand replacement=\&]
0\arrow[r]\& \mathcal{O}(-3\eta-2K_B)\arrow[r]\& \mathcal{H}^{-1}\arrow[r]\arrow[d, phantom, ""{coordinate, name=Z}]\& \mathcal{O}(-\eta-K_B) \arrow[dll,to path={ -- ([xshift=2ex]\tikztostart.east)
|- (Z) [near end]\tikztonodes
-| ([xshift=-2ex]\tikztotarget.west)
-- (\tikztotarget)}] \& \\
\& \mathcal{O}(\eta+K_B)\arrow[r] \& \mathcal{H}^{0}\arrow[r]\& \mathcal{O}(-\eta)\arrow[r]\& 0.
\end{tikzcd}
\end{eqnarray}
Using the relations mentioned above, we can see that this is in complete agreement with our heuristic result in $SU(5)$ example. 

\begin{itemize}
    \item Using Pontrjagin Product For Double Cover With A Vertical Component.
\end{itemize}
Without loss of generality,
\begin{eqnarray}
\begin{tikzcd}[ampersand replacement=\&, column sep=small]
 \& \& \&0\arrow[d] \& \\
 \& \& \&\mathcal{L}'_1\arrow[d] \& \\
0\arrow[r] \& \mathcal{A}\arrow[r] \& \mathcal{L}_2\arrow[r] \&\mathcal{L}_1\arrow[r]\arrow[d] \& 0 \\
 \& \& \&\mathcal{L}_V \arrow[d] \& \\
 \& \& \&0 \& \\
\end{tikzcd}
\end{eqnarray}
where $\mathcal{A}$ and $\mathcal{L}'_1$ are line bundles supported over $\sigma$ and $\mathcal{L}_V$ is a line bundle over $V$. The parameters needed to define $\mathcal{L}$ are given by the extension groups,
\begin{eqnarray}
R^*Hom_{D(X)}(\mathcal{L}_1,\mathcal{A}) = R^* Hom_{D(B)}(Li^*_{\sigma}\mathcal{L}_1, \mathcal{A}) =  R^* Hom_{D(B)} (\mathcal{L}'_1\otimes K_B^{-1} [+1]\oplus \mathcal{L}_1|_{\sigma},\mathcal{A})\\
R^*Hom_{D(X)}(\mathcal{L}_V,\mathcal{L}'_1)= R^*Hom_{D(c)} (\mathcal{L}_V|_c,\mathcal{L}'_1|_c\otimes \mathcal{O}_c(c)[-1]),
\end{eqnarray}
where $c$ in the last line is the intersection of the vertical and horizontal components, and this extension group parameterizes  the gluing. There are three parameters in $Ext^1(\mathcal{L}_1,\mathcal{A})$,
\begin{eqnarray}
&\mathcal{E}xt\in  Hom_{D(B)}(\mathcal{L}'_1\otimes K_B^{-1},\mathcal{A})&,\nonumber \\
&Ext^1_{D(B)}(\mathcal{L}'_1,\mathcal{A})&,\nonumber \\
&Ext^1_{D(B)}(i_{\sigma}^*\mathcal{L}_V,\mathcal{A})=Hom_{D(c)} (\mathcal{L}_V,\mathcal{A}\otimes \mathcal{O}_c(c))&
\end{eqnarray}
The first group parameterizes a line bundle over the non-reduced components. The second one gives a rank two bundle over $\sigma$, and the third group corresponds to gluing on the intersection. We turn the second factor off, so we don't get the a vector bundle over $B$. Also for finding $R\pi_*V$ we need to restricted the spectral data on the zero section,

\begin{eqnarray}
\begin{tikzcd}[ampersand replacement=\&, column sep=small]
0\arrow[r]\& \mathcal{A} \otimes K_B^{-1} \arrow[r]\& L^{-1}i^*_{\sigma}\mathcal{L}_2 \arrow[r] \& \mathcal{L}'_1 \otimes K_B^{-1} \\
\arrow[r,hook]\& \mathcal{A}\arrow[r]\&  L^{0}i^*_{\sigma}\mathcal{L}_2\arrow[r] \& \mathcal{L}_1|_{\sigma} \arrow[r] \& 0.
\end{tikzcd}
\end{eqnarray}

\begin{eqnarray}
R\pi_* V = \mathcal{A} \otimes K_B^{-1} \oplus \mathcal{R}[-1], \\
\begin{tikzcd}[ampersand replacement=\&, column sep=small]
0\arrow[r]\& \mathcal{A}|_{\mathcal{E}xt=0} \arrow[r]\& \mathcal{R}2\arrow[r]\& \mathcal{L}_1|_{\sigma} \arrow[r] \& 0. 
\end{tikzcd}
\end{eqnarray}
The Extension for this sequence is induced by $Ext_{D(B)}^1(\mathcal{L}'_1,\mathcal{A})$ restricted on $\mathcal{E}xt=0$. But we chose this to be zero everywhere. Also note that there is a topological constraint on this as before,
\begin{eqnarray}
c_1(\mathcal{A})+c_1(\mathcal{L}'_1)+[V]=2 c_1(K_B).
\end{eqnarray}
There is also a smoothness condition, that is $\pi_*V \sim i_{\sigma}^*\mathcal{L}_2$ must be locally free (smooth coherent sheaf). Also from the reasons mentioned before $i_V^*\mathcal{L}_2$ must have relative +1. Putting these requirements together one gets $\mathcal{L}_V$,
\begin{eqnarray}
\mathcal{L}_V = \mathcal{O}_V(\sigma+K_B-c_1(\mathcal{A})).
\end{eqnarray}

Now we can proceed to find the Fourier transform of $\Lambda^2V$. To compute $\mathcal{L}_2\star_A\mathcal{L}_2$ we should be careful. This is because of the existence of the vertical components, which is already mentioned. To see how to deal with this, first note that,

\begin{eqnarray}
\mathcal{L}_V\star_A\mathcal{L}_V \simeq_{Qis} 0\longrightarrow \mathcal{L}'_1\star_S \mathcal{L}'_1 \longrightarrow \mathcal{L}'_1\star \mathcal{L}_1 \longrightarrow \mathcal{L}_1 \star_A \mathcal{L}_1 \longrightarrow 0.
\end{eqnarray}
Instead of computing $\mathcal{L}_V\star_A\mathcal{L}_V$ directly, we can compute the antisymmetrized product of the inverse Fourier transform. In other words, we use, 

\begin{eqnarray}
\mathcal{L}_V\star_A\mathcal{L}_V = i_{V*} \mathcal{L}_V \star \mathcal{L}_V \otimes \mathcal{O}(-V) [+1] = i_{V*}\Phi \left(\mathcal{O}_V \right)(-2\sigma+c_1(\mathcal{L}'_1)-c_1(\mathcal{A}))\otimes K_B[+2]. 
\end{eqnarray}
It is easy to compute the right hand side,
\begin{eqnarray}
\begin{tikzcd}[ampersand replacement=\&, column sep=small]
0\arrow[r]\& K_B^{-1}\arrow[r]\& \mathcal{G}\arrow[r]\& \mathcal{O}(\sigma) \arrow[r]\& 0
\end{tikzcd}\\
\mathcal{L}_V\star_A\mathcal{L}_V = i_{V*} \mathcal{G}\otimes \mathcal{O}(+K_B+c_1(\mathcal{L}'_1)-c_1(\mathcal{A})) [+1]
\end{eqnarray}
Therefore one gets a diagram as follows,

\begin{eqnarray}
\begin{tikzcd}[ampersand replacement=\&, column sep=small]
 \& 0\arrow[d] \& 0\arrow[d] \& \& \\
 \& \mathcal{L}'_1\star_S\mathcal{L}'_1\arrow[r, "\simeq"]\arrow[d] \& \mathcal{L}'_1\star\mathcal{L}'_1 \arrow[d] \& \& \\
0\arrow[r] \& Q \arrow[d]\arrow[r] \& \mathcal{L}'_1\star\mathcal{L}_1 \arrow[d]\arrow[r] \& \mathcal{L}_1\star_A\mathcal{L}_1 \arrow[r] \& 0 \\
 \& \mathcal{L}_V\star_A \mathcal{L}_V \arrow[d]\arrow[r]\& \mathcal{L}'_1\otimes \mathcal{L}_V \arrow[d]\arrow[r] \& 0 \& \\
 \& 0 \& 0 \& \&
\end{tikzcd}
\end{eqnarray}

Putting every thing together we get the result for $\mathcal{L}_1\star_A \mathcal{L}_1$,
\begin{eqnarray}
\mathcal{L}_1\star_A \mathcal{L}_1 = \mathcal{O}_V(c_1(\mathcal{L}'_1)-c_1(\mathcal{A}))[+1].
\end{eqnarray}

The idea that we want to emphasize is that we can continue this approach for finding the ``antisymmetric" Pontrjagin product of spectral sheaves of higher degrees. So when we add another sheet, one gets the following quasi-isomorphism,

\begin{eqnarray}
&&\mathcal{L}_1 \star_A \mathcal{L}_1 \simeq_{Qis} \nonumber \\
&&\begin{tikzcd}[ampersand replacement=\&, column sep=small]
0\arrow[r] \& \mathcal{A} \star_S \mathcal{A} \arrow[r]\& \pi^*\pi_*\mathcal{A}\otimes \mathcal{L}_2 \arrow[r]\& \mathcal{L}_2\star_A\mathcal{L}_2 \arrow[r]\& 0.
\end{tikzcd}
\end{eqnarray}
Similar to the previous part, this quasi isomorphism is equivalent to the following diagram,
\begin{eqnarray}
\begin{tikzcd}[ampersand replacement=\&, column sep=small]
 \& 0\arrow[d] \& 0\arrow[d] \& \& \\
 \& \mathcal{A}\star_S\mathcal{A}\arrow[r, "\simeq"]\arrow[d] \& \mathcal{A}\star\mathcal{A} \arrow[d] \& \& \\
0\arrow[r] \& Q \arrow[d]\arrow[r] \& \mathcal{A}\star\mathcal{L}_2 \arrow[d]\arrow[r] \& \mathcal{L}_2\star_A\mathcal{L}_2 \arrow[r] \& 0 \\
 \& \mathcal{L}_1\star_A \mathcal{L}_1 \arrow[d]\arrow[r]\& \mathcal{A}\otimes \mathcal{L}_1 \arrow[d] \& \& \\
 \& 0 \& 0 \& \&
\end{tikzcd}
\end{eqnarray}
Again by putting the pieces together and chasing the diagrams, one gets the final result, (if $c_1(\mathcal{A})-c_1(\mathcal{L}'_1+c_1(K_B))=0$, which is induced again by the stability of $V_2$) one gets,

\begin{eqnarray}
\mathcal{L}_2\star_A\mathcal{L}_2 = \mathcal{O}_{\sigma}\otimes K_B^{\otimes 2}.
\end{eqnarray}
This consistent with the expectation of $\Lambda^2V=\mathcal{O}_X$.

\end{document}